\documentclass[a4paper,11pt]{article}

\usepackage[margin=1in]{geometry}
\usepackage[T1]{fontenc}
\DeclareFontShape{T1}{cmr}{m}{scit}{<->ssub*cmr/m/scsl}{}
\usepackage{microtype}
\usepackage{amsmath}
\usepackage{amssymb}
\usepackage{mathtools}
\usepackage{amsthm}
\usepackage{thmtools}
\usepackage{thm-restate}

\usepackage{graphicx}
\usepackage{xcolor}
\usepackage{caption}
\usepackage{subcaption}
\usepackage{tabularx}
\usepackage{tcolorbox}
\usepackage{booktabs}
\usepackage{array}
\usepackage{algorithm}
\usepackage{algpseudocode}

\usepackage{enumitem}
\setlist[itemize,enumerate]{noitemsep}

\usepackage{tikz}
\usetikzlibrary{
  arrows.meta,
  backgrounds,
  calc,
  decorations.pathreplacing,
  fit,
  positioning,
  shapes.geometric, bending
}

\definecolor{linkblue}{HTML}{1A4E8A}
\definecolor{citegreen}{HTML}{2F6B3F}
\definecolor{urlmagenta}{HTML}{7A3E7A}

\usepackage[
  colorlinks=true,
  linkcolor=linkblue,
  citecolor=citegreen,
  urlcolor=urlmagenta
]{hyperref}

\usepackage[nameinlink]{cleveref}

\theoremstyle{plain}
\newtheorem{theorem}{Theorem}[section]
\newtheorem{lemma}[theorem]{Lemma}
\newtheorem{proposition}[theorem]{Proposition}
\newtheorem{observation}[theorem]{Observation}
\newtheorem{corollary}[theorem]{Corollary}
\newtheorem{claim}[theorem]{Claim}
\newtheorem{example}[theorem]{Example}
\newtheorem{remark}[theorem]{Remark}

\theoremstyle{definition}
\newtheorem{definition}[theorem]{Definition}

\Crefname{observation}{Observation}{Observations}
\Crefname{claim}{Claim}{Claims}

\newenvironment{claimproof}[1][\proofname]
  {\proof[#1]}
  {\endproof}

\algnewcommand\Input{\item[\textbf{Input:}]}
\algnewcommand\Output{\item[\textbf{Output:}]}
\algtext*{EndIf}
\algrenewtext{EndIf}{\unskip}

\makeatletter
\def\namedlabel#1#2{%
  \begingroup
  #2%
  \def\@currentlabel{#2}%
  \phantomsection\label{#1}%
  \endgroup
}
\makeatother

\usepackage{xspace}
\newcommand{\mcm}[3]{%
  \newcommand{#1}[#2]{\ensuremath{#3}}%
}

\newcommand{\bigoh}{\mathcal{O}}

\newcommand{\Nbb}{\mathbb{N}}

\newcommand{\Dcal}{\mathcal D}

\newcommand{\Ical}{\mathcal I}

\newcommand{\Pcal}{\mathcal P}

\newcommand{\Rcal}{\mathcal R}

\newcommand{\Xcal}{\mathcal X}

\newcommand{\CC}{\mathcal C}

\newcommand{\np}{\textsf{NP}}

\newcommand{\arcDoct}{\textsc{Arc DOCT}\xspace}
\newcommand{\arcdoct}{\arcDoct{}\xspace}
\newcommand{\vertexDoct}{\textsc{Vertex DOCT}\xspace}

\newcommand{\yes}{\textsc{Yes}\xspace}
\newcommand{\no}{\textsc{No}\xspace}

\newcommand{\tail}{\mathsf{tail}}
\newcommand{\head}{\mathsf{head}}

\newcommand{\safeExit}[2]{\ensuremath{\mathsf{safe}_{#1}(#2)}\xspace}
\newcommand{\blockedExit}[2]{\ensuremath{\mathsf{blocked}_{#1}(#2)}\xspace}
\newcommand{\contact}[2]{\ensuremath{\mathsf{contact}_{#1}(#2)}\xspace}

\newcommand{\defprob}[3]{%
  \begin{tcolorbox}[colback=white,boxrule=0.3pt,arc=0.3pt]
    \begin{minipage}{0.99\textwidth}
      \begin{tabular*}{\textwidth}{@{\extracolsep{\fill}}ll}
        #1
      \end{tabular*}

      \textbf{Input:} #2

      \textbf{Question:} #3
    \end{minipage}
  \end{tcolorbox}
}

\tikzset{
  thickred/.style={
    draw=red,
    line width=.20pt,
    double distance=.65pt,
    -Stealth
  },
  startstop/.style={
    rounded corners=9pt,
    draw=black!75,
    fill=blue!7,
    line width=0.8pt,
    align=center,
    text width=8.7cm,
    inner sep=6pt,
    font=\small
  },
  process/.style={
    draw=black!75,
    fill=black!4,
    line width=0.7pt,
    align=center,
    text width=8.7cm,
    inner sep=6pt,
    font=\small
  },
  nestbox/.style={
    rounded corners=4pt,
    draw=green!45!black,
    fill=green!7,
    line width=0.7pt,
    align=left,
    text width=5.3cm,
    inner sep=6pt,
    font=\small
  },
  crossbox/.style={
    rounded corners=4pt,
    draw=blue!55!black,
    fill=blue!6,
    line width=0.7pt,
    align=left,
    text width=5.3cm,
    inner sep=6pt,
    font=\small
  },
  decision/.style={
    diamond,
    aspect=2.4,
    draw=black!75,
    fill=orange!12,
    line width=0.7pt,
    align=center,
    text width=3.4cm,
    inner sep=2pt,
    font=\small
  },
  flow/.style={
    ->,
    >=Stealth,
    line width=0.8pt,
    draw=black!70
  },
  ihflow/.style={
    ->,
    >=Stealth,
    line width=0.7pt,
    draw=red!65!black,
    dashed
  },
  elab/.style={
    font=\scriptsize,
    fill=white,
    inner sep=1.5pt
  }
}

\title{Important Cuts with Logical Constraints: A Meta-Theorem for Eulerian Digraphs}
\author{V\'aclav Bla\v{z}ej\thanks{vaclavblazej@seznam.cz}
  \and Satyabrata Jana\thanks{IISER Berhampur (satyabrataj@iiserbpr.ac.in).}
  \and J\k{e}drzej Olkowski\thanks{University of Warwick (Jedrzej.Olkowski@warwick.ac.uk, R.Maadapuzhi-Sridharan@warwick.ac.uk, Peter.Strulo@warwick.ac.uk).}
  \and M. S. Ramanujan\footnotemark[3]
  \and Peter Strulo\footnotemark[3]}
\date{}

\hypersetup{pageanchor=false}
\graphicspath{{figures/}}

\begin{document}

\maketitle
\begin{abstract}
Important separators are a standard tool in parameterized cut algorithms.
An $S$-$T$ separator is important if it is inclusion-wise minimal and no
separator of at most the same size leaves a strict superset of the vertices
reachable from $S$ after deletion. 
Their use relies on a boundedness theorem: for fixed $k$, the number of 
 important $S$-$T$ separators of size at most $k$ is at most a constant ($4^k$ to be precise) independent of the input size, even in directed
graphs, and they can be enumerated efficiently. The classical theory,
however, controls only the size of the separator, not the structure of the
source side cut off by it. Once the source side is required to satisfy a
condition such as acyclicity, bipartiteness, or another logical property,
the corresponding maximal source sides can be unbounded in general
digraphs. This obstruction is fundamentally related to hardness phenomena in
directed deletion problems such as \textsc{Directed Odd Cycle
Transversal}. Prior to this work, even in undirected graphs, boundedness of such maximal constrained source sides was known only in more specialized settings, not at the level of generality considered in our work.

In this paper, we prove a source-side analogue of the important-separator bound for
balanced digraphs. A digraph is $\alpha$-balanced if, for every vertex set,
the numbers of arcs entering and leaving the set differ by at most $\alpha$.
Eulerian digraphs are exactly the $0$-balanced digraphs. We prove that for every fixed cut budget $k$, balance parameter $\alpha$,
and Counting Monadic Second Order (CMSO) sentence $\pi$ expressing a hereditary digraph property, there are
only boundedly many inclusion-wise maximal source-side regions whose outgoing
arcs form an inclusion-wise minimal $S$-$T$ cut of size at most $k$, and
whose induced subgraph satisfies $\pi$. We call these regions important
chips. When $\pi$ is the trivial property, their outgoing arc-sets recover the usual important $S$-$T$ cuts.

Under a natural testability assumption on $\pi$, our proof also gives a
non-uniform fixed-parameter tractable (FPT) algorithm for enumerating all
important chips: for each
fixed choice of parameters, it runs in polynomial time with a
parameter-independent exponent and may use parameter-specific advice.  
We further isolate a sufficient condition that makes this  algorithm uniform. 
As an application, we prove that the arc-deletion variant of
\textsc{Directed Odd Cycle Transversal} (called \arcdoct{}) is fixed-parameter tractable on
balanced digraphs parameterized by the solution size and the balance,
and hence on Eulerian digraphs parameterized by the solution size. This contrasts with the W[1]-hardness of the problem on general digraphs. Finally, we show that the framework also yields boundedness and enumeration results for analogous notions of  important edge
chips and vertex chips in undirected graphs.
\end{abstract}

\newpage
\tableofcontents
\newpage
\newcommand{\mk}{\pi}
 \newcommand{\stpik}{\ensuremath{(S,T,\mk,k)}\xspace}
 \pagenumbering{arabic}
 \hypersetup{pageanchor=true}
 \section{Introduction}\label{sec:intro}

Important separators, introduced by Marx~\cite{Marx06}\footnote{Marx's
paper was later recognized with an EATCS-IPEC Nerode Prize.}, are one of the
standard tools in parameterized cut algorithms. An $S$-$T$
separator is important if it is inclusion-wise minimal and no separator
of at most the same size leaves a strict superset of the vertices reachable
from $S$ after deletion. The key
fact is that, for any two terminal sets $S$ and $T$, there are at most
$4^k$ important $S$-$T$ separators of size at most $k$, even in directed
graphs, and all of them can be enumerated in time $\bigoh(4^kk(m+n))$.

Many fixed-parameter algorithms (FPT algorithms) use this fact in the same way. After identifying
suitable terminal sets $S$ and $T$, one proves that some optimal solution contains, or
intersects, a separator from this bounded family. One can then enumerate the
bounded family and branch on its elements. This mechanism underlies algorithms for problems such as
\textsc{Multiway Cut}~\cite{Marx06},
\textsc{Directed Feedback Vertex Set}~\cite{ChenLLOR08}, and
\textsc{2-SAT Deletion}~\cite{RazgonO09}. It also appears in more
sophisticated forms, including the shadow-removal and random-sampling
frameworks used for \textsc{Multicut}~\cite{MarxR14},
\textsc{Directed Multiway Cut}~\cite{ChitnisHM13}, and
\textsc{Directed Subset Feedback Vertex Set}~\cite{ChitnisCHM15}. We point the reader to the surveys \cite{Marx2011ImportantSeparators,Chitnis2016} and textbook \cite{CyganFKLMPPS2015ParameterizedAlgorithms} for more examples. 

The classical important-separator theorem controls the size of the separator,
but not the structure of the source side behind it. This distinction becomes
critical in some directed deletion problems. There, one often needs the reachable
side of a cut to satisfy an additional condition, such as acyclicity or more generally, a
property definable in a logic fragment. Once such a condition is
imposed, the number of maximal reachable sides of cut size at most $k$ can
be unbounded in general digraphs. This unboundedness is exploited by hardness constructions
for directed deletion problems such as \textsc{Directed Odd Cycle
Transversal} on general digraphs, for instance, in the form of the forward clock
gadget in~\cite{LokshtanovRSZ20}. Several recent works have also generalized important separators in other
directions, mostly in undirected settings, and we compare these developments
with our framework in \Cref{subsec:relatedImportantSeparators}.

\subsection{Our results}
The main contribution of this paper is a source-side analogue of the
important-separator bound for balanced digraphs. A digraph is
$\alpha$-balanced if, for every vertex set, the numbers of arcs entering and
leaving the set differ by at most $\alpha$. Eulerian digraphs are exactly
the $0$-balanced digraphs.

Counting monadic second-order logic (CMSO) allows quantification over
vertices, arcs, and sets of vertices or arcs, together with modulo-counting
predicates on set sizes. A digraph property is hereditary if it is closed
under taking induced subgraphs.

In this setting we prove a bounded-family theorem for reachable regions 
behind small outgoing cuts, even when those regions are required to satisfy
an arbitrary hereditary CMSO constraint. A single proof applies to acyclicity,
bipartiteness of the underlying graph, edgelessness, $H$-freeness for any
fixed graph or digraph $H$, absence of directed odd cycles, and so on. We call the resulting objects important chips.

The results have three levels. First, we prove a combinatorial bound on the
number of important chips, showing that their number is independent of the input size. Second, we turn the proof into an enumeration
algorithm, initially non-uniform because it 
assumes bounded advice to the algorithm, meaning a finite list of graphs that depends only on $k$, $\alpha$, and $\pi$ and is supplied to the algorithm rather than computed from the input. Third, we apply the framework to a directed deletion problem,
namely the arc-deletion version of \textsc{Directed Odd Cycle Transversal}.
The application first gives a non-uniform FPT algorithm on balanced
digraphs, which then becomes uniform using computable representatives satisfying
weaker preservation conditions. The undirected edge-cut and
vertex-separator results at the end are consequences and variants of the
same framework.

\paragraph{Important chips in balanced digraphs.}
Given a digraph $D$, disjoint vertex sets $S$ and $T$ of $D$, a cut budget $k$, and a CMSO
sentence $\pi$, an $(S,T,\pi,k)$-chip is a vertex set $Z$ such that
$S\subseteq Z$, $Z\cap T=\emptyset$, every vertex of $Z$ is reachable from
$S$ in $D[Z]$, the outgoing boundary $\partial^+(Z)$, which is the set of arcs with tail in $Z$ and head outside $Z$, is an inclusion-wise
minimal $S$-$T$ cut of size at most $k$, and $D[Z]\models \pi$, that is, $D[Z]$ satisfies $\pi$. Here an $S$-$T$ cut is a set of arcs whose deletion leaves no directed path from $S$ to $T$. An
\emph{important} $(S,T,\pi,k)$-chip is one that is inclusion-wise maximal
subject to not increasing the cut size. 
Thus, just as classical important separators can be viewed through their
source sides as maximal elements of a poset, important chips are the
maximal chips under the order $Z\preceq Z'$ if
$Z\subseteq Z'$ and
$|\partial^+(Z')|\le |\partial^+(Z)|$. So, when $\pi$ expresses the trivial property
satisfied by all digraphs, the outgoing boundaries of important chips
recover the usual important $S$-$T$ cuts, which are the arc analogue of important separators.

Our central combinatorial result is the following.

\begin{restatable}{theorem}{thmboundImpChips}
\label{thm:boundImpChips}
For every $k,\alpha\in \mathbb{N}_0$ and hereditary CMSO sentence $\pi$,
there is a constant $c_{k,\pi,\alpha}\in \mathbb{N}$ such that, for every
$\alpha$-balanced digraph $D$ and disjoint non-empty vertex sets $S$ and
$T$, the number of important $(S,T,\pi,k)$-chips in $D$ is at most
$c_{k,\pi,\alpha}$.
\end{restatable}
%

Since Eulerian digraphs are $0$-balanced, the theorem
implies the Eulerian case.

The proof uses finite-state replacement for CMSO on boundaried digraphs.
A boundaried digraph has distinguished boundary vertices with distinct
labels. Two such digraphs are compatible if their boundary label sets
coincide. Their gluing $D_1\oplus D_2$ takes disjoint copies, identifies
equally labeled boundary vertices, and retains the arcs of both inputs. For
a fixed boundary size and a fixed CMSO sentence $\pi$, boundaried digraphs
fall into finitely many equivalence classes according to their effect on
the truth of $\pi$ after gluing to an arbitrary compatible boundaried digraph. We fix
a smallest representative from each relevant class. Once this finite list is
fixed, the proof becomes constructive. The enumeration algorithm is
therefore non-uniform. For fixed $k,\alpha,\pi$, the representative list is
treated as advice to the algorithm rather than computed from the input.
  This gives the following non-uniform enumeration theorem.

\begin{restatable}{theorem}{thmenumerateBoundImpChips}
\label{thm:enumerateBoundImpChips}
For every $k,\alpha\in \mathbb{N}_0$ and hereditary CMSO sentence $\pi$
that is efficiently testable parameterized by the balance, there is an
algorithm that, given an $\alpha$-balanced digraph $D$ and disjoint
non-empty vertex sets $S$ and $T$, runs in time
$c'_{\pi,k,\alpha}\cdot n^{\bigoh(1)}$ and outputs the set of important
$(S,T,\pi,k)$-chips in $D$. Here, the exponent of $n$ is independent of $k$ and $\alpha$.
\end{restatable}

Here, efficient testability means that, given a candidate set or subgraph, we can decide whether it satisfies $\pi$ with a (not necessarily uniform) FPT algorithm parameterized by the balance. \Cref{def:efficientTestability} gives the precise running-time condition, together with the uniform version used in \Cref{thm:enumerateChipsWeakRepresentatives}.
The efficient testability assumption is used to recognize candidate chips during
the enumeration. The cut and reachability conditions are checked by standard
polynomial-time routines.

\paragraph{Directed Odd Cycle Transversal.}
We next instantiate the framework for \textsc{Directed Odd Cycle
Transversal}. The arc-deletion variant, \textsc{Arc DOCT}, asks whether one can delete at most
$k$ arcs so that the resulting digraph has no directed odd cycle. The
vertex-deletion variant is defined analogously, with vertices deleted
instead of arcs. The vertex-deletion problem is $\mathrm{W}[1]$-hard on
general digraphs~\cite{LokshtanovRSZ20}, and this hardness persists even
on Eulerian digraphs. The arc-deletion problem is also
$\mathrm{W}[1]$-hard on general digraphs. In contrast, our framework shows
that the arc-deletion problem becomes fixed-parameter tractable on balanced digraphs, and specifically Eulerian digraphs.

\begin{restatable}{theorem}{thmgeneralBalancedDOCT}
\label{thm:generalBalancedDOCT}
For every $k,\alpha \in \mathbb{N}_0$, there is an algorithm that, given an
$\alpha$-balanced digraph $D$, runs in time
$c''_{k,\alpha}\cdot n^{\bigoh(1)}$ and decides whether $(D,k)$ is a
yes-instance of \textsc{Arc DOCT}.
\end{restatable}

\paragraph{Weak representatives and uniformity.}
The enumeration theorem uses finite-state CMSO representatives as advice.
They exist, but may not be computable. 

For the uniform algorithm we isolate the weaker replacement property actually
used in our proof. The finite-state property is used only for two one-sided checks. After replacing
a part of a candidate chip, the chip must still satisfy $\pi$, and any
larger candidate chip found after the replacement must correspond to a larger
candidate before the replacement. 

We now define weak representatives, which provide a computable replacement for the advice.
Call a boundaried digraph boundary-clean if it has no arc with both
endpoints in its boundary. For $b\in\mathbb N_0$, call it $b$-boundaried if
its boundary labels belong to $\{1,\ldots,b\}$.
A hereditary property $\pi$ has computable weak representatives if there
are a computable function $f_\pi$ and an algorithm that, given only $b$,
outputs in time $f_\pi(b)$ a finite family $\mathcal R_{\pi,b}$ of
$b$-boundaried digraphs, each with at most $f_\pi(b)$ vertices and at most
$f_\pi(b)$ arcs, with the following property. For every compatible pair
$D_1,D_2$ of $b$-boundaried digraphs such that $D_1$ is boundary-clean and
$D_1\oplus D_2\models\pi$, there is a compatible $R\in\mathcal R_{\pi,b}$
such that $D_1\oplus R\models\pi$ and, for every compatible boundary-clean
$b$-boundaried digraph $D_1'$,
\[
D_1'\oplus R\models\pi \quad\Longrightarrow\quad D_1'\oplus D_2\models\pi.
\]

\begin{restatable}{theorem}{thmenumerateChipsWeakRepresentatives}
\label{thm:enumerateChipsWeakRepresentatives}
Let $\pi$ be a hereditary CMSO-definable digraph property that has
computable weak representatives and is uniformly efficiently testable
parameterized by the balance. Then there is
an algorithm that, given $k,\alpha\in\mathbb{N}_0$, an
$\alpha$-balanced digraph $D$ and disjoint non-empty vertex sets
$S,T\subseteq V(D)$, runs in time $g_\pi(k,\alpha)\cdot n^{\bigoh(1)}$ and
outputs the set of important $(S,T,\pi,k)$-chips in $D$, where $g_\pi$ is a computable function.
\end{restatable}

Thus weak representatives separate the boundedness meta-theorem from the
property-specific work needed for uniformity. We carry out this
property-specific step for \arcDoct{}.

For \arcDoct{}, we care about the family of properties
$\{\pi_\ell \mid \ell\in{\mathbb N}_0\}$, where $\pi_\ell$ states that at
most $\ell$ arcs can be deleted to remove all directed odd cycles. We prove
in \Cref{sec:weakRep} that these properties have computable weak
representatives, uniformly in $\ell$ and in the boundary size.

The weak-representative theorem also needs a uniform test for $\pi_\ell$
on balanced digraphs. For \arcDoct{},
this test is obtained at the same time as the decision algorithm, by
induction on the deletion budget. A call with current budget $k$ only
enumerates chips for properties $\pi_\ell$ with $\ell<k$.

\begin{restatable}{theorem}{thmuniformBalancedArcDoct}
\label{thm:uniformBalancedArcDoct}
There is an algorithm which, given an $\alpha$-balanced digraph $D$ and an
integer $k$, decides whether $D$ has an arc directed odd cycle transversal
of size at most $k$ in time $f(k,\alpha)\cdot n^{\bigoh(1)}$ for a
computable function $f$.
\end{restatable}

In particular, setting $\alpha=0$ gives a uniform FPT algorithm for
\arcDoct{} on Eulerian digraphs, parameterized by the solution size.


%
%

\paragraph{Undirected edge cuts and vertex separators.}
We also obtain two undirected variants of the framework. These results are
not needed for the \arcDoct{} application, but show that the important-chip
viewpoint is not tied to balanced digraphs. 

 For edge cuts, an
\stpik-edge chip is defined analogously to the directed version, except with the cut size measured by the
number of undirected edges leaving the source side. For vertex separators,
an \stpik-vertex chip is defined analogously, requiring the open
neighborhood of the source side to be an inclusion-wise minimal $S$-$T$
vertex separator disjoint from $S\cup T$, and measuring the cut size by
the cardinality of that neighborhood.  Importance is always defined by
inclusion-wise maximality subject to not increasing the relevant cut size.

For edge chips, the result follows by replacing every undirected edge by
two opposite arcs. The resulting digraph is Eulerian, and CMSO sentences
over undirected graphs can be translated to CMSO sentences over the
bidirected graph.

\begin{restatable}{corollary}{corundirectedEdgeChipBound}
\label{cor:undirectedEdgeChipBound}
For every $k\in\mathbb{N}_0$ and every hereditary CMSO sentence $\pi$ over
undirected graphs, there is a constant $c^e_{k,\pi}\in\mathbb N$ such that,
for every undirected graph $G$ and disjoint non-empty vertex sets
$S,T\subseteq V(G)$, the number of important $(S,T,\pi,k)$-edge chips in
$G$ is at most $c^e_{k,\pi}$.
\end{restatable}

For vertex chips there is no analogous bidirected reduction. We retain the
counting scheme of the main theorem, using vertex neighborhoods in place of
outgoing arc boundaries. The additional charging and lifting arguments,
which preserve a separate strictness witness for each charged boundary
vertex, are given in \Cref{subsec:undirectedVertexSeparators}.


\begin{restatable}{claim}{thmundirectedVertexChipBound}
\label{thm:undirectedVertexChipBound}
For every $k\in\mathbb{N}_0$ and every hereditary CMSO sentence $\pi$ over
undirected graphs, there is a constant $c^v_{k,\pi}\in\mathbb N$ such that,
for every undirected graph $G$ and disjoint non-empty vertex sets
$S,T\subseteq V(G)$, the number of important $(S,T,\pi,k)$-vertex chips in
$G$ is at most $c^v_{k,\pi}$.
\end{restatable}

%
In both cases, the important chips can be enumerated in time
$f(k,\pi)\cdot n^{\bigoh(1)}$ whenever $\pi$ can be tested in polynomial time
on undirected graphs.
\subsection{Why balanced digraphs?}

Eulerian digraphs are the most natural starting point. They include all
bidirected undirected graphs. Arc deletion need not preserve Eulerianness,
so reachability and parity conditions in the resulting digraph remain
genuinely directed.
Thus, Eulerian digraphs lie between undirected and fully directed cut
problems, and some cut problems are easier on them than on general digraphs. For example, \textsc{Edge-Disjoint Paths} is FPT on Eulerian
digraphs~\cite{CavallaroKK24}, while it is NP-hard already for two paths on general digraphs. Our results add a new meta-algorithmic
framework to this picture.

Our proof uses the more general class of balanced digraphs. In a general digraph, a set $Z$ may have small outgoing boundary
$\partial^+(Z)$ and arbitrarily large incoming boundary $\partial^-(Z)$ or vice-versa.
Thus, a small directed cut need not give a bounded interface between the two
sides of the cut. Balance controls this asymmetry. If the outgoing boundary
of a set is small, then its incoming boundary is also bounded up to the
additive balance parameter. This gives the bounded interface needed before any finite-state replacement
argument can be attempted.

However, this observation does not by itself make the  argument go through.
The proof repeatedly passes to auxiliary and reduced instances, and these
operations need not preserve the original balance parameter. A crucial part
of the proof is to show that the relevant imbalance remains bounded
throughout the induction. In this sense, balanced digraphs are not merely a
convenient generalization of Eulerian digraphs. They provide the invariant
needed to make the source-side analogue of important separators stable under
the reductions used in the proof.

\subsection{Roadmap of the proofs}

\label{subsec:technicalRoadmap}

The counting argument seeks a bounded set $F$ of arcs that intersects the
outgoing arc set of every important chip, apart from at most one chip counted
separately. We call $F$ a \emph{hitting set}. A chip is \emph{handled} by $F$
when one of its outgoing arcs belongs to $F$. Deleting such an arc reduces
the cut budget, so a bounded hitting set reduces the counting problem to
boundedly many smaller-budget instances.

To construct $F$, we compare chips with a nested sequence of minimum-cut
source sides. The two recursive cases either increase the minimum cut
value or reduce the budget. Only the second case uses finite-state CMSO
replacement.

\subsubsection{The combinatorial bound}
The important-chip theorem is stated for ordinary digraphs, but the proof is
more naturally carried out in a slightly richer setting. We use
\emph{routed digraphs}, whose arcs are divided into \emph{real} arcs and
\emph{auxiliary} arcs. Real arcs come from the original input. Paths and
reachability may use both kinds of arcs, while cuts, cut sizes, boundaries,
hitting sets, balance, and the sentence $\pi$ are evaluated only on real
arcs. Auxiliary arcs are bookkeeping devices introduced by the inductive
reductions.

An ordinary digraph is a routed digraph with no auxiliary arcs, so the main
theorem can be read with real arcs interpreted simply as arcs. 


We prove the bound through a slightly more general statement, on augmented
routed instances
\[
(\widehat D,\allowbreak S,\allowbreak T_1,\allowbreak T_2,\allowbreak Q,\allowbreak \pi,\allowbreak k).
\]
Here $\widehat D$ is the routed digraph and $T_1$ is the target set for the cut, playing the role of $T$ in
\Cref{thm:boundImpChips}. The sets $T_2$ and $Q$ are empty in the initial
ordinary instance and are used only by the recursion. Vertices in $T_2$ are
forbidden to be part of the chip. Vertices in $Q$ are \emph{requests}: after the
chip is deleted, they must still be able to reach $T_1$.

These two sets record reachability information that may be lost when the
proof restricts to a local region or replaces part of a chip by a
representative. This bookkeeping preserves the requirement that the outgoing
boundary of the chip is an inclusionwise minimal cut from the source to the target. Equivalently, after the chip is removed, the heads of its outgoing arcs must
still have paths to the target.

The induction is lexicographic. The primary parameter is the cut budget $k$.
For fixed $k$, progress is measured by the current minimum cut value
$\lambda_{\widehat D}(S,T_1)$, where only real arcs may be cut. Larger
values of $\lambda_{\widehat D}(S,T_1)$ are better, so the secondary
parameter is written as an upper bound on
$k-\lambda_{\widehat D}(S,T_1)$. The two reductions below will make progress
in one of the following two ways. Either the budget decreases, or the budget stays fixed and the relevant minimum cut value increases. This is standard in important separator arguments.

The next object is a nested list of minimum-cut source sides. More
precisely, we compute sets
$X_1\subset \cdots \subset X_q$
such that each $X_i$ is the source side of a minimum $S$-$T_1$ cut. This is
the minimum cut sequence, defined formally later and computed from a
maximum-flow residual graph. In routed instances we compute it by treating
auxiliary arcs as infinite-capacity arcs. Hence no auxiliary arc leaves any
$X_i$, and the outgoing real boundary of each $X_i$ has size
$\lambda_{\widehat D}(S,T_1)$. Let $X_\mu$ be the last member of this sequence satisfying the chip
conditions together with the forbidden-vertex and request conditions, if
such a member exists. That is, the index $\mu$ marks where the sequence changes from satisfying the augmented chip conditions to failing them. There is at most one such transition, since $\pi$ is hereditary and the forbidden and request conditions are monotone, meaning that they continue to hold when the chip is replaced by a subset of itself.

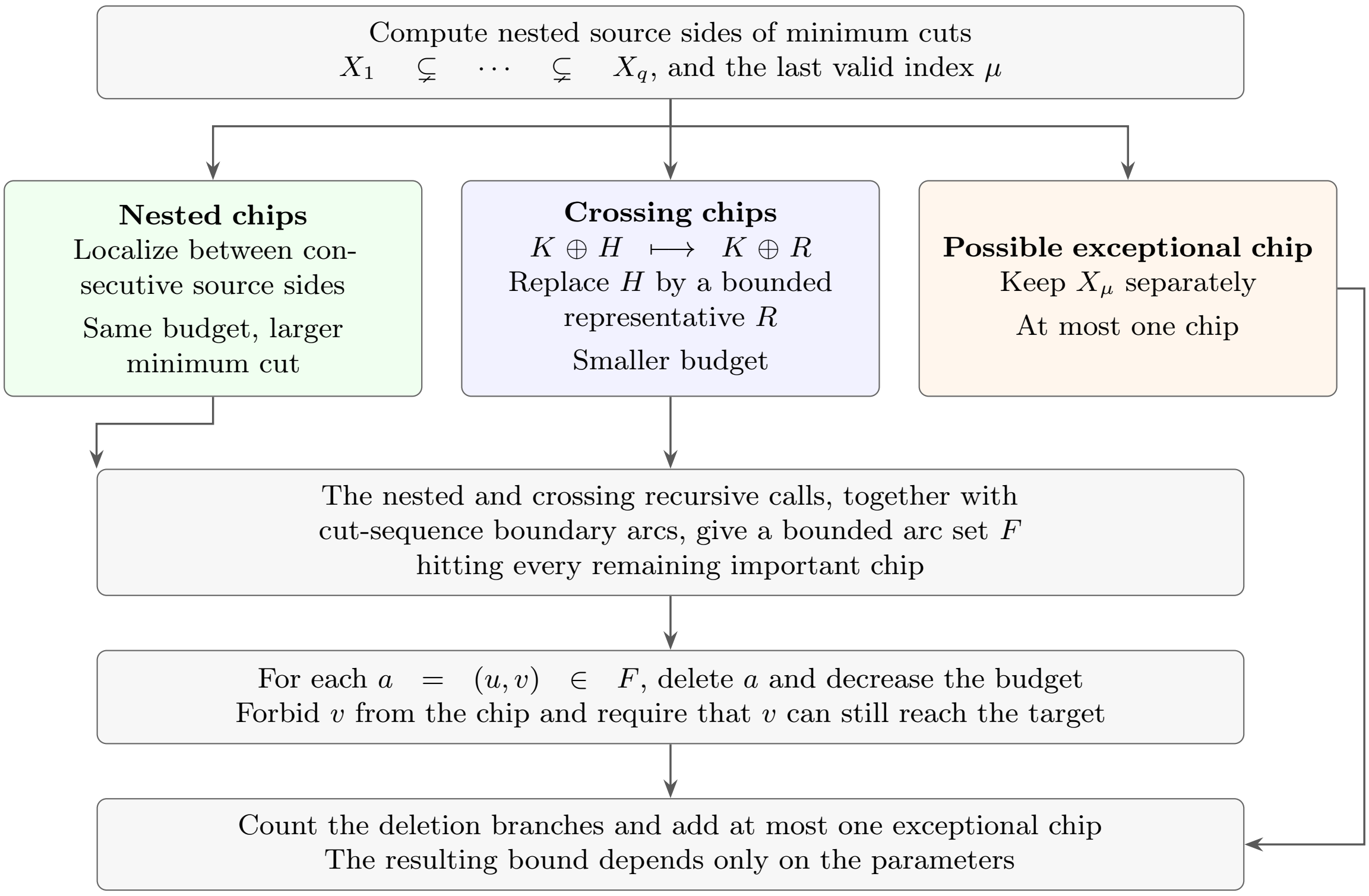
\begin{figure}[t]
\centering
\begin{tikzpicture}[
  >=Stealth,
  overviewbox/.style={draw=black!60,rounded corners=3pt,fill=black!3,
    align=center,font=\footnotesize,inner sep=5pt,text width=.78\textwidth},
  branchbox/.style={overviewbox,text width=.27\textwidth,minimum height=24mm},
  flow/.style={->,draw=black!65,line width=.65pt},
  node distance=6mm
]
\node[overviewbox] (sequence) {Compute nested source sides of minimum cuts\\
$X_1\subsetneq\cdots\subsetneq X_q$, and the last valid index $\mu$};
\node[branchbox,below=9mm of sequence,xshift=-.32\textwidth,
  fill=green!6] (nested) {\textbf{Nested chips}\\
Localize between consecutive source sides\\[1mm]
Same budget, larger minimum cut};
\node[branchbox,below=9mm of sequence,fill=blue!5] (crossing) {
\textbf{Crossing chips}\\
$K\oplus H\ \longmapsto\ K\oplus R$\\
Replace $H$ by a bounded representative $R$\\[1mm]
Smaller budget};
\node[branchbox,below=9mm of sequence,xshift=.32\textwidth,
  fill=orange!7] (exception) {\textbf{Possible exceptional chip}\\
Keep $X_\mu$ separately\\[1mm]
At most one chip};
\draw[flow] (sequence.south) -- ++(0,-3mm) -| (nested.north);
\draw[flow] (sequence) -- (crossing);
\draw[flow] (sequence.south) -- ++(0,-3mm) -| (exception.north);
\node[overviewbox,below=8mm of crossing] (hit) {The nested and crossing recursive calls, together with\\
cut-sequence boundary arcs, give a bounded arc set $F$\\
hitting every remaining important chip};
\draw[flow] (nested.south) -- ++(0,-3mm) -| (hit.north west);
\draw[flow] (crossing) -- (hit);
\node[overviewbox,below=of hit] (delete) {For each $a=(u,v)\in F$, delete $a$ and decrease the budget\\
Forbid $v$ from the chip and require that $v$ can still reach the target};
\draw[flow] (hit) -- (delete);
\node[overviewbox,below=of delete] (count) {Count the deletion branches and add at most one exceptional chip\\
The resulting bound depends only on the parameters};
\draw[flow] (delete) -- (count);
\draw[flow] (exception.east) -- ++(3mm,0) |- (count.east);
\end{tikzpicture}
\caption{The counting induction has two recursive cases and one possible
exceptional chip. In the crossing case, $K$ is the kept part of the real
graph on the chip, including the labeled attachment vertices, $H$ is the
outside part, and $R$ is its bounded CMSO representative. Gluing means
identifying equally labeled attachment vertices. This replacement preserves
$\pi$ and reduces the budget. The nested case instead increases the minimum
cut value. The exceptional chip $X_\mu$ is counted separately.}
\label{fig:counting-overview}
\end{figure}

There are three cases, as shown in \Cref{fig:counting-overview}.
The chip may equal $X_\mu$, in which case we count it separately.
Every other important chip either lies between $X_\mu$ and its successor,
or crosses one of the relevant minimum-cut source sides. Here two sets
\emph{cross} if neither contains the other. Thus there are two recursive
cases after the possible exceptional chip has been set aside.
If no sequence member satisfies the augmented chip conditions, or all do,
then the region before $X_1$ or after $X_q$, respectively, takes the place of the region between $X_\mu$ and its successor.

In the nested case, where the chip lies between $X_\mu$ and its successor, we localize the instance to the region between these two sets.
The source and target are enlarged in a controlled way so that the local
minimum cut value is strictly larger than $\lambda_{\widehat D}(S,T_1)$. The budget $k$
is unchanged, but the secondary induction parameter decreases.

In the crossing case, let $Z$ be the chip and let $X$ be the minimum-cut
source side that it crosses. Only two sequence members need to be considered,
namely $X_\mu$ and its successor when both exist, with the analogous endpoint
choice in the extreme cases. First add to $F$ all real arcs crossing between
$X$ and its complement, in either direction. We need only consider chips
whose outgoing arcs avoid this set. If the current real digraph is
$\alpha$-balanced, then $X$ has at most $k$ outgoing and $k+\alpha$ incoming
real arcs. Consequently, the vertices of $Z\setminus X$ adjacent by a real
arc to $Z\cap X$ form a set $B$ of size at most $2k+\alpha$.

This is where CMSO finite state is used. Write the real graph on $Z$ as
$K\oplus H$, where $H$ is the outside part $Z\setminus X$, $K$ is the kept
part together with $B$, and $\oplus$ identifies equally labeled vertices
of $B$. For fixed $B$ and $\pi$, there are only finitely many different
effects that $H$ can have on the truth of $\pi$ when glued to any kept graph.
We replace $H$ by a bounded representative $R$ with the same effect.
Thus $K\oplus H\models\pi$ if and only if $K\oplus R\models\pi$.
Auxiliary arcs separately preserve the required paths. The resulting
reduced chip has an outgoing real arc set that is a nonempty proper subset
of the original one, so the recursive budget is $k-1$.

The nested and crossing analyses together produce, from boundedly many
recursive instances, a bounded set of real arcs that hits the outgoing
boundary of every remaining important chip, with $X_\mu$ counted separately
if it is a chip. For every real arc in
the bounded hitting set, we delete that arc, decrease the budget by one, and
add its head to both the forbidden set and the request set. This gives the
recurrence for the number of important augmented chips. The original theorem
is obtained by viewing an ordinary digraph as a routed digraph with no
auxiliary arcs and taking $T_2=Q=\emptyset$.

\subsubsection{The enumeration algorithm}
The enumeration theorem follows the same proof as the combinatorial bound.
Almost every step of the argument is constructive. Minimum cut sequences can
be computed by a max-flow computation, and the local instances used in the
nested case are explicitly defined. In the crossing case, the proof later
constructs a bounded family of reduced routed instances. Algorithmically,
this is done by trying all bounded choices of boundary vertices, shortcut
arcs, request entries, and representatives. The final hitting set then
consists of explicitly generated real arcs.

The only non-constructive ingredient is the finite set of CMSO
representatives for the CMSO sentence $\pi$. This is harmless for
\Cref{thm:enumerateBoundImpChips}, since the theorem is non-uniform in
$\pi,k,\alpha$. For fixed parameters, we may regard the required representatives as advice
to the algorithm.

The additional assumption that $\pi$ is efficiently testable parameterized by
the balance is used when the algorithm has to decide which members of the
minimum cut sequence are chips, in particular to compute the index $\mu$.
All other chip conditions are polynomial-time verifiable. 
 Thus the constructive version of the inductive proof
enumerates all important augmented chips in time
$c'_{\pi,k,\alpha}\cdot n^{\bigoh(1)}$, and the ordinary enumeration theorem
follows by setting $T_2=Q=\emptyset$ and using no auxiliary arcs.

\subsubsection{Application to \textsc{Arc DOCT}}
We now explain, at a high level, how the important chip framework is used in
the algorithm for \arcDoct{} on balanced digraphs. For this discussion, take
\Cref{thm:enumerateBoundImpChips} as a black box. Namely, for fixed
$k,\alpha$ and a hereditary CMSO sentence $\pi$, it gives an FPT procedure
for enumerating all important $(S,T,\pi,k)$-chips in an $\alpha$-balanced
digraph.

The algorithm starts by computing a small vertex modulator $W$ such that
$D-W$ has no directed odd cycle. For this, we use the constant-factor FPT-approximation from \cite{LokshtanovRSZ20}. If no such modulator of size $O(k)$ exists,
then the instance is a no-instance, since the endpoints of the arcs in any
arc solution of size at most $k$ form a vertex set of size at most $2k$ that
hits all directed odd cycles. Thus, after this step, all remaining difficulty
is concentrated around the bounded set $W$.

Consider a hypothetical solution $F$ of size at most $k$. In $D-F$, the
vertices of $W$ are distributed among strongly connected components. Since
$|W|=O(k)$, we can guess the ordered partition
$W_1,\ldots,W_t$ 
of $W$ induced by these components, in reverse topological order in $D-F$. Thus $D-F$ has no path from a vertex of $W_i$ to a vertex of $W_j$ for $i<j$. If $t=1$, then all of $W$ lies in one strongly connected component
of $D-F$. Using the standard characterization that, in a strongly connected digraph,
absence of directed odd cycles is equivalent to bipartiteness of the
underlying undirected graph, one guesses the bipartition of $W$ in that
component and uses the subroutine of \Cref{lem:specialCaseDOCT}. Given
$(D,k)$ and a vertex set $W$ such that $D-W$ has no directed odd cycles,
the subroutine either solves the instance or correctly concludes that
there is no solution $F$ of at most $k$ arcs for which $W$ intersects
exactly one strongly connected component of $D-F$.

The interesting case is $t>1$. Then $F$ must hit all paths from the
first block $W_1$ to the later blocks $W\setminus W_1$. Let $F'$ be a
minimal subset of $F$ that is a $W_1$-$W\setminus W_1$ cut, and let $Z'$ be the set of vertices reachable from $W_1$ in $D-F'$. By the minimality of $F'$, the outgoing boundary of $Z'$ is exactly the cut
$F'$, so $Z'$ has boundary size $|F'|$. Let
\[
F_{\mathrm{in}}=F\cap A(D[Z'])
\qquad\text{and}\qquad
\ell=|F_{\mathrm{in}}|.
\]
Then $F_{\mathrm{in}}$ is an arc directed odd cycle transversal of $D[Z']$. Equivalently,
$D[Z']$ satisfies the CMSO property $\pi_\ell$ saying that at most $\ell$
arcs can be deleted to remove all directed odd cycles. Hence $Z'$ is a
$
(W_1,W\setminus W_1,\pi_\ell,|F'|)
$-chip. Moreover,
$\ell+|F'|\le k$.

The chip $Z'$ need not be important. However, the standard important-separator pushing argument
shows that it can be replaced by an important chip. If $Z'$ is dominated by
a larger chip $Z''$, then $Z''$ has no larger outgoing boundary and still
satisfies $\pi_\ell$. We may therefore replace the part of the solution used
inside $D[Z']$ and across $\partial^+(Z')$ by an arc directed odd cycle transversal of $D[Z'']$ of
size at most $\ell$, together with the boundary $\partial^+(Z'')$. This does
not increase the solution size. Consequently, there is a solution of size at
most $k$ that contains an arc of $\partial^+(Z'')$ for some important
$
(W_1,W\setminus W_1,\pi_\ell,k')
$-chip $Z''$, where $\ell+k'\le k$ and $k'>0$.

We can now branch. For every pair $\ell,k'$ with
$\ell+k'\le k$, $\ell<k$, $k'>0$, 
we enumerate all important
$
(W_1,W\setminus W_1,\pi_\ell,k')
$-chips and branch on the arcs in their outgoing boundaries. The preceding
paragraph guarantees that at least one branch deletes an arc belonging to
some solution. Each such branch reduces the budget from $k$ to $k-1$.
Deleting one arc increases the balance by at most one, so throughout a
root-to-leaf path the balance remains bounded by a function of the initial
balance and $k$.

One case has to be handled separately. It may already be that there is no
path from $W_1$ to $W\setminus W_1$ before any arc is deleted. Then the cut
$F'$ above is empty, so there is no boundary arc on which to branch. Let
$U$ be the set of vertices reachable from $W_1$ in the current digraph.
Since no vertex of $W\setminus W_1$ is reachable from $W_1$, we have
$U\cap W=W_1$. Also, no arc leaves $U$, by the definition of reachability.
Hence no directed cycle can use vertices both inside and outside $U$. We may
therefore solve the two induced subinstances on $U$ and on
$V(D)\setminus U$, trying all splits of the remaining arc budget.

Since no arc leaves $U$, in the $\alpha$-balanced case at most $\alpha$
arcs enter $U$. In particular, no arc enters $U$ in the Eulerian case.
Therefore the two induced subinstances are separated
by a bounded interface, and their balance parameters remain bounded by a
function of the original balance. Thus the recursive calls stay within the
same parameterized framework. Moreover, the remaining modulator strictly
decreases, since the first subinstance contains only the block $W_1$ and the
second contains the remaining blocks.

Thus every recursive step either decreases the arc budget, by branching on
the boundary of an important chip, or decreases the number of modulator
vertices still under consideration, by splitting off a closed reachable
region. Since the important chip theorem bounds the number of chips by a
function of the parameters alone, this yields an FPT branching algorithm on
balanced digraphs, and in particular on Eulerian digraphs.

\subsubsection{Why weak representatives suffice}
We only use CMSO representatives in a very restricted way.
Suppose that an important chip $Z$ crosses a set $X$ from the
minimum cut sequence. The reduction keeps the part $X\cap Z$ explicitly and
replaces the part $Z\setminus X$ by a CMSO representative. The
boundary along which the replacement happens consists of the vertices of $Z\setminus X$ that are adjacent to the
kept part $X\cap Z$. Thus, up to the convention of assigning arcs
with both endpoints on the boundary to the replaced side, the original real
graph induced by the chip can be viewed as a gluing
  $D_1\oplus D_2$,
where $D_1$ is the kept side together with the boundary and $D_2$, the subgraph induced by $Z\setminus X$, is the part being replaced. A vertex of $X\setminus Z$ may be adjacent to the replaced part $Z\setminus X$. Such vertices are forbidden to be part of the reduced chip when recursing, so that when a reduced chip ``grows'' inside $X\setminus Z$, it cannot acquire new adjacencies to the forgotten piece
$Z\setminus X$.  

A full CMSO representative for $D_2$ would preserve the truth of
$\pi$ after gluing to every compatible boundaried digraph, in both directions. The
actual proof does not need this much. If $R$ is the bounded graph used in
place of $D_2$, then two one-sided requirements suffice.

First, the reduced chip must still satisfy the logical constraint. Since the
original chip satisfies
  $D_1\oplus D_2\models \pi$,
we need
  $D_1\oplus R\models \pi$.
Second, if the reduced chip can be enlarged in the reduced instance, then
that enlargement must lift back to an enlargement of the original chip. In
that lifting step the kept side of the chip may have grown from $D_1$ to a larger
boundaried digraph $D_1'$. What is needed is only the implication
  $D_1'\oplus R\models \pi
  \quad\Longrightarrow\quad
  D_1'\oplus D_2\models \pi$.


For \textsc{Arc DOCT}, weak representatives can be made explicit. The key
observation is that, for hitting directed odd cycles, it is enough to
remember which boundary vertices are connected by even and odd directed
paths after deleting a small number of arcs. This stems from the fact that a digraph has a directed odd-length closed walk if and only if it has a directed odd cycle. Thus an \textsc{Arc DOCT} weak
representative only has to preserve the finite family of parity profiles
obtainable after deleting at most $r$ arcs, for some $r\leq k$. Here a parity profile states, for each ordered pair of boundary vertices, whether there is an even and whether there is an odd directed path from the first to the second. Since the
boundary size is bounded in terms of $k$ and the balance parameter there are only finitely many such profiles. This is what
allows the representatives needed for \textsc{Arc DOCT} to be computed
uniformly, rather than supplied non-uniformly through full CMSO
representatives.

\subsection{Related generalizations of important separators}\label{subsec:relatedImportantSeparators} 

\begin{table}[t]
\centering
\small
\renewcommand{\arraystretch}{1.3}
\begin{tabular}{@{}l l >{\raggedright\arraybackslash}p{3.3cm} l@{}}
\toprule
\textbf{Framework} & \textbf{Setting} & \textbf{Constraint} & \textbf{\# objects} \\
\midrule
Imp. separators \cite{Marx06}
  & undirected \& directed
  & none 
  & $4^{k}$ \\
Secluded subgraphs \cite{JansendeKroonWlodarczyk2025}
  & undirected
  & $\mathcal F$-free, connected
  & $2^{\mathcal O(k)}$ \\
Imp. bal. subgraphs \cite{DabrowskiJOOW25}
  & biased graphs
  & balanced
  & \begin{tabular}[t]{@{}l@{}}$4^{k}$ (dominating\\ family)\end{tabular} \\
\begin{tabular}[t]{@{}l@{}}Conn.-preserving\\ imp.\ separators \cite{Kenig25cpis}\end{tabular}
  & undirected
  & connectivity to a terminal set
  & $2^{\bigoh(k^{2}\log k)}$ \\
Parity imp. seps. \cite{LokshtanovR12ParityMWC}
  & undirected
  & given parity of terminal paths
  & $2^{k^{\bigoh(1)}}$ (implicit) \\
\midrule
\textbf{Important chips (this paper)}
  & $\alpha$-balanced digraphs
  & hereditary CMSO property $\pi$
  & $c_{k,\pi,\alpha}$ \\
\bottomrule
\end{tabular}

\vspace{0.5em}
\begin{minipage}{\linewidth}
\footnotesize

\end{minipage}
\caption{Generalizations of important separators, their constraints, and bounds on
associated families. Our framework is the only one that is simultaneously
directed and applies to arbitrary hereditary CMSO constraints, but the tradeoff is a
non-explicit constant.}
\label{tab:related-important-separators}
\end{table}

Our work is closely related to other attempts (see \Cref{tab:related-important-separators}) to extend the important
separator paradigm beyond plain separation. These works differ
mainly in the kind of additional structure imposed on the object being
enumerated.

\medskip
\noindent{\bf Secluded subgraphs.}
A vertex set $C$ is \emph{$k$-secluded} if its open neighborhood has size
at most $k$. Jansen, de~Kroon, and W\l{}odarczyk~\cite{JansendeKroonWlodarczyk2025} showed
that, for every fixed finite family $\mathcal F$, the number of maximal
$k$-secluded connected sets $C$ containing a prescribed source set, avoiding
a prescribed terminal set, and satisfying that $G[C]$ has no induced
subgraph isomorphic to a member of $\mathcal F$, is bounded by
$2^{\bigoh(k)}$. They also gave a single-exponential FPT algorithm to
enumerate these subgraphs. Enumerating over the root gives the corresponding
$2^{\bigoh(k)}n$ bound. This is close to our undirected vertex-chip result.
An $\mathcal F$-free $k$-secluded connected subgraph containing a vertex
$v$ can often be expressed as a $(\{v\},W,\pi_\mathcal F,k)$-chip for a
suitable set $W$, where $\pi_\mathcal F$ is the hereditary property of having no induced subgraph isomorphic to a member of $\mathcal F$. The correspondence is not exact in all cases, for example
when the closed neighborhood of the secluded subgraph is the entire graph.

\medskip
\noindent{\bf Important balanced subgraphs.}\par\noindent
Dabrowski, Jonsson, Ordyniak, Osipov, and Wahlstr\"om~\cite{DabrowskiJOOW25}
introduced important balanced subgraphs in the setting of biased graphs. A biased graph $\Psi=(G,\mathcal B)$ is an undirected graph together with
a distinguished class $\mathcal B$ of ``balanced'' cycles satisfying the
following restriction: if two of the three cycles formed by three internally
vertex-disjoint paths with the same two endpoints belong to $\mathcal B$,
then the third does as well. A subgraph is balanced if all its cycles belong to
$\mathcal B$. Here ``balanced'' is unrelated to the 
cut-balance condition used in this paper. This
framework captures separation problems such as Odd Cycle Transversal, Group
Feedback Vertex Set, Multiway Cut, and Multicut. In the unit-weight case,
the cost of a subgraph $H$ counts edges leaving $V(H)$ and edges of
$G[V(H)]$ omitted from $H$. A balanced subgraph dominates another if its
vertex set contains the other's and its cost is no greater. Domination is
strict if either comparison is strict. A connected balanced subgraph is
important if no balanced subgraph strictly dominates it. A family is
dominating if each subgraph under consideration is dominated by a member.
Their main structural 
statement is that, for a fixed root, the connected balanced subgraphs of
cost at most $k$ admit a dominating family of important balanced subgraphs
of size at most $4^k$, computable in FPT time. Their bound matches the classical 
important-separator bound, but the objects and the domination
notion are tailored to biased graphs and the cycle restriction above.

\medskip
\noindent{\bf Connectivity-preserving important separators.}
Kenig~\cite{Kenig25cpis} introduced connectivity-preserving important
separators, where a separator must disconnect prescribed terminal sets while
also preserving specified terminal connectivity constraints. These
constraints can express requirements such as keeping a terminal set connected
to one side of the separation. The number of such
separators of size at most $k$ is shown to be bounded by
$2^{\bigoh(k^2\log k)}$, with a corresponding enumeration algorithm, and
the framework yields applications to cut-uncut problems such as
\textsc{Node Multiway Cut-Uncut} with a fixed number of terminal classes.
Note that connectivity is not hereditary. 

\medskip
\medskip
\noindent{\bf Parity constraints.}
Lokshtanov and Ramanujan~\cite{LokshtanovR12ParityMWC} introduced a
parity-sensitive generalization of important separators in their algorithm for
\textsc{Parity Multiway Cut}. The input in this problem is an undirected graph $G$, terminal
sets $T_e$ and $T_o$ of even and odd terminals, and an integer $k$. The task
is to delete at most $k$ vertices so as to hit every even path from a
terminal $v\in T_e$ to another terminal in
$(T_e\cup T_o)\setminus\{v\}$, and every odd path from a terminal
$v\in T_o$ to another terminal in $(T_e\cup T_o)\setminus\{v\}$. Their
notion refines ordinary importance by combining reachability domination with
a parity condition. In broad terms, a separator may be replaced by a
dominating one only if this does not make the residual parity-hitting task
harder. Their work is close to ours in proof strategy: it also uses a
sequence of minimum separators and a nested-versus-crossing dichotomy.
However, their framework is tailored to undirected graphs and one parity-cut problem, whereas our
framework applies to hereditary CMSO properties of source-side regions in
balanced digraphs.

These results are complementary. Our results apply to arbitrary hereditary
CMSO properties of source-side regions in balanced digraphs. The general
enumeration theorem is non-uniform. Computable weak representatives give a
uniform version under the corresponding uniform testability assumption.

\subsection{Organization and reading guide}

The rest of the paper is organized as follows.

\begin{center}
\begin{tabularx}{\textwidth}{@{}p{0.26\textwidth}X@{}}
\textbf{Where to look} & \textbf{What is there} \\
\hline
\Cref{sec:prelim} &
Notation, CMSO, finite-state properties, and boundaried digraphs. \\

\Cref{sec:genImpCuts} &
Boundedness and enumeration of important chips on balanced digraphs. \\

\Cref{sec:arcDoctApplication} &
Use of chip enumeration to obtain FPT algorithms for \arcDoct{} on balanced
and Eulerian digraphs. \\

\Cref{sec:weakRep} &
Weak representatives and the uniform versions of the enumeration and
\arcDoct{} algorithms. \\

\Cref{sec:undirectedPort} &
Important edge chips and vertex chips in undirected graphs. \\

\Cref{sec:conclusion} &
Open questions and possible extensions. \\

\Cref{app:deferred-proofs} &
Deferred proofs, including the NP-completeness of \arcDoct{} on Eulerian digraphs.
\end{tabularx}
\end{center}


The technical core of the paper is \Cref{sec:genImpCuts}. It proves the
combinatorial bound on important chips and its enumeration counterpart. 
%
%
\Cref{sec:arcDoctApplication} applies important-chip enumeration to \arcDoct{}. It can be read
independently by treating \Cref{thm:enumerateBoundImpChips} as a black box.
\Cref{sec:weakRep} proves uniform chip enumeration using computable weak
representatives, then applies it to the properties $\pi_\ell$ to obtain
uniform \arcDoct{} algorithms on balanced and Eulerian digraphs. Finally, \Cref{sec:undirectedPort} explains how the framework transfers to
undirected graphs. The edge-cut version follows by bidirecting every edge.
The vertex-separator version uses the same counting scheme, with separate
charging and lifting arguments for vertex boundaries. We conclude with open
questions and possible extensions.

\section{Preliminaries}\label{sec:prelim}

For $i\in {\mathbb N}_0$, we use $[i]$ to denote $\{1,\dots,i\}$, so $[0]=\emptyset$, and $[i]_0$ to denote $\{0\}\cup [i]$.

\paragraph{Directed graphs.} 
We only work with simple digraphs, which have no parallel arcs and no self-loops. Paths and cycles do not contain repeated vertices, while walks and closed walks can contain repeated vertices. 
Unless stated otherwise, paths, cycles, and (closed) walks in a digraph are directed. We allow paths of length zero. In particular, every vertex is reachable from itself, and every vertex of a set $U$ has a path to $U$.
We write $X\subset Y$ and $X\subsetneq Y$ interchangeably for proper inclusion.
The {\em undirected graph underlying} a directed graph $D$ is the simple undirected graph $G$ where $V(G)=V(D)$ and $(u,v)\in E(G)$ if and only if $(u,v)\in A(D)$ or $(v,u)\in A(D)$.

Given a vertex set $X\subseteq V(D)$, we let $D[X]$ denote the subgraph of $D$ induced by $X$, and we define $D-X=D[V(D)\setminus X]$. For a set $Z\subseteq V(D)\cup A(D)$, we denote by $D-Z$ the digraph obtained from $D$ by deleting the vertices and arcs in $Z$. Given an arc $(u,v)\in A(D)$, we refer to $u$ as the {\em tail} of the arc and to $v$ as the {\em head} of the arc. For a set of arcs $A$, we denote by $\tail(A)$ the set of all tails of arcs in $A$, and by  $\head(A)$ the set of all heads of arcs in $A$. For a single arc $a$, we write $\head(a)$ and $\tail(a)$ instead of $\head(\{a\})$ and $\tail(\{a\})$. 
Given a vertex set $X\subseteq V(D)$, we use $N^+(X)$ to denote the set of out-neighbors of $X$, and $N^-(X)$ to denote the set of in-neighbors of $X$. 
 Formally, $N^+(X)=\{y\notin X\mid \exists x\in X, (x,y)\in A(D)\}$. 
 Similarly, $N^-(X)=\{y\notin X\mid \exists x\in X,(y,x)\in A(D)\}$. 
We use $N^i_Z(X)$ to denote the set $N^i(X) \cap Z$, where $i \in \{+, -\}$.
 A {\em strongly connected component} of $D$ is a maximal subgraph in which every vertex has a directed path to every other vertex. We say that a strongly connected component is {\em non-trivial} if it consists of at least two vertices and {\em trivial} otherwise. For vertex sets $X$ and $Y$, the set $Y$ is said to be
 {\em reachable from $X$} if for {every} vertex  $y\in Y$, there exists a vertex $x\in X$ such that $D$ contains a directed path from $x$ to $y$. 
  
For disjoint subsets $X,Y,Z\subseteq V(D)$, we call $Z$ an $X$-$Y$-{\em separator} if there is no path from a vertex of $X$ to a vertex of $Y$ (also called an $X$-$Y$ path) in $D-Z$.
We say that $(X, Y)$ is a separation in a graph $G$ if $X \cup Y = V(G)$ and there is no edge between $X \setminus Y$ and $Y \setminus X$. For a digraph $G$ and vertex sets $A,B\subseteq V(G)$,
we write
$A\leadsto_G B$ or $A\leadsto B$ in $G$ 
to mean that some vertex of $B$ is reachable from some vertex of $A$ in $G$ and we write $A\not\leadsto_G B$ or $A\not \leadsto B$ in $G$ otherwise. 
When $A=\{a\}$ or $B=\{b\}$, we write $a$ instead of $A$ and $b$ instead of $B$.
%

A set $F\subseteq A(D)$ is an $X$-$Y$-{\em cut} if there is no $X$-$Y$ path in $D-F$. 
The size of a minimum $X$-$Y$-cut is denoted by $\lambda_D(X,Y)$, with the reference to $D$ omitted when clear from the context. For a set $Z\subseteq V(D)\cup A(D)$, we denote by $R(X,Z)$ the set of vertices reachable from $X$ in $D-Z$.  
We use $\partial^{+}(v)$ or $\partial^{+}(S)$ to denote arcs leaving a vertex $v$ or a vertex set $S$. The notation $\partial^{-}$ is defined symmetrically for incoming arcs.
We also define $\partial(S)$ as $\partial^+(S) \cup \partial^-(S)$.
For a set $\Xcal$ of vertex sets of $D$, we denote by $\partial^{+}(\Xcal)$ the set $\bigcup_{X\in \Xcal}\partial^{+}(X)$.
A digraph is {\em Eulerian} if every vertex has equal indegree and outdegree.

\paragraph{Counting monadic second-order logic and representatives.}
We use Counting Monadic Second-Order Logic, abbreviated CMSO, on directed
graphs. Thus formulas may quantify over vertices, arcs, sets of vertices, and
sets of arcs, and may use the incidence predicates $\tail(e,v)$ and
$\head(e,v)$, which state that $v$ is the tail or the head of the arc $e$, respectively. They may also use the usual logical connectives and predicates testing whether $|S|\equiv q \pmod r$ for a set variable $S$ and integers $0\le q<r$ with $r\ge 2$. This is the variant often denoted ${\rm CMSO}_2$. A digraph
property is CMSO-definable if it is defined by a CMSO sentence. We write
$D\models \pi$ when the digraph $D$ satisfies the sentence $\pi$.

A boundaried digraph is a digraph $D$ together with a distinguished
set $\partial(D)\subseteq V(D)$ of boundary vertices and an injective
labeling $\beta_D\colon \partial(D)\to\mathbb N$. Its label set is
$\Lambda(D)=\{\beta_D(v)\mid v\in\partial(D)\}$. For a boundaried digraph, $\partial(D)$ is a set of vertices. This differs from $\partial(S)$ for a vertex set $S$, which is a set of arcs. For a finite set
$I\subseteq\mathbb N$, let $\Dcal_I$ be the class of boundaried digraphs with
label set $I$, and let
$\Dcal_{\subseteq I}=\bigcup_{I'\subseteq I}\Dcal_{I'}$.
A boundaried digraph in $\Dcal_{\subseteq [t]}$ is called $t$-boundaried.

Two boundaried digraphs are \emph{compatible} if they have the same label
set. Their gluing $D_1\oplus D_2$ identifies equally labeled boundary
vertices. Arcs present in either input graph are present in the glued graph.

Let $\sigma$ be a digraph property. For two boundaried digraphs
$D_1,D_2$, we write $D_1\equiv_\sigma D_2$ if they are compatible and, for every
boundaried digraph $H$ compatible with them,
$\sigma(D_1\oplus H)=\mathsf{true}$ if and only if 
$\sigma(D_2\oplus H)=\mathsf{true}$. 
If $\pi$ is a CMSO sentence, we write $\equiv_\pi$ for $\equiv_\sigma$, where $\sigma$ is the property defined by $\pi$, that is, $\sigma(D)=\mathsf{true}$ if and only if $D\models\pi$.

We use the following standard consequence of Courcelle's theorem, see
\cite{Courcelle90,Courcelle92a,Courcelle97,Courcelle:2012book}.

\begin{proposition}\label{prop:finiteState}
For every CMSO sentence $\pi$ and every finite set
$I\subseteq\mathbb N$, the restriction of $\equiv_\pi$ to
$\Dcal_{\subseteq I}$ has finitely many equivalence classes.
\end{proposition}

Since compatible digraphs have the same label set, every class of $\equiv_\pi$
lies in a single $\Dcal_{I'}$. We fix one representative of minimum encoding
length for every class once and for all, where encodings list all vertices and arcs, so the size of a digraph is at most the length of its encoding. Let
$\Rcal_{\equiv_\pi}[\Dcal_{\subseteq I}]$ be the set of representatives of the
classes contained in $\Dcal_{\subseteq I}$. For $t\in\mathbb N_0$, let
$r_\pi(t)$ be the largest of the numbers $|\Rcal_{\equiv_\pi}[\Dcal_{\subseteq I}]|$ and the encoding sizes of the graphs in $\Rcal_{\equiv_\pi}[\Dcal_{\subseteq I}]$, taken over all $I\subseteq [t]$. By \Cref{prop:finiteState}, $r_\pi(t)$ is finite and depends only on $\pi$ and $t$.
When $\pi$ is clear from the context, we write $r(t)$.

\paragraph{Testability.}
The enumeration theorems test $\pi$ on digraphs whose balance may exceed the
balance of the input. They therefore use the following condition.

\begin{definition}[Efficient testability parameterized by the balance]
\label{def:efficientTestability}
Let $\pi$ be a CMSO sentence. We say that $\pi$ is \emph{efficiently testable
parameterized by the balance} if there is a function
$f_\pi$ such that, for every $\gamma\in\mathbb N_0$, some algorithm decides
whether $D\models\pi$ for every digraph $D$ of balance at most $\gamma$ in time
$f_\pi(\gamma)\cdot n^{d}$, where $n=|V(D)|$ and $d$ is independent of $\pi$ and $\gamma$. We say that $\pi$ is
\emph{uniformly efficiently testable parameterized by the balance} if, in
addition, a single algorithm takes $\gamma$ and $D$ as input and achieves this
bound with a computable function $f_\pi$.
\end{definition}

\newcommand{\sizeof}[1]{|#1|}
\mcm{\pplus}{2}{\partial_{#1}^+(#2)}
\mcm{\sttqpk}{0}{(S, T_1, T_2, Q, \pi, k)}
\mcm{\parto}{2}{\partial^{+}_{#1}(#2)}
\mcm{\parout}{2}{\partial^{+}_{\overline{#1}}(#2)}
\mcm{\pthout}{2}{\ensuremath{\mathsf{cap}_{#1}(#2)}\xspace}
\mcm{\pthin}{2}{\ensuremath{\mathsf{sm}_{#1}(#2)}\xspace}

\mcm{\first}{3}{\rho^{#1}_{#2}(#3)}
\mcm{\dispath}{3}{\Pcal^{#1}_{#2}(#3)}
\newcommand{\sttreach}[3]{$#1$-$#2#3$}
\newcommand{\reach}[2]{\text{$#1$-$#2$}}

\section{Important chips in balanced digraphs}\label{sec:genImpCuts}

This section proves the main important-chip theorem and its enumeration
version. After the basic definitions, we move to the routed and augmented
setting used by the induction. The proof then analyzes important chips
relative to a fixed nested sequence of minimum cuts, separating the nested case
from the crossing case where bounded representatives are used.

\providecommand{\Dhat}{\widehat{D}}
    \providecommand{\real}{\mathrm{real}}
    \providecommand{\aux}{\mathrm{aux}}
    \providecommand{\Aaux}{A_{\aux}}
    \providecommand{\dreal}{\partial_{\real}}
    \providecommand{\dpreal}{\partial^{+}_{\real}}
    \providecommand{\BX}{B_X(Z)}
    \providecommand{\GX}{\Gamma_X(Z)}
    \providecommand{\CX}{C_X(Z)}
    \providecommand{\tail}{\mathsf{tail}}      \providecommand{\Nbb}{\mathbb{N}}

\subsection{Basic definitions}

\begin{definition}
We say that a digraph property is {\em hereditary} if it is closed under the induced subgraph relation. We say that a CMSO sentence $\phi$ is {\em hereditary} if the digraph property defined by $\phi$ is hereditary.
\end{definition}

\begin{definition}\label{def:reachabilityset}
Let $D$ be a digraph and let $S,T$ be disjoint non-empty vertex sets.
We say that a set $Z\supseteq S$ is an \emph{$S$-$T$ reachability set}
if $Z$ is disjoint from $T$ and reachable from $S$ in $D[Z]$.
We say that $Z$ is \emph{strict} if $\partial^{+}(Z)$ is an
inclusion-wise minimal $S$-$T$ cut in $D$. In addition, we say that $Z$ is
\emph{compact} if $|\partial^+(Z)|=\lambda(S,T)$.
\end{definition}

We shall use the following elementary facts about reachability sets. They
are standard consequences of the definitions.

\begin{observation}\label{obs:transreachability}
For any sets $X \subseteq Y \subseteq Z$, if $Y$ is reachable from $X$ in $D[Y]$ and
$Z$ is reachable from $Y$ in $D[Z]$, then $Z$ is reachable from $X$ in $D[Z]$.
\end{observation}

\begin{lemma}[Basic facts about strict reachability sets]
\label{lem:basicStrictFacts}
Let $D$ be a digraph and let $S,T\subseteq V(D)$ be disjoint non-empty
vertex sets. The following statements hold.
\begin{enumerate}
\item Every compact $S$-$T$ reachability set is strict.
\item An $S$-$T$ reachability set $Z$ is strict if and only if, for every
$v\in N^+(Z)$, there is a $v$-$T$ path in $D-Z$.
\item If $Z$ is a strict $S$-$T$ reachability set and
$a\in\partial^+(Z)$, then $Z$ is a strict $S$-$T$ reachability set in
$D-a$.
\end{enumerate}
\end{lemma}

The proof is deferred to \Cref{app:basic-strictness-balance}. We similarly defer routine verifications, while keeping the main inductive arguments in the text.

All CMSO sentences and formulas discussed in this section are for digraphs.
In what follows, let $D$ be a digraph, let $S,T$ be disjoint non-empty vertex
sets, let $\pi$ be a CMSO sentence, and let $k\in {\mathbb N}_0$.

\begin{definition}[Chips]\label{def:chip}
An {\em $(S,T,\pi,k)$-chip} is a vertex set $Z$ such that:
\begin{enumerate}
\item $Z$ is a strict $S$-$T$ reachability set in $D$.
\item $|\partial^{+}(Z)|\leq k$.
\item $D[Z]\models \pi$.
\end{enumerate}
\end{definition}

\begin{definition}[Important chips]
An \stpik-chip $Z$ is {\em important} if there is no \stpik-chip $Z'$ such
that $Z'\supset Z$ and
$|\partial^{+}(Z')|\leq |\partial^{+}(Z)|$.
\end{definition}

\begin{example}
If $G\models\pi$ for every digraph $G$, then for every important
$(S,T,\pi,k)$-chip $Z$, the cut $\partial^{+}(Z)$ is an important
$S$-$T$ cut of size at most $k$.
\end{example}

This recovers the classical important-cut setting when no structural
condition is imposed on the source side. We now introduce the balance
condition under which the number of important chips becomes bounded.

\begin{definition}\label{def:stability}
For $\alpha\in {\mathbb N}_0$ and directed graph $D$, we say that $Z \subseteq V(D)$ is {\em $\alpha$-balanced} in $D$ if $\big||\partial^{-}(Z)| - |\partial^{+}(Z)| \big| \leq \alpha$. 
We say that $D$ is {\em $\alpha$-balanced} if every $Z\subseteq V(D)$ is $\alpha$-balanced in $D$. 
 The smallest value of $\alpha$ such that $D$ is $\alpha$-balanced is called the {\em balance of} $D$.
\end{definition}

\begin{example}
    Eulerian digraphs are 0-balanced. 
\end{example}

%
%
%
%
%
%
%
%
%
%
%
%

\subsection{Combinatorial bound on number of important chips 
}\label{subsec:combinatorial_bound}

\subsubsection{Basics of routed digraphs}

\begin{definition}[Routed digraphs]
{\rm A
\emph{routed digraph} is a triple $\widehat D=(U,A_{\mathrm{real}},A_{\mathrm{aux}})$,
where $A_{\mathrm{real}}$ and $A_{\mathrm{aux}}$ are disjoint sets of
real and auxiliary arcs, respectively. We write
$A(\widehat D)\coloneqq A_{\mathrm{real}}\cup A_{\mathrm{aux}}$.  We also use $V(\widehat D)$ to refer to the vertex set $U$ of this routed digraph. 
Paths and reachability in $\widehat D$ are computed using all arcs in
$A(\widehat D)$. However, cuts, boundaries, cut sizes, hitting sets, balance,
and the CMSO sentence $\pi$ see only the real arcs.
A set
$F\subseteq A_{\mathrm{real}}$ is an $S$-$T$ cut in $\widehat D$ if there is
no $S$-$T$ path in the digraph with vertex set $U$ and arc-set
$A(\widehat D)\setminus F$. We write $\lambda_{\widehat D}(S,T)$ for the
minimum size of such a real $S$-$T$ cut, with the convention that
$\lambda_{\widehat D}(S,T)=\infty$ if no such real cut exists. We use $\widehat D_{\mathrm{real}}$ to denote the digraph $(U,A_{\mathrm{real}})$, which has all vertices of $\widehat D$ and only the real arcs. We call $\widehat D$ \emph{$\alpha$-balanced} if $\widehat D_{\mathrm{real}}$ is $\alpha$-balanced. 
}
\end{definition}

\begin{example}
	Every ordinary digraph $D$ is a routed digraph 
$(V(D),A(D),\emptyset)$.
\end{example}

Thus routed digraphs do not change the original problem. They only allow
recursive instances to remember paths through regions that have been
contracted or replaced. Whenever an auxiliary arc appears, it should be read
as a stored route rather than as an arc that may be paid for by a cut.

\begin{definition}
	{\rm For a vertex set $Y\subseteq V(\widehat D)$, we define
\[
\partial_{\mathrm{sub}}^+(Y)
\coloneqq
\{(u,v)\in A_{\mathrm{sub}}\mid u\in Y,\ v\notin Y\},
\]
 where $\mathrm{sub}\in \{\mathrm{real, aux}\}$. 
We define $\partial_{\mathrm{real}}^-(Y)$ and $\partial_{\mathrm{aux}}^-(Y)$ symmetrically and define 
$\partial_{\mathrm{real}}(Y)
\coloneqq
\partial_{\mathrm{real}}^+(Y)\cup
\partial_{\mathrm{real}}^-(Y)$.
We say that $Y$ is \emph{auxiliary-respecting} if
$\partial_{\mathrm{aux}}^+(Y)=\emptyset$.
}
\end{definition}

\subsubsection{Balance and interfaces}

We first show that balance is stable under the local operations used in
the induction. 

\begin{lemma}[Stability of balance]
\label{lem:balanceStability}
\label{lem:localChanges}
\label{stable}
\label{lem:gluing_preserves_balanced_prop}
Let $D$ be a digraph and $Z\subseteq V(D)$. 
The following statements hold.
\begin{enumerate}
\item If $Z$ is 
$\alpha$-balanced in $D$, then, for every $F\subseteq A(D)$, the set $Z$
is $(\alpha+|F|)$-balanced in $D-F$.

\item Suppose that 
$|\partial_D^+(Z)|\le r$ and $Z$ is $\beta$-balanced in
$D$. Let $D'\coloneqq D[Z]$. Then, for every $W\subseteq Z$, if $W$ is
$\alpha$-balanced in $D$, then $W$ is $(\alpha+2r+\beta)$-balanced in
$D'$.

\item Let $D_a$ and $D_b$ be two boundaried digraphs with the same boundary
$B$. Assume that $D_a$ is $\alpha$-balanced, and let
$D\coloneqq D_a\oplus D_b$.
Then $D$ is $(\alpha+|A(D_b)|)$-balanced.
\end{enumerate}
\end{lemma}

We next define a notion of {\em interfaces}. In an
undirected graph, a small cut gives a small set of vertices (i.e., their endpoints) through which the two sides of the cut interact. In a directed graph this need not be true based on the outgoing arc set alone being small. However, the balance assumption gives control of both
incoming and outgoing boundary arcs, and hence we can define a relevant ``interface''.

\newcommand{\intf}{\mathsf{intf}}
\newcommand{\outf}{\mathsf{outf}}

\begin{definition}[Interface and outerface]
Let $D$ be a digraph and let $Z\subseteq V(D)$. The \emph{interface} of
$Z$, denoted by $\intf(Z)$, is the set of vertices of $Z$ incident with an
arc crossing between $Z$ and $V(D)\setminus Z$. Thus
\[
\intf(Z)=
\{u\in Z \mid \text{there exists } v\notin Z \text{ such that }
(u,v)\in A(D) \text{ or } (v,u)\in A(D)\}.
\]
The \emph{outerface} of $Z$, denoted by $\outf(Z)$, is the set of vertices
outside $Z$ incident with an arc crossing between $Z$ and
$V(D)\setminus Z$. Thus
\[
\outf(Z)=
\{u\notin Z \mid \text{there exists } v\in Z \text{ such that }
(u,v)\in A(D) \text{ or } (v,u)\in A(D)\}.
\]
\end{definition}

\newcommand{\auxv}{\mathsf{auxv}}
\newcommand{\rintf}{\mathsf{rintf}}

\begin{definition}[Routed interface]
Let $\widehat D$ be a routed digraph. 

Define
$\auxv(\widehat D)\coloneqq
\tail(A_{\mathrm{aux}})\cup \head(A_{\mathrm{aux}})$. For $X\subseteq V(\widehat D)$, define the \emph{routed interface} of $X$ by
$\rintf_{\widehat D}(X)\coloneqq
\intf_{\widehat D_{\mathrm{real}}}(X)
\cup
(X\cap \auxv(\widehat D))$.
\end{definition}

The above definition is deliberately more generous than only adding endpoints
of auxiliary arcs {\em crossing} $X$, but it keeps the definition simple. Later we
only need boundedness, not a minimum interface. Since the number of auxiliary
arcs is always bounded by the parameters, this loose approximation only changes
the constants by a bounded amount.

The next observation explains why the interface can be treated as an
ordinary separator in the underlying undirected graph.

\begin{observation}\label{obs:InterfaceSeparator}
Let $D$ be a digraph, and let $Z,Y\subseteq V(D)$ be non-empty disjoint
vertex sets. Let
$X=\intf(Z)$.
If $X\subset Z$, then $X$ is a $(Z\setminus X)$-$Y$ separator in the
undirected graph underlying $D$.
\end{observation}

\subsubsection{Minimum cut sequences and augmented chips}
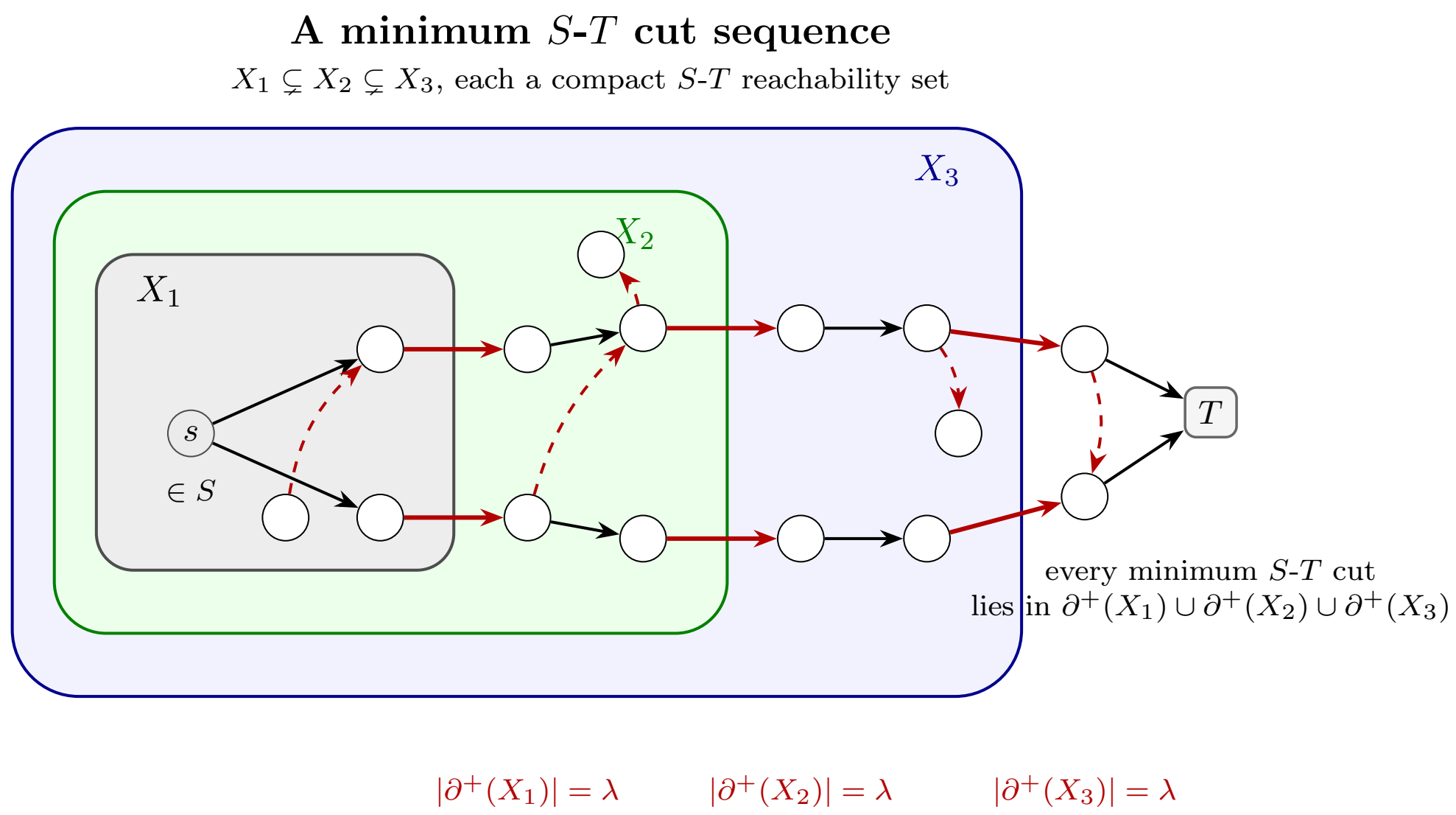
\begin{figure}[t]
\centering
\resizebox{0.82\textwidth}{!}{%
\begin{tikzpicture}[
  >={Stealth[length=2.2mm]},
  vtx/.style={circle, draw, fill=white, inner sep=0.4pt, minimum size=4.4mm, font=\footnotesize},
  svtx/.style={circle, draw=black!70, fill=black!8, inner sep=0.4pt, minimum size=4.4mm, font=\footnotesize},
  X1/.style={rounded corners=10pt, draw=black!70, fill=black!7, line width=0.8pt},
  X2/.style={rounded corners=14pt, draw=green!50!black, fill=green!8, line width=0.8pt},
  X3/.style={rounded corners=18pt, draw=blue!55!black, fill=blue!5, line width=0.8pt},
  tbox/.style={rounded corners=3pt, draw=black!60, fill=black!4, line width=0.7pt, font=\footnotesize},
  real/.style={->, thick, black},
  cut/.style={->, very thick, red!70!black},
  aux/.style={->, thick, red!70!black, dashed},
  setn/.style={font=\small\bfseries},
  tag/.style={font=\scriptsize},
  title/.style={font=\bfseries},
]

\draw[X3] (0.0,-0.3) rectangle (9.6,5.1);
\draw[X2] (0.4,0.3) rectangle (6.8,4.5);
\draw[X1] (0.8,0.9) rectangle (4.2,3.9);

\node[setn] at (1.4,3.55) {$X_1$};
\node[setn, green!50!black] at (5.9,4.1) {$X_2$};
\node[setn, blue!55!black]  at (8.8,4.7) {$X_3$};

\node[svtx] (s) at (1.7,2.2) {$s$};
\node[tag]  at (1.7,1.65) {$\in S$};

\node[tbox] (T) at (11.4,2.4) {$T$};

\node[vtx] (a1) at (3.5,3.0) {};
\node[vtx] (a2) at (3.5,1.4) {};
\node[vtx] (b1) at (6.0,3.2) {};
\node[vtx] (b2) at (6.0,1.2) {};
\node[vtx] (c1) at (8.7,3.2) {};
\node[vtx] (c2) at (8.7,1.2) {};

\node[vtx] (h1) at (4.9,3.0) {};
\node[vtx] (h2) at (4.9,1.4) {};
\node[vtx] (g1) at (7.5,3.2) {};
\node[vtx] (g2) at (7.5,1.2) {};
\node[vtx] (e1) at (10.2,3.0) {};
\node[vtx] (e2) at (10.2,1.6) {};

\node[vtx] (d1) at (2.6,1.4) {};
\node[vtx] (f1) at (5.6,3.9) {};
\node[vtx] (k1) at (9.0,2.2) {};

\draw[real] (s)  -- (a1);
\draw[real] (s)  -- (a2);
\draw[real] (h1) -- (b1);
\draw[real] (h2) -- (b2);
\draw[real] (g1) -- (c1);
\draw[real] (g2) -- (c2);

\draw[cut] (a1) -- (h1);
\draw[cut] (a2) -- (h2);
\draw[cut] (b1) -- (g1);
\draw[cut] (b2) -- (g2);
\draw[cut] (c1) -- (e1);
\draw[cut] (c2) -- (e2);

\draw[real] (e1) -- (T);
\draw[real] (e2) -- (T);

\draw[aux] (d1) to[bend left=20] (a1);          
\draw[aux] (b1) to[bend right=18] (f1);         
\draw[aux] (h2) to[bend left=15] (b1);          
\draw[aux] (c1) to[bend left=18] (k1);          
\draw[aux] (e1) to[bend left=18] (e2);          

\node[tag, red!70!black] at (4.9,-1.2) {$|\partial^+(X_1)|=\lambda$};
\node[tag, red!70!black] at (7.5,-1.2) {$|\partial^+(X_2)|=\lambda$};
\node[tag, red!70!black] at (10.2,-1.2) {$|\partial^+(X_3)|=\lambda$};

\node[title] at (5.5,6.0) {A minimum $S$-$T$ cut sequence};
\node[tag] at (5.5,5.55)
  {$X_1\subsetneq X_2\subsetneq X_3$, each a compact $S$-$T$ reachability set};
\node[tag, align=center] at (11.4,0.7)
  {every minimum $S$-$T$ cut\\ lies in
   $\partial^+(X_1)\cup\partial^+(X_2)\cup\partial^+(X_3)$};

\end{tikzpicture}%
}
\caption{A minimum $S$-$T$ cut sequence $X_1\subsetneq X_2\subsetneq X_3$. Each
set $X_i$ is a compact $S$-$T$ reachability set, so its outgoing boundary
$\partial^+(X_i)$ has the minimum size $\lambda=\lambda(S,T)$, drawn here as the
solid red arcs leaving each band. The source $S$ lies inside the innermost set
and the target $T$ lies outside the outermost set. The dashed arcs are auxiliary
arcs in the routed setting, treated as infinite-capacity arcs, and none of them
leaves any $X_i$, so they never appear in a finite minimum cut. The defining
coverage property is that every minimum $S$-$T$ cut is contained in
$\partial^+(X_1)\cup\partial^+(X_2)\cup\partial^+(X_3)$.}
\label{fig:minCutSequence}
\end{figure}

We next recall the notion of a minimum cut sequence. This sequence decomposes the reachable side of the graph into layers between consecutive compact
reachability sets. The key property of this decomposition that we shall use
is that a cut using an arc strictly inside one layer cannot be minimum.
\Cref{fig:minCutSequence} shows an example.

\begin{definition}
\label{def:minCutSequence}
Let $D$ be a digraph and let $S,T$ be disjoint non-empty vertex sets.
An ordered collection $\mathcal X=(X_1,\dots,X_q)$ of vertex sets is a
\emph{minimum $S$-$T$ cut sequence} if the following hold.
\begin{enumerate}
\item $X_1\varsubsetneq X_2\varsubsetneq \cdots \varsubsetneq X_q$.
\item For each $i\in [q]$, the set $X_i$ is a compact $S$-$T$
reachability set.
\item Every minimum $S$-$T$ cut is contained in
$\bigcup_{i=1}^q \partial^+(X_i)$.
\item $X_1$ is inclusion-wise minimal among compact $S$-$T$ reachability sets 
and $X_q$ is inclusion-wise maximal among compact $S$-$T$ reachability sets.
\end{enumerate}
This definition extends in the natural way to routed digraphs, by treating the auxiliary arcs as infinite capacity arcs.  
\end{definition}

The following standard proposition follows from the residual graph structure in the maximum flow computation of the Ford-Fulkerson algorithm. 


\begin{proposition}[Minimum cut sequences in routed digraphs]
\label{prop:routedCutSequenceExists}
For every routed digraph $\widehat D$ and disjoint vertex sets $S$ and $T$ such that $\lambda_{\widehat D}(S,T)$ is finite, a minimum $S$-$T$ cut sequence exists and can be computed in polynomial time. Moreover, every set in the sequence is auxiliary-respecting.
\end{proposition}

\begin{proof}
Give every real arc capacity $1$ and every auxiliary arc capacity
$|A_{\mathrm{real}}|+1$, and apply the standard residual-graph construction
of a minimum cut sequence. Since a real cut exists, every minimum cut has
value at most $|A_{\mathrm{real}}|$. Thus no auxiliary arc crosses any of
its source sides. The resulting sets are auxiliary-respecting compact
reachability sets and form the required sequence.
\end{proof}

\begin{lemma}[Layer arcs force a larger cut]
\label{lem:layerArcRaisesCut}
Let $\widehat D$ be a routed digraph, and let $S,T$ be disjoint non-empty
vertex sets such that there is an $S$-$T$ path in $\widehat D$ and
$\lambda_{\widehat D}(S,T)$ is finite. Let
$\mathcal X=(X_1,\dots,X_q)$ be a minimum $S$-$T$ cut sequence in
$\widehat D$, and let $\ell=\lambda_{\widehat D}(S,T)$.

Set $X_0=\emptyset$ and $X_{q+1}=R(S,T)$, where $R(S,T)$ denotes the set of vertices reachable from $S$ in $\widehat D-T$. For each $i\in [q+1]$, let
$L_i\coloneqq X_i\setminus X_{i-1}$. Then every real $S$-$T$ cut in
$\widehat D$ that contains a real arc with both endpoints in $L_i$ for
some $i\in [q+1]$ has size at least $\ell+1$.
\end{lemma}

\begin{proof}
Let $a=(u,v)$ be a real arc with both endpoints in $L_i$ for some
$i\in [q+1]$. By the definition of the layers, the arc $a$ is not contained
in
$
\bigcup_{j=1}^q \partial^+_{\mathrm{real}}(X_j)$.

Hence no minimum $S$-$T$ cut contains $a$, by the definition of a minimum
cut sequence. Therefore every real $S$-$T$ cut containing $a$ has size at
least $\ell+1$.
\end{proof}

For the induction it is convenient to use an {\em augmented version} of a chip.
The original target set $T$ is henceforth denoted by $T_1$. We also keep two auxiliary
sets. The set $T_2$ contains vertices that the chip is forbidden to contain,
and the set $Q$ contains vertices which we call requests, since they are required to remain connected to
$T_1$ after the chip is removed.

\begin{definition}[Augmented chips]\label{def:qchip}
Let $\widehat D$ be a routed digraph, and let
$S,T_1,T_2,Q\subseteq V(\widehat D)$. A set
$Z\subseteq V(\widehat D)$ is an $\sttqpk$-chip in $\widehat D$ if the
following conditions hold.
\begin{enumerate}
\item $Z$ is an auxiliary-respecting strict $S$-$T_1$ reachability set in
$\widehat D$.
\item $|\partial^+_{\mathrm{real}}(Z)|\leq k$.
\item $\widehat D_{\mathrm{real}}[Z]\models \pi$.
\item $T_2\cap Z=\emptyset$.
\item For every $q\in Q$, there is a $q$-$T_1$ path in $\widehat D-Z$.
\end{enumerate}
\end{definition}

Notice that every
$(S,T,\emptyset,\emptyset,\pi,k)$-chip is an ordinary
$(S,T,\pi,k)$-chip, since the forbidden and request conditions
are vacuous. We allow $T_1=\emptyset$. In this case $\emptyset$ is the unique
minimal $S$-$\emptyset$ cut, so a reachability set $Z$ is strict exactly when
$\partial^+(Z)=\emptyset$.

\begin{definition}[Guarded routed instances]
Let $\eta\in\mathbb N_0$. 
\begin{itemize}\item We say that
$(\widehat D,S,T_1,T_2,Q)$ is {\em guarded} if $\tail(A_{\mathrm{aux}}(\widehat D))\subseteq T_2$.
\item Moreover, $(\widehat D,S,T_1,T_2,Q)$ is \emph{$\eta$-guarded} if it is guarded and 
$|A_{\mathrm{aux}}(\widehat D)|\le \eta$.
\end{itemize}
\end{definition}

We maintain guardedness throughout the induction. The number of auxiliary
arcs stays bounded by the parameters, and their tails remain forbidden.

\begin{remark}[Requests and guarded auxiliary arcs]
{\rm
The set $Q$ records vertices outside the chip that must still reach $T_1$
after the chip is removed. This is needed to preserve strictness through the
recursive operations. For example, if
$a=(u,v)\in\partial^+_{\real}(Z)$ is deleted, then the smaller instance no
longer contains the arc witnessing that $v$ is just outside $Z$. Adding $v$
to $Q$ requires $v$ to retain a path to $T_1$
avoiding $Z$.

Auxiliary arcs have a separate role. They remember routes through parts of
the graph that have been abstracted away. The guardedness condition
$\tail(A_{\aux}(\widehat D))\subseteq T_2$, together with
$Z\cap T_2=\emptyset$, implies that no chip contains the tail of an
auxiliary arc. Hence every chip in a guarded instance is automatically
auxiliary-respecting:
\[
  \partial^+_{\aux}(Z)=\emptyset
  \qquad\text{and hence}\qquad
  \partial^+_{\widehat D}(Z)=\partial^+_{\real}(Z).
\]
Thus, for chips in guarded instances, the subscript $\real$ on the outgoing
boundary is mathematically redundant. We keep it in formal statements and
recurrences to emphasize that cuts, cut sizes, hitting sets, balance, and
budgets count only real arcs. In prose, we may simply refer to the outgoing
boundary of a chip.

The original instance has no auxiliary arcs. They are introduced only by the
crossing reduction. The nested reduction is nevertheless stated for routed
instances so that it can be applied after earlier crossing reductions.
}
\end{remark}

For the rest of this subsection, fix an integer $k$, a CMSO sentence $\pi$, an  $\alpha$-balanced routed digraph
$\widehat D$ and vertex sets $S,T_1,T_2,Q\subseteq V(\widehat D)$ with
$S\cap (T_1\cup T_2\cup Q)=\emptyset$. 

\begin{definition}[Important augmented chips]
An $\sttqpk$-chip $Z$ in $\widehat D$ is said to be \emph{important} if
there is no $\sttqpk$-chip $Z'$ in $\widehat D$ such that $Z'\supset Z$ and
$|\partial^+_{\mathrm{real}}(Z')|\le |\partial^+_{\mathrm{real}}(Z)|$.
We denote the set of all important $\sttqpk$-chips in $\widehat D$ by
$\Ical_{\widehat D}\sttqpk$. For a real arc
$a\in A_{\mathrm{real}}(\widehat D)$, we denote by
$\Ical_{\widehat D}^a\sttqpk$ the set of all important $\sttqpk$-chips
$Z$ in $\widehat D$ such that
$a\in\partial^+_{\mathrm{real}}(Z)$. The reference to $\widehat D$ is
omitted when it is clear from the context.
\end{definition}

Let us make a simple observation regarding some edge cases. 

\begin{observation}\label{ex:trivialChip}
If there is no $S$-$T_1$ path in $\widehat D$, then either there is no
$\sttqpk$-chip in $\widehat D$, or the unique $\sttqpk$-chip is the set
$R(S,\emptyset)$ of vertices reachable from $S$ in $\widehat D$. Moreover,
if $\lambda_{\widehat D}(S,T_1)>k$, then there is no $\sttqpk$-chip in
$\widehat D$.
\end{observation}

Let us next consider the standard deletion step from important separator arguments. If we know that a real arc $a=(u,v)$ leaves our target important chip, then we delete $a$,
reduce the budget by one, and remember that $v$ must stay outside the chip
and must still reach the target without going through the chip.

For a real arc $a\in A_{\mathrm{real}}(\widehat D)$, we write
$\widehat D-a$ for the routed digraph obtained by deleting $a$ from
$A_{\mathrm{real}}(\widehat D)$ and leaving the auxiliary arcs unchanged.

\begin{lemma}[Deletion step]\label{lem:inducedChips}
Suppose $Z$ is an $\sttqpk$-chip in $\widehat D$ and let
$a=(u,v)\in \partial^+_{\mathrm{real}}(Z)$. Then $Z$ is an
$
(S,T_1,T_2\cup \{v\},Q\cup \{v\},\pi,k-1)
$-chip in $\widehat D-a$.

Moreover, if $Z$ is an important $\sttqpk$-chip in $\widehat D$, then $Z$
is an important
 $
(S,T_1,T_2\cup \{v\},Q\cup \{v\},\pi,k-1)
 $-chip in $\widehat D-a$.
\end{lemma}

As a consequence, once we find a bounded set of arcs hitting the outgoing
boundary of every important chip under consideration, counting those chips
reduces to instances with budget $k-1$. The rest of the proof is devoted to
finding such bounded hitting sets from the minimum cut sequence.

\begin{corollary}[Boundary-hitting reduction]\label{cor:inducedChips}
Let $a=(u,v)\in A_{\mathrm{real}}(\widehat D)$. Then
\[
\left|\Ical_{\widehat D}^a\sttqpk\right|
\le
\left|
\Ical_{\widehat D-a}(S,T_1,T_2\cup\{v\},Q\cup\{v\},\pi,k-1)
\right|.
\]

\end{corollary}

We next use the cut-sequence dichotomy to find a bounded set of candidate
arcs for the deletion step.

\subsubsection{The cut-sequence dichotomy}

\begin{figure}[t]
\centering
\begin{tikzpicture}[
  reg/.style={rounded corners=10pt, line width=0.8pt},
  rl/.style={font=\footnotesize},
  tg/.style={font=\scriptsize},
]
\draw[reg, draw=blue!55!black,  fill=blue!3]            (0,0)     rectangle (10.5,4.8);
\draw[reg, draw=red!65!black,   dashed, fill=red!4]     (1.8,0.9) rectangle (8.3,4.1);
\draw[reg, draw=green!45!black, fill=green!8]           (2.6,1.4) rectangle (6.7,3.6);
\draw[reg, draw=black!70,       fill=black!5]           (3.5,2.0) rectangle (5.5,3.0);
\draw[reg, draw=blue!60!black, dotted, line width=1pt]  (2.2,1.15) rectangle (7.5,3.85);
\node[rl, blue!55!black] at (3.1,4.5) {$X_{q+1}=R(S,T_1)$};
\node[rl] at (4.5,2.5) {$S=X_0$};
\node[tg, green!45!black] at (5.0,0.4) {$X_\mu$: last $\sttqpk$-chip};
\draw[green!45!black, ->, >=Stealth] (5.0,0.6) -- (4.9,1.45);
\node[tg, red!65!black] at (9.6,4.5) {$X_{\mu+1}$: first non-chip};
\draw[red!65!black, ->, >=Stealth] (8.6,4.4) -- (8.0,4.05);
\node[tg, blue!60!black] at (9.9,2.3) {Case 1: $X_\mu\!\subset\!Z\!\subseteq\!X_{\mu+1}$};
\draw[blue!60!black, ->, >=Stealth] (8.5,2.3) -- (7.55,2.3);
\end{tikzpicture}
\caption{The case distinction of \Cref{lem:possibleBranchings}. The chain
$S=X_0\subset X_1\subset\cdots\subset X_{q+1}$ consists of compact
$S$-$T_1$ reachability sets, except for the first and the last set
$X_{q+1}=R(S,T_1)$. The set $X_\mu$ is the last member that is a chip.
An important chip $Z$ either is nested in the band
$X_\mu\subset Z\subseteq X_{\mu+1}$ or crosses one of
$X_1,X_\mu,X_{\mu+1}$.}
\label{fig:dichotomy}
\end{figure}
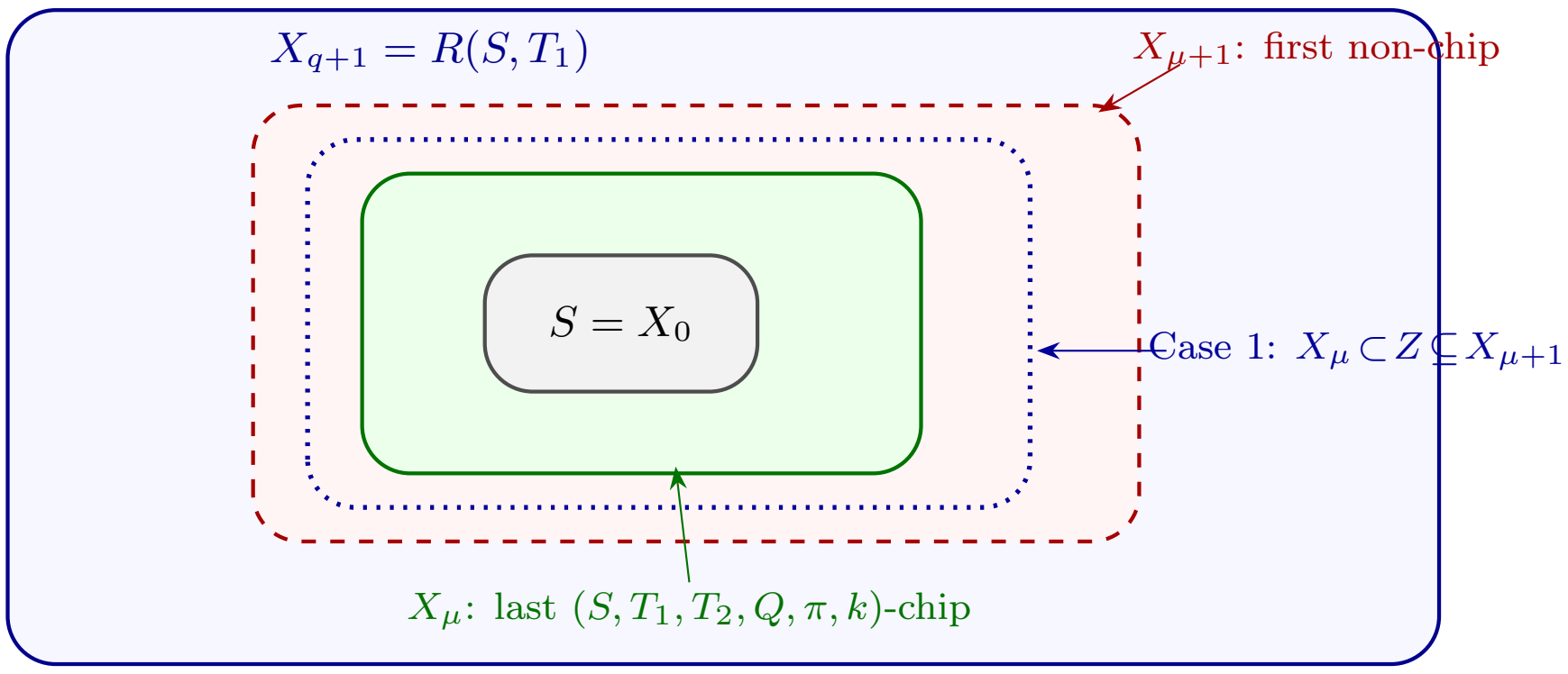

The nesting terminology below separates chips lying between consecutive
compact sets from chips crossing a compact set. This distinction leads to
the dichotomy in \Cref{lem:possibleBranchings}, which \Cref{fig:dichotomy} illustrates.

\begin{definition}[Nesting and crossing]\label{def:SSideAndTSideNesting}
Let $X$ and $Y$ be $S$-$T$ reachability sets. We say that $X$ and $Y$ are \emph{nested} if $X\subseteq Y$ or $Y\subseteq X$. Otherwise, we say that they \emph{cross}. If $X\subseteq Y$, then we say that $X$ is nested with $Y$ on the $S$-side, and that $Y$ is nested with $X$ on the $T$-side. If $X\subset Y$ and $Z$ is an $S$-$T$ reachability set satisfying $X\subset Z\subset Y$, then we say that $Z$ is \emph{nested between} $X$ and $Y$.
\end{definition}

\begin{figure}[t]
\centering
\resizebox{\textwidth}{!}{%
\begin{tikzpicture}[
  >={Stealth[length=2.2mm]},
  vtx/.style={circle, draw, fill=white, inner sep=0.4pt, minimum size=4.6mm, font=\footnotesize},
  svtx/.style={circle, draw=black!70, fill=black!8, inner sep=0.4pt, minimum size=4.6mm, font=\footnotesize},
  Xset/.style={rounded corners=12pt, draw=blue!55!black, fill=blue!6, line width=0.8pt},
  Zset/.style={rounded corners=12pt, draw=green!50!black, fill=green!12, line width=0.8pt},
  tbox/.style={rounded corners=3pt, draw=black!60, fill=black!4, line width=0.7pt, font=\footnotesize},
  real/.style={->, thick, black},
  cutx/.style={->, very thick, blue!60!black},
  cutz/.style={->, very thick, green!50!black},
  lead/.style={gray!55, line width=0.4pt},
  reg/.style={font=\footnotesize},
  setn/.style={font=\small\bfseries},
  tag/.style={font=\scriptsize},
  title/.style={font=\bfseries},
]

\begin{scope}
\draw[Xset] (0.5,0.4) rectangle (7.6,4.4);
\draw[Zset] (1.1,1.0) rectangle (4.9,3.8);
\node[setn, blue!55!black]  at (0.8,4.75) {$X$};
\node[setn, green!50!black] at (4.65,4.05) {$Z$};
\node[title] at (4.0,5.35) {Heredity makes a strict subset a chip};

\node[tag] at (6.2,3.9) {$Z\subseteq X$};

\node[svtx] (s)  at (1.9,2.4) {$s$};
\node[tag]  at (1.9,1.85) {$\in S$};
\node[vtx]  (z1) at (3.2,3.1) {};
\node[vtx]  (z2) at (3.5,1.8) {};
\node[vtx]  (xw) at (6.3,2.4) {};

\draw[real] (s)  -- (z1);
\draw[real] (z1) -- (z2);
\draw[real] (s)  -- (z2);

\node[tbox] (T1) at (9.0,2.4) {$T_1$};
\node[vtx]  (h1) at (8.0,2.9) {};
\node[vtx]  (h2) at (8.0,1.9) {};
\draw[cutz] (z1) to[out=10,in=170] (xw);
\draw[real] (xw) -- (h1);
\draw[real] (xw) -- (h2);
\draw[real] (h1) -- (T1);
\draw[real] (h2) -- (T1);

\node[tag, blue!55!black] at (3.0,0.7) {$\widehat D_{\mathrm{real}}[X]\models\pi$};

\node[tag] at (3.5,-0.2)
  {$Z$ strict with $|\partial^+_{\mathrm{real}}(Z)|\le k$, hence $Z$ is a chip};
\end{scope}

\draw[gray!45, dotted, line width=0.7pt] (10.6,-0.6) -- (10.6,5.2);

\begin{scope}[xshift=12.6cm]
\draw[Xset] (0.5,0.4) rectangle (8.2,4.4);
\draw[Zset] (1.1,1.0) rectangle (4.6,3.8);
\node[setn, blue!55!black]  at (0.8,4.75) {$X$};
\node[setn, green!50!black] at (4.35,4.05) {};
\node[title] at (4.5,5.35) {Domination forces an important nested chip to equal $X$};

\node[tag] at (6.0,3.95) {$Z\subsetneq X$, $X$ a compact chip};

\node[svtx] (s) at (1.9,2.4) {$s$};
\node[tag]  at (1.9,1.85) {$\in S$};

\node[vtx] (zt) at (4.0,3.2) {};
\node[vtx] (zm) at (4.0,2.4) {};
\node[vtx] (zb) at (4.0,1.6) {};
\node[vtx] (wt) at (5.8,3.2) {};
\node[vtx] (wm) at (5.8,2.4) {};
\node[vtx] (wb) at (5.8,1.6) {};

\draw[cutz] (zt) -- (wt);
\draw[cutz] (zm) -- (wm);
\draw[cutz] (zb) -- (wb);

\node[tbox] (T1b) at (9.6,2.4) {$T_1$};
\node[vtx] (xt) at (7.6,3.0) {};
\node[vtx] (xb) at (7.6,1.8) {};
\draw[cutx] (xt) -- (8.6,3.0);
\draw[cutx] (xb) -- (8.6,1.8);
\draw[real] (8.6,3.0) -- (T1b);
\draw[real] (8.6,1.8) -- (T1b);

\node[tag, green!45!black] (Ztag) at (3.6,0.65)
  {$|\partial^+_{\mathrm{real}}(Z)|\ge\lambda_{\widehat D}(S,T_1)$};

\node[tag, blue!60!black] (Xtag) at (7.6,0.7)
  {$|\partial^+_{\mathrm{real}}(X)|=\lambda_{\widehat D}(S,T_1)$};

\node[tag, align=left] at (4.2,-0.4)
  {$X\supset Z$ with $|\partial^+_{\mathrm{real}}(X)|\le|\partial^+_{\mathrm{real}}(Z)|$,
   so $X$ dominates $Z$};
\end{scope}

\end{tikzpicture}%
}
\caption{An illustration of the two statements of   \Cref{lem:monotonicityOfChips}. \\ \noindent
{\bf (a)} On the left,
$X$ is a compact $S$-$T_1$ chip and $Z\subseteq X$ is a strict reachability set
of real cut size at most $k$. The graph $\widehat D_{\mathrm{real}}[Z]$ is an
induced subgraph of $\widehat D_{\mathrm{real}}[X]$, so heredity of $\pi$ carries
the chip property down to $Z$. As the contrapositive, if $Z$ is not a chip, then $X$ cannot be a chip either.
\\ \noindent
{\bf (b)} On the right, the outgoing boundary of the
compact set $X$ has the minimum value $\lambda_{\widehat D}(S,T_1)$, while every proper
reachability subset $Z\subsetneq X$ has outgoing boundary of size at least
$\lambda_{\widehat D}(S,T_1)$. Since $X$ properly contains $Z$ and has no larger outgoing
boundary, $X$ dominates $Z$, so an important chip nested with $X$ on the
$S$-side must equal $X$.}
\label{fig:monotonicity}
\end{figure}
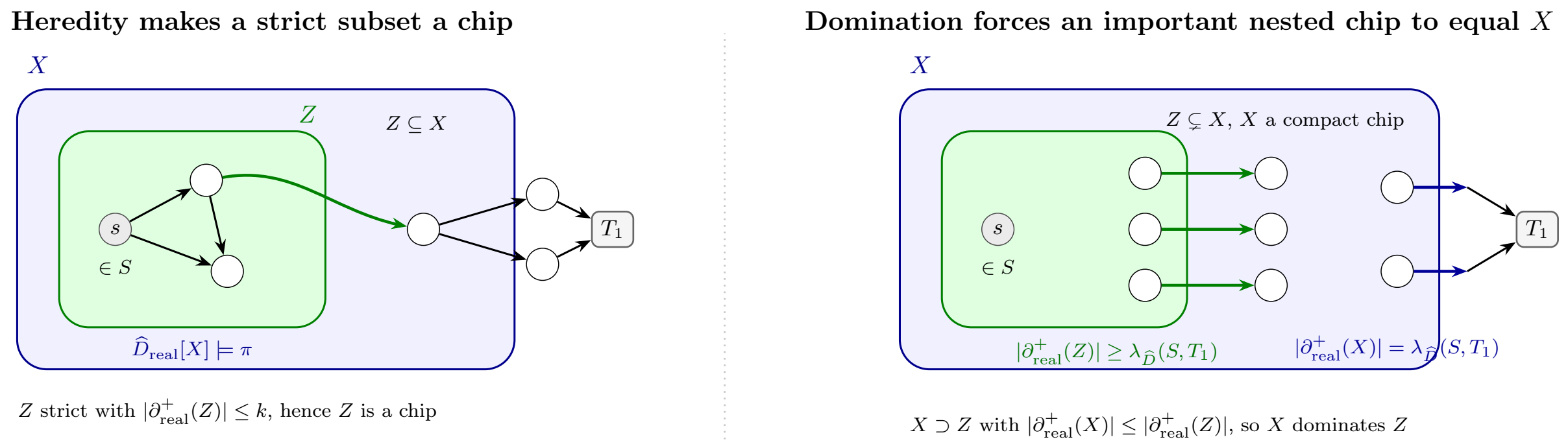

We next observe a simple but crucial monotonicity property. If a compact reachability
set is already a chip, then every subset of it that is a  strict reachability set with cut size at most $k$ is also a chip. As compact sets have the minimum possible cut size, an important chip cannot sit properly contained inside such a compact chip. \Cref{fig:monotonicity} illustrates both statements of \Cref{lem:monotonicityOfChips}.

\begin{lemma}[Monotonicity inside compact chips]
\label{lem:monotonicityOfChips}
\label{lem:trivialConsequenceOfMonotonicityLemma}
Suppose $\pi$ is hereditary, and let $X$ be a compact $S$-$T_1$
reachability set in $\widehat D$ that is an $\sttqpk$-chip. The following
statements hold.
\begin{enumerate}
\item If $Z\subseteq X$ is an auxiliary-respecting strict $S$-$T_1$
reachability set in $\widehat D$ and
$|\partial_{\mathrm{real}}^+(Z)|\le k$, then $Z$ is an $\sttqpk$-chip in
$\widehat D$.  
\item If $Z$ is an important $\sttqpk$-chip in $\widehat D$ and is nested
with $X$ on the $S$-side, then $Z=X$.
\end{enumerate}
\end{lemma}

\begin{figure}[t]
\centering
\begin{tikzpicture}[
  >={Stealth[length=2.4mm]},
  vtx/.style={circle, draw, fill=white, inner sep=0.5pt, minimum size=4.6mm, font=\footnotesize},
  arc/.style={->, thick, black},
  pth/.style={->, thick, orange!85!black},
  rl/.style={font=\small\bfseries},
  tg/.style={font=\scriptsize},
]
\draw[draw=blue!55!black,  fill=blue!7,  line width=0.8pt] (3.0,2.5) ellipse [x radius=2.6, y radius=1.8];
\draw[draw=green!45!black, fill=green!7, fill opacity=0.6, line width=0.8pt] (6.5,2.5) ellipse [x radius=2.8, y radius=1.8];
\node[rl, blue!55!black]  at (1.3,4.2) {$X$};
\node[rl, green!45!black] at (8.7,4.0) {$Z$};
\node[vtx] (s) at (4.3,2.7) {$s$};  \node[tg] at (4.3,3.25) {$\in S$};
\node[vtx] (p) at (5.0,1.9) {$p$};
\node[vtx] (h) at (6.2,1.7) {$h$};
\node[vtx] (z) at (7.7,1.4) {$z$};
\node[vtx] (g) at (8.1,3.0) {$g$};
\draw[rounded corners=6pt, draw=black!55, fill=black!3] (10.4,2.2) rectangle (11.8,3.8);
\node[vtx] (y) at (11.0,3.0) {$y$};
\node[rl] at (11.1,3.55) {$T$};
\draw[pth] (s) to[out=-40,in=120] (p);
\draw[pth] (p) -- (h);
\draw[pth] (h) to[out=-30,in=170] (z);
\draw[pth, dashed] (h) to[out=40,in=230] (g);
\draw[arc] (g) -- (y);
\node[tg, orange!75!black] at (5.35,2.45) {$b=(p,h)$};
\node[tg] at (9.6,3.55) {$a\in\partial^+(Z)$};
\node[tg] at (3.0,0.4) {path in $D[Z]$ from $S$ to $z\in Z\setminus X$};
\end{tikzpicture}
\caption{The crossing case (\Cref{lem:crossingCaseBudgetDrop}). The sets $X$
and $Z$ cross, with $S\subseteq X\cap Z$. A path in $D[Z]$ from $S$ to
$z\in Z\setminus X$ leaves $X$ through $b=(p,h)\in\partial^+(X)$, where
$p\in\intf(X)\cap Z$ and $h\in Z\setminus X$. This witnesses
$\intf(X)\cap Z\neq\emptyset$. A path in $D-X$ from $h$ to $T$ (dashed) first
leaves $Z$ through an arc $a=(g,y)$ with $g\in Z\setminus X$ and
$y\notin X\cup Z$, so $a\in\partial^+(Z)\setminus\partial^+(X)$.}
\label{fig:crossing}
\end{figure}
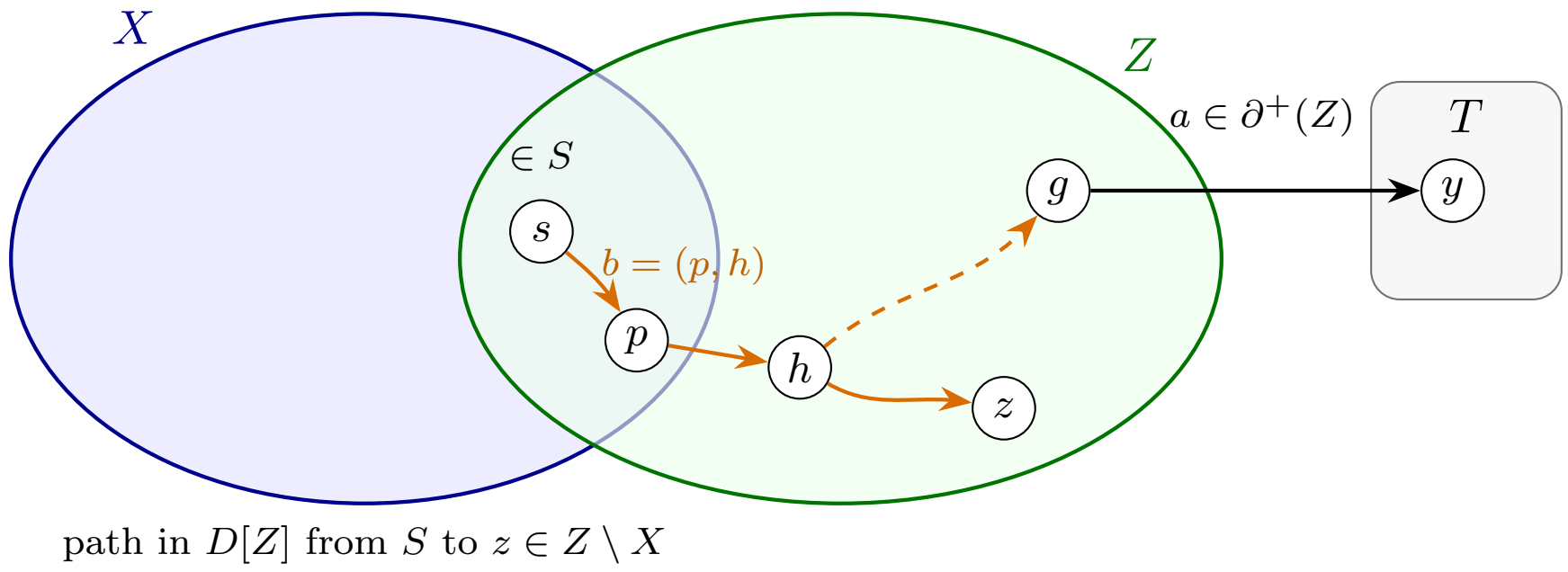

The next lemma gives the two properties of crossing that we need later. \Cref{fig:crossing} illustrates both properties.

\begin{lemma}[Crossing forces a budget drop]\label{lem:crossingCaseBudgetDrop}
Let $Z$ and $X$ be strict $S$-$T$ reachability sets that cross. Suppose
that $\partial^+(Z)\cap \partial^+(X)=\emptyset$. Then the following hold.
\begin{enumerate}
\item There is an arc $a=(u,v)$ such that $u\in Z\setminus X$ and
$v\notin X\cup Z$. In particular, $a\in \partial^+(Z)\cap \partial^+(X\cup Z)$, and $a\notin \partial^+(X)$.
\item  $\intf(X)\cap Z\neq \emptyset$.
\end{enumerate}\end{lemma}

\begin{proof}
Choose an $S$-$z$ path in $D[Z]$ for some $z\in Z\setminus X$, and let
$b=(p,h)$ be its first arc leaving $X$. Then
$p\in\intf(X)\cap Z$ and $h\in Z\setminus X$, proving the second item.
By strictness of $X$ and \Cref{lem:basicStrictFacts}, there is an $h$-$T$
path in $D-X$. Its first arc $a=(u,v)$ leaving $Z$ satisfies
$u\in Z\setminus X$ and $v\notin X\cup Z$, as required in the first item.
\end{proof}

\begin{observation}[Routed crossing budget drop]
\label{obs:routedCrossingBudgetDrop}
Let $\widehat D$ be a routed digraph, and let $X$ and $Z$ be strict
$S$-$T$ reachability sets in $\widehat D$ that are auxiliary-respecting and cross. Suppose that $\partial_{\mathrm{real}}^+(Z)\cap
\partial_{\mathrm{real}}^+(X)=\emptyset$.
Then there is a real arc $a=(u,v)$ such that
$
u\in Z\setminus X$ and 
$v\notin X\cup Z$.
In particular,
$a\in \partial_{\mathrm{real}}^+(Z)
\setminus
\partial_{\mathrm{real}}^+(X)$.
\end{observation}

\begin{proof}
Apply \Cref{lem:crossingCaseBudgetDrop} to the digraph containing all real
and auxiliary arcs. Since $X$ and $Z$ are auxiliary-respecting, their full
outgoing boundaries equal their real outgoing boundaries. Minimality of
these cuts is therefore the same in the ordinary and routed settings.
The resulting arc leaves $Z$, so it is real.
\end{proof}

This is the formal reason why the budget when dealing with a crossing chip can later be made smaller.
After boundary arcs of $X$ have been put into the temporary hitting set, any
remaining crossing chip has an outgoing real arc with tail in $Z\setminus X$ and head outside $X\cup Z$. That arc will disappear in the reduced instance.

The second conclusion of the lemma will be used to keep the number of
possible reduced instances bounded. Since a crossing chip must meet the
interface of $X$, and the interface has bounded size in a balanced graph
when the cut is small, only boundedly many attachment patterns have to be
remembered.

We now turn the preceding monotonicity and crossing facts into the case
distinction used in the induction. We compare an important chip with a fixed
minimum $S$-$T_1$ cut sequence. Let $X_\mu$ be the last member of the
sequence that is itself an augmented chip. Then an important chip is either
equal to $X_\mu$, lies between $X_\mu$ and the next compact set, or crosses
one of the relevant compact sets.

The last compact chip rules out smaller important chips by domination. Here a chip $Z'$ dominates a chip $Z$ if $Z'\supset Z$ and $|\partial^+_{\mathrm{real}}(Z')|\le|\partial^+_{\mathrm{real}}(Z)|$, so a chip is important exactly when no chip dominates it.
If the next compact set exists, it is not a chip. Heredity of $\pi$, the
forbidden set $T_2$, and the request condition then rule out chips containing it.

\begin{lemma}[Cut sequence dichotomy]\label{lem:possibleBranchings}
Suppose $\pi$ is hereditary and there is an $S$-$T_1$ path in
$\widehat D$. Let $Z$ be an important $\sttqpk$-chip in $\widehat D$, and
let $\mathcal X=(X_1,\dots,X_q)$ be a minimum $S$-$T_1$ cut sequence in
$\widehat D$. Set $X_0=S$ and $X_{q+1}=R(S,T_1)$.
Let $\mu$ be the largest index in $[q]$ such that $X_\mu$ is an
$\sttqpk$-chip in $\widehat D$, and set $\mu=0$ if no such index exists.
Then one of the following holds.
\begin{enumerate}
\item $Z=X_\mu$ or $X_\mu\subset Z\subseteq X_{\mu+1}$.
\item $\mu=0$ and $Z$ crosses $X_1$.
\item $\mu\geq 1$ and $Z$ crosses at least one of $X_\mu$ and
$X_{\mu+1}$.
\end{enumerate}
\end{lemma}

\begin{proof}
Every reachability set is contained in $R(S,T_1)$, since its source paths
avoid $T_1$. Also, for $i\in[q]$, if $X_i$ is not a chip, no chip can contain it. Indeed,
$X_i$ is compact and auxiliary-respecting, and its boundary size is at
most $k$ because $Z$ is a chip. Failure can therefore only concern $\pi$,
avoidance of $T_2$, or a request path. Each failure persists for every superset of $X_i$, by heredity of $\pi$
for the first condition and because deleting a larger set leaves fewer
paths for the other two.

If $\mu=0$ and $Z$ does not cross $X_1$, then $Z\subseteq X_1$ by the
preceding observation. Since $S=X_0\subseteq Z$, outcome 1 holds.

Suppose $\mu\ge1$ and $Z$ crosses neither $X_\mu$ nor $X_{\mu+1}$.
If $Z\subseteq X_\mu$, then \Cref{lem:monotonicityOfChips} gives
$Z=X_\mu$. Otherwise $X_\mu\subset Z$. When $\mu<q$, the non-chip
$X_{\mu+1}$ cannot be contained in $Z$, so nesting gives
$Z\subseteq X_{\mu+1}$. When $\mu=q$, this containment follows from
$X_{q+1}=R(S,T_1)$. Again outcome 1 holds.
\end{proof}

The rest of the argument treats the outcomes of the decomposition
separately. The nested case is handled by zooming into the region between two
consecutive members of the cut sequence, which increases the relevant
minimum cut value. The crossing case is handled by replacing the part of the
chip outside a compact set by a bounded representative, which decreases the
cut budget.

\subsubsection{The nested case}

We introduce {\em entry points} to track requests when a chip lies between
two consecutive compact $S$-$T_1$ reachability sets $X\subset X'$ and we
restrict the instance to the region between their outgoing arc sets. 
 However, a request vertex may lie outside this region, while its chosen path to $T_1$ may enter the region before leaving again. To ensure that we do not lose such paths, we keep track of the first point where this happens.

\newcommand{\avoidpath}[4]{P_{#1,#2}^{#3}(#4)}
\newcommand{\entry}[4]{\operatorname{ent}_{#2}^{#1,#3}(#4)}

\begin{definition}[Target paths and entry points]\label{def:suffixPath}
Let $\widehat D$ be a routed digraph, and let
$U,Z,W\subseteq V(\widehat D)$. For every vertex
$v\in V(\widehat D)$, fix once and for all a path $\avoidpath{Z}{W}{U}{v}$ from $v$ to $U$ in
$\widehat D-Z$, if such a path exists. The same path is used whenever
$\avoidpath{Z}{W}{U}{v}$ or $\entry{Z}{W}{U}{v}$ appears. Paths may use both
real and auxiliary arcs. If there is such a path that avoids $W$, then we
choose $\avoidpath{Z}{W}{U}{v}$ to avoid $W$. If no $v$-$U$ path exists in
$\widehat D-Z$, set
\[
\avoidpath{Z}{W}{U}{v}=\emptyset .
\]

If $\avoidpath{Z}{W}{U}{v}$ contains a vertex of $W$, then
$\entry{Z}{W}{U}{v}$ denotes the singleton set containing the first vertex
of $\avoidpath{Z}{W}{U}{v}$ (in the traversal from $v$ to $U$) that belongs to $W$. If
$\avoidpath{Z}{W}{U}{v}$ contains no vertex of $W$, then
\[
\entry{Z}{W}{U}{v}=\emptyset .
\]

For a set $Y\subseteq V(\widehat D)$, define
\[
\entry{Z}{W}{U}{Y}\coloneqq \bigcup_{v\in Y}\entry{Z}{W}{U}{v}.
\]
In words, $\entry{Z}{W}{U}{v}$ is where the chosen $v$-$U$ path avoiding
$Z$ first touches $W$. No entry point exists if such a path can avoid $W$
altogether. 
\end{definition}

The following immediate observation expresses the only property of entry
points that we will need.

\begin{observation}[First entries lie in the routed interface]
\label{obs:routedPrefPath}
Let $\widehat D$ be a routed digraph, let
$U,Z,W\subseteq V(\widehat D)$, and let $v\in V(\widehat D)\setminus W$.
\begin{enumerate}
	\item If $\entry{Z}{W}{U}{v}\neq \emptyset$, then $\entry{Z}{W}{U}{v}\subseteq \rintf_{\widehat D}(W)$.
	\item Moreover, if $\entry{Z}{W}{U}{v}=\{x\}$, then the prefix of
$\avoidpath{Z}{W}{U}{v}$ from $v$ to $x$ is contained in
$\widehat D-(W\setminus \{x\})$.
\end{enumerate}

\end{observation}

\begin{proof}[Proof of \Cref{obs:routedPrefPath}]
Let $\entry{Z}{W}{U}{v}=\{x\}$. Since $v\notin W$ and $x$ is the first
vertex of $\avoidpath{Z}{W}{U}{v}$ that belongs to $W$, the vertex
immediately preceding $x$ on this path lies outside $W$.

If the arc entering $x$ is real, then $x$ is incident with a real arc
crossing between $W$ and $V(\widehat D)\setminus W$. Hence
$x\in \intf_{\widehat D_{\mathrm{real}}}(W)$. If the arc entering $x$ is
auxiliary, then $x\in W\cap \auxv(\widehat D)$. In both cases,
$x\in \rintf_{\widehat D}(W)$.

The second statement follows from the choice of $x$ as the first vertex of
the path in $W$.
\end{proof}

We now handle the nested case for two consecutive members of the minimum
cut sequence. Let $X\subset X'$ be consecutive members of a minimum
$S$-$T_1$ cut sequence. We want to view an important chip
$Z$ with $X\subset Z\subset X'$ as a chip in a local instance defined
inside $X'$.

The initial band between $S$ and $X_1$, and the final band above $X_q$, are
handled separately later. They use the same idea, but their local sources and  targets
are slightly different.

For vertex sets $X\subseteq X'$ appearing consecutively in this sequence, define a new source-target pair as follows.  
\[S_X^\star\coloneqq X\cup \head(\partial^+_{\mathrm{real}}(X))
\qquad\text{and}\qquad
T_{X'}^\star\coloneqq \tail(\partial^+_{\mathrm{real}}(X')).
\]



\begin{remark}
{\em In the nested case, the arcs in $\partial^+_{\mathrm{real}}(X)\cup\partial^+_{\mathrm{real}}(X')$ will be added to the hitting set. Hence, when we analyze a chip $Z$ that has not already been accounted for through this hitting set, we may assume that $\partial^+_{\mathrm{real}}(Z)\cap\bigl(\partial^+_{\mathrm{real}}(X)\cup\partial^+_{\mathrm{real}}(X')\bigr)=\emptyset$.
This assumption has two useful consequences. First, the head of any arc
leaving $X$ belongs to $Z$, and hence $S_X^\star\subseteq Z$. Second, no
vertex of $Z$ is the tail of an arc leaving $X'$, and hence
$Z\cap T_{X'}^\star=\emptyset$.
}
\end{remark}

It remains to address the request set. A request $q\in Q$ that already
lies in $X'$ can be kept as it is since any relevant $q$-$T_1$ path in $\widehat D$ would have to escape
  $X'$ precisely through $T_{X'}^\star$. On the other hand, for a request $q\in Q\setminus X'$, we
look at the chosen path $\avoidpath{Z}{X'}{T_1}{q}$ from $q$ to $T_1$ in
$\widehat D-Z$. If this path enters $X'$, we keep only its first entry vertex in
$X'$. Thus we define
\[
\Delta_Z \coloneqq \entry{Z}{X'}{T_1}{Q\setminus X'}
\text{, }
\quad 
Q^\star \coloneqq (Q\cap X')\cup \Delta_Z .
\]
In words, $Q^\star$ consists of the old requests  already inside the local
region, together with the first points where paths to $T_1$ from the exterior requests enter the sandwiched region.

The point of this definition is that a path from an exterior request to
$T_1$ can now be split into two parts. The prefix reaches a vertex of
$Q^\star$ without using the local region earlier, and the new request
condition will preserve a path from this vertex to the local target
$T_{X'}^\star$. Once a path reaches $T_{X'}^\star$, it can be extended to
the original target $T_1$ outside $X'$. This extension is guaranteed by
strictness of $X'$.

\begin{figure}[t]
\centering
\begin{tikzpicture}[
  >={Stealth[length=2.4mm]},
  vtx/.style={circle, draw, fill=white, inner sep=0.5pt, minimum size=4.6mm, font=\footnotesize},
  arc/.style={->, thick, black},
  wit/.style={->, thick, red!70!black, dashed},
  reg/.style={rounded corners=9pt, line width=0.8pt},
  rl/.style={font=\small\bfseries},
  tg/.style={font=\scriptsize},
]
\draw[reg, draw=blue!55!black, fill=blue!4]   (0,0)     rectangle (12,5);
\node[rl, blue!55!black]  at (11.3,4.6) {$X'$};
\draw[reg, draw=green!45!black, fill=green!6] (0.6,0.6) rectangle (6.5,4.4);
\node[rl, green!45!black] at (5.9,4.0) {$Z$};
\draw[reg, draw=black!70, fill=black!5]       (1.2,1.6) rectangle (3.4,3.4);
\node[rl] at (1.75,3.05) {$X$};
\node[vtx] (s) at (2.3,2.5) {$s$};   \node[tg] at (2.3,1.95) {$\in S$};
\node[vtx] (w) at (5.1,2.5) {$w$};   \node[tg] at (5.1,1.95) {$\in N^+(X)$};
\node[vtx] (t) at (9.3,2.2) {$t^\star$}; \node[tg] at (9.3,1.65) {$\in T_{X'}^\star$};
\node[vtx] (r) at (8.0,5.0) {$r$};   \node[tg] at (8.0,5.5) {$\in Q^\star$};
\node[vtx] (q) at (2.0,6.0) {$q$};   \node[tg] at (2.0,6.5) {$\in Q\setminus X'$};
\node[vtx] (h) at (13.0,2.2) {};
\draw[reg, draw=black!55, fill=black!3] (12.4,1.4) rectangle (13.8,3.0);
\node[rl] at (13.1,2.75) {$T_1$};
\draw[arc] (s) -- (w); \node[tg] at (4.2,2.85) {{$\partial^+_{\mathrm{real}}(X)$}};
\draw[arc] (t) -- (h); \node[tg] at (10.8,2.55) {{$\partial^+_{\mathrm{real}}(X')$}};
\draw[wit] (q) -- (r);
\draw[wit] (r) to[out=-80,in=70] (t);
\node[tg, red!70!black] at (5.2,6.0) {chosen path $P_Z(q)$};
\end{tikzpicture}

\caption{The nested case (\Cref{lem:nestedWithin}). Here $X\subset Z\subset X'$ with
$\partial^+_{\mathrm{real}}(Z)$ disjoint from
$\partial^+_{\mathrm{real}}(X)\cup\partial^+_{\mathrm{real}}(X')$. The local source
$S_X^\star=X\cup\head(\partial^+_{\mathrm{real}}(X))$ lies inside $Z$.
Thus the head $w$ of any real arc leaving $X$ belongs to $Z$. The local target
$T_{X'}^\star$ consists of the tails $t^\star$ of real arcs leaving $X'$, so
$T_{X'}^\star\cap Z=\emptyset$. For a request $q\in Q\setminus X'$, the chosen
path $P_Z(q)=\avoidpath{Z}{X'}{T_1}{q}$ first meets $X'$ at $r\in Q^\star$ and then leaves $X'$ through
$\partial^+_{\mathrm{real}}(X')$ to reach $T_1$.}

\label{fig:nested}
\end{figure}
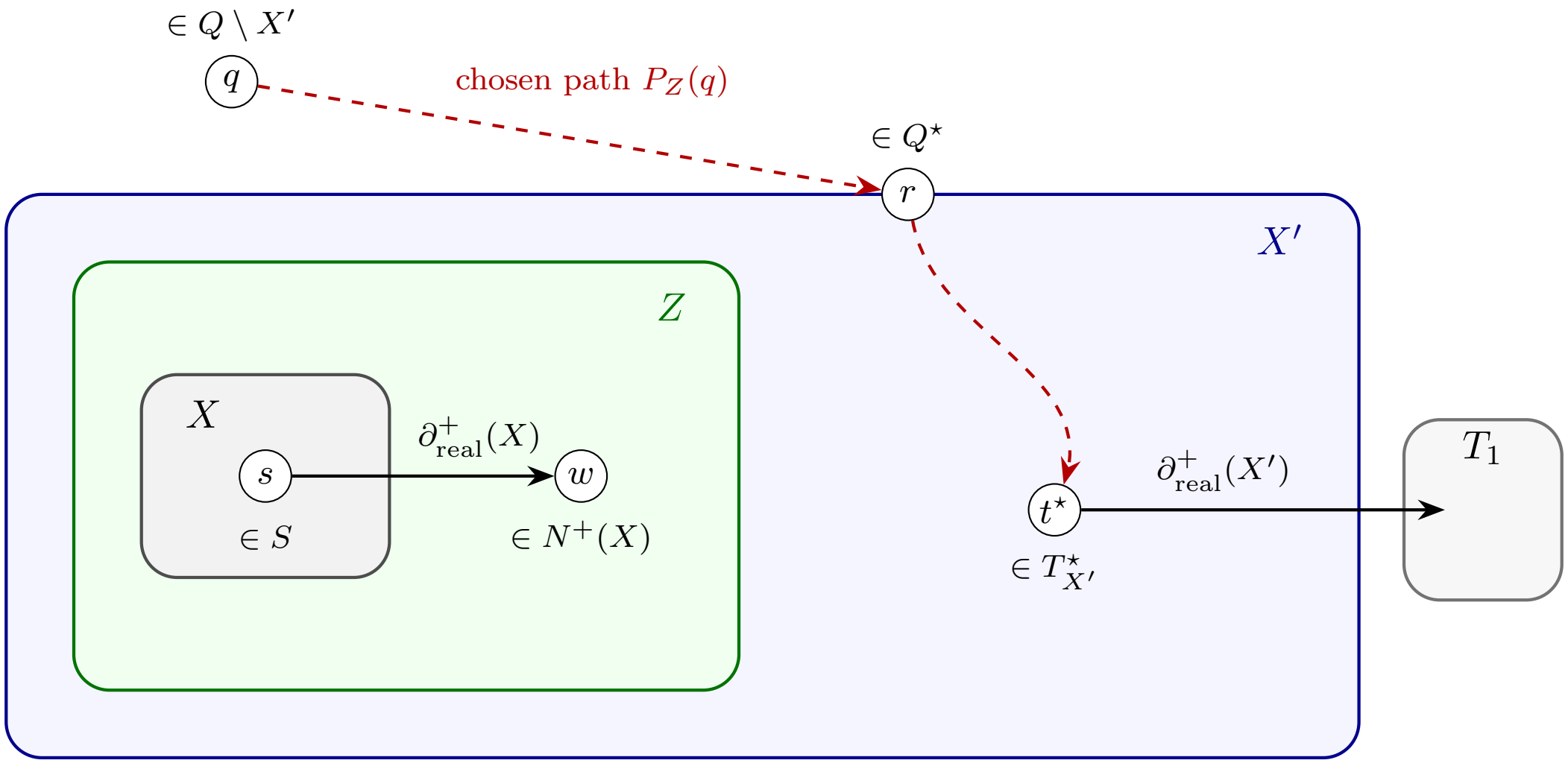

\begin{observation}[Exiting the outer region]\label{obs:exitFromXprime}
Let $X'$ be an auxiliary-respecting strict $S$-$T_1$ reachability set
in a routed digraph $\widehat D$. For every
$t\in T_{X'}^\star$, there is a $t$-$T_1$ path in $\widehat D$ whose first
arc is a real arc leaving $X'$ and whose remaining vertices lie in
$V(\widehat D)\setminus X'$.
\end{observation}

The reason for enlarging the local source and target is the following
increase in the minimum cut value. Any local cut is also a global
$S$-$T_1$ cut, but it must use an arc lying strictly between the two
consecutive regions. Such a cut cannot be minimum in the original instance.
This gives the required decrease of the secondary induction parameter, which is an upper bound on $k-\lambda_{\widehat D}(S,T_1)$.

\begin{observation}[Local cut increase]\label{obs:localCutIncrease}
Let $\widehat D$ be a routed digraph, and suppose that there is an
$S$-$T_1$ path in $\widehat D$. Let
$\mathcal X=(X_1,\ldots,X_q)$ be a minimum $S$-$T_1$ cut sequence in
$\widehat D$. Let $X$ and $X'$ be consecutive members of $\mathcal X$. In particular,
both $X$ and $X'$ are auxiliary-respecting by
\Cref{prop:routedCutSequenceExists}. 
Suppose that there is an auxiliary-respecting set $Z$ such that
$X\subseteq Z\subseteq X'$ and
\[
\partial^+_{\mathrm{real}}(Z)\cap
\bigl(
\partial^+_{\mathrm{real}}(X)\cup
\partial^+_{\mathrm{real}}(X')
\bigr)
=
\emptyset .
\]
Then
\[
\lambda_{\widehat D}(S_X^\star,T_{X'}^\star)
>
\lambda_{\widehat D}(S,T_1).
\]
\end{observation}

The next lemma also preserves importance in the local instance. \Cref{fig:nested} illustrates the local instance.

\begin{lemma}[Nested localization]\label{lem:nestedWithin}
Suppose $\pi$ is hereditary, and suppose that there is an $S$-$T_1$ path
in the routed digraph $\widehat D$. Let
$\mathcal X=(X_1,\ldots,X_q)$ be a minimum $S$-$T_1$ cut sequence in
$\widehat D$. Let $X$ and $X'$ be consecutive members of $\mathcal X$.
%
Let $Z$ be an important $\sttqpk$-chip in $\widehat D$ such that
$X\subset Z\subset X'$ and suppose that
\[
\partial^+_{\mathrm{real}}(Z)\cap
\bigl(
\partial^+_{\mathrm{real}}(X)\cup
\partial^+_{\mathrm{real}}(X')
\bigr)
=\emptyset .
\]
Set
\[
\Delta_Z \coloneqq \entry{Z}{X'}{T_1}{Q\setminus X'}
\qquad\text{and}\qquad
Q^\star \coloneqq (Q\cap X')\cup \Delta_Z .
\]
Then $Z$ is an important
$(S_X^\star,\allowbreak T_{X'}^\star,\allowbreak T_2,\allowbreak Q^\star,\allowbreak \pi,\allowbreak k)$-chip in
$\widehat D$. Moreover,
$\lambda_{\widehat D}(S_X^\star,T_{X'}^\star)>
\lambda_{\widehat D}(S,T_1)$.
\end{lemma}

The crossing case, treated next, gives progress in the primary
parameter, by lowering the budget.

\subsubsection{The crossing reduction}
\label{sec:routedInstances}

We fix a compact $S$-$T_1$ reachability set $X$ and an important
$\sttqpk$-chip $Z$ crossing $X$. As usual, we first add all arcs of
$\partial_{\mathrm{real}}(X)$ to the hitting set. If this hits
$\partial^+_{\mathrm{real}}(Z)$, then $Z$ is already handled. Hence, for
the chips that remain, we may assume that
\[
\partial_{\mathrm{real}}^+(Z)\cap \partial_{\mathrm{real}}(X)=\emptyset .
\]
This assumption will be used throughout the reduction. In particular, there is no real arc leaving $Z\setminus X$ and entering
$X\setminus Z$. Real arcs in the opposite direction may still exist and they need to be accounted for in our arguments.

We now define the local data used by the reduced instance. The attachment
boundary records how $Z\setminus X$ attaches to $X\cap Z$. The exits and
contacts determine the new target and forbidden sets. The reduced graph then
replaces $Z\setminus X$ by a representative and records exterior routes by
auxiliary shortcuts.

\begin{definition}[Attachment boundary and visible side]
\label{def:attachmentBoundary}
Let $\widehat D$ be a routed digraph, and let
$X,Z\subseteq V(\widehat D)$. The \emph{attachment boundary} of
$Z\setminus X$ towards $X\cap Z$ is
\[
B_X(Z)\coloneqq
\{b\in Z\setminus X \mid
\text{there exists } x\in X\cap Z
\text{ such that } (b,x)\in A_{\mathrm{real}}(\widehat D)
\text{ or } (x,b)\in A_{\mathrm{real}}(\widehat D)\}.
\]
The \emph{visible side} of $Z$ with respect to $X$ is
$L_X(Z)\coloneqq (X\cap Z)\cup B_X(Z)$.
\end{definition}

Labeling $B_X(Z)$ lets us replace the real graph on $Z\setminus X$ by a
boundaried representative and glue it to the visible side without changing
$\pi$. The attachment boundary is small when $X$ has a small outgoing
boundary.

\begin{observation}[Bounded attachment boundary]\label{obs:boundedAttachmentBoundary}
Let $\widehat D$ be a routed digraph whose real part is $\alpha$-balanced, and let
$X,Z\subseteq V(\widehat D)$. If $|\partial^+_{\mathrm{real}}(X)|\le k$,
then $B_X(Z)\subseteq \outf_{\widehat D_{\mathrm{real}}}(X)$ and
$|B_X(Z)|\le 2k+\alpha$.
\end{observation}

The proof is given in \Cref{app:crossingReductionDetails}.

\begin{definition}[Safe and blocked exits]\label{def:safeBlockedExits}
Let $\widehat D$ be a routed digraph, and let
$X,Z\subseteq V(\widehat D)$ be such that
$T_1\cap (X\cup Z)=\emptyset$.
A vertex $t\in X\setminus Z$ is a \emph{safe exit} for $(Z,X)$ if there is
a real arc $(t,h)\in\partial_{\mathrm{real}}^+(X)$ with $h\notin Z$ such
that there is an $h$-$T_1$ path in $\widehat D-(X\cup Z)$, which we write as $h \leadsto T_1$ in $\widehat D-(X\cup Z)$. Let
\[
\safeExit{Z}{X}\coloneqq
\{t\in X\setminus Z \mid
\text{$t$ is a safe exit for $(Z,X)$}\}.
\]

A vertex $t\in X\setminus Z$ is an \emph{exit} for $(Z,X)$ if there is a
real arc $(t,h)\in\partial_{\mathrm{real}}^+(X)$ with $h\notin Z$. The
\emph{blocked exits} are the exits that are not safe exits:
\[
\blockedExit{Z}{X}\coloneqq
\{t\in X\setminus Z \mid
\text{$t$ is an exit for $(Z,X)$}\}
\setminus
\safeExit{Z}{X}.
\]
\end{definition}

Safe exits become local targets. Blocked exits are forbidden, since
including them in a recursive chip could destroy strictness after lifting.

\begin{definition}[Contact vertices]\label{def:contactVertices}
Let $\widehat D$ be a routed digraph, and let
$X,Z\subseteq V(\widehat D)$. Define
\[
\begin{aligned}
\contact{Z}{X}\coloneqq\{x\in X\setminus Z \mid {}&
\text{there exists } b\in Z\setminus X\text{ such that}\\
&(x,b)\in A_{\mathrm{real}}(\widehat D)
\text{ or }(b,x)\in A_{\mathrm{real}}(\widehat D)\}.
\end{aligned}
\]
\end{definition}

Contact vertices are also forbidden, so that a larger reduced chip cannot gain
new adjacency to the replaced part $Z\setminus X$.

The reduced graph $\widehat D_X(Z)$ keeps $X$, replaces $Z\setminus X$ by
a representative $R$, and records exterior routes by auxiliary shortcuts.
We put all of $R$ in the new source, forcing every recursive chip to
contain it.

\begin{definition}[Reduced graph in the crossing case]
\label{def:crossingReducedGraph}
Let $\widehat D$ be a routed digraph. Let $X$ be a compact
$S$-$T_1$ reachability set in $\widehat D$ that is auxiliary-respecting,
and let $Z$ be an $\sttqpk$-chip in $\widehat D$ that crosses $X$. Assume
that
\[
\partial_{\mathrm{real}}^+(Z)\cap \partial_{\mathrm{real}}(X)=\emptyset .
\]

We first define the real part of the reduced graph. Set $B\coloneqq
B_X(Z)$. Fix an injective labeling of $B$ with label set $I$. We view
$\widehat D_{\mathrm{real}}[Z\setminus X]$ as a boundaried digraph with
boundary $B$ and this labeling. Let $R$ be the chosen representative of
$\widehat D_{\mathrm{real}}[Z\setminus X]$ in
$\Rcal_{\equiv_\pi}[\Dcal_{\subseteq I}]$. Thus the boundary of $R$ is
identified with $B$ using the same labels.

The representative $R$ replaces the real graph on $Z\setminus X$. Boundary
arcs inside $B$ are taken from the representative rather than from the old
graph on the $X$ side. Let
$A_B\coloneqq
\{(u,v)\in A_{\mathrm{real}}(\widehat D)\mid u,v\in B\}$, and define
\[
D_{\mathrm{real}}'\coloneqq
\bigl(\widehat D_{\mathrm{real}}[X\cup B]-A_B\bigr)\oplus R,
\]
where the gluing takes place along $B$.

We next define the auxiliary arcs. Let
$A_{\mathrm{old}}^X\coloneqq
\{(u,v)\in A_{\mathrm{aux}}(\widehat D)\mid u,v\in X\}$ be the old
auxiliary arcs whose two endpoints remain in $X$. We also add shortcuts for
routes through the exterior that has been discarded. Let $\Gamma_X(Z)$ be the set of ordered pairs $(u,v)$ with
$u,v\in X\setminus Z$ for which there is a directed path
\[
u,p_1,\ldots,p_r,v
\]
in $\widehat D-Z$, where $r\geq 1$ and
$p_i\in V(\widehat D)\setminus (X\cup Z)$ for every $i\in [r]$.
Some of these ordered pairs may already be real arcs in the reduced graph.
We keep only the genuinely auxiliary ones:
\[
A_{\mathrm{aux}}'
\coloneqq
(A_{\mathrm{old}}^X\cup \Gamma_X(Z))
\setminus A(D_{\mathrm{real}}').
\]

The \emph{reduced graph of $X$ with respect to $Z$} is the routed digraph
$\widehat D_X(Z)$ defined by
\[
\begin{aligned}
V(\widehat D_X(Z))&\coloneqq V(D_{\mathrm{real}}'),\\
A_{\mathrm{real}}(\widehat D_X(Z))&\coloneqq A(D_{\mathrm{real}}'),\\
A_{\mathrm{aux}}(\widehat D_X(Z))&\coloneqq A_{\mathrm{aux}}'.
\end{aligned}
\]

Finally, we describe the chip and associated data in the reduced instance:
\[
\begin{aligned}
C_X(Z)&\coloneqq (X\cap Z)\cup V(R),&
S_X(Z)&\coloneqq S\cup V(R),\\
T_X^1(Z)&\coloneqq \safeExit{Z}{X},&
T_X^2(Z)&\coloneqq
(T_2\cap X)\cup
\blockedExit{Z}{X}\cup
\contact{Z}{X}\cup
\tail(A_{\mathrm{aux}}').
\end{aligned}
\]
The request set is
\[
Q_X(Z)\coloneqq
(Q\cap X)
\cup
\entry{Z}{X}{T_1}{Q\setminus X}
\cup
\entry{Z}{X}{T_1}{\head(\partial_{\mathrm{real}}^+(Z))\setminus X}.
\]
\end{definition}

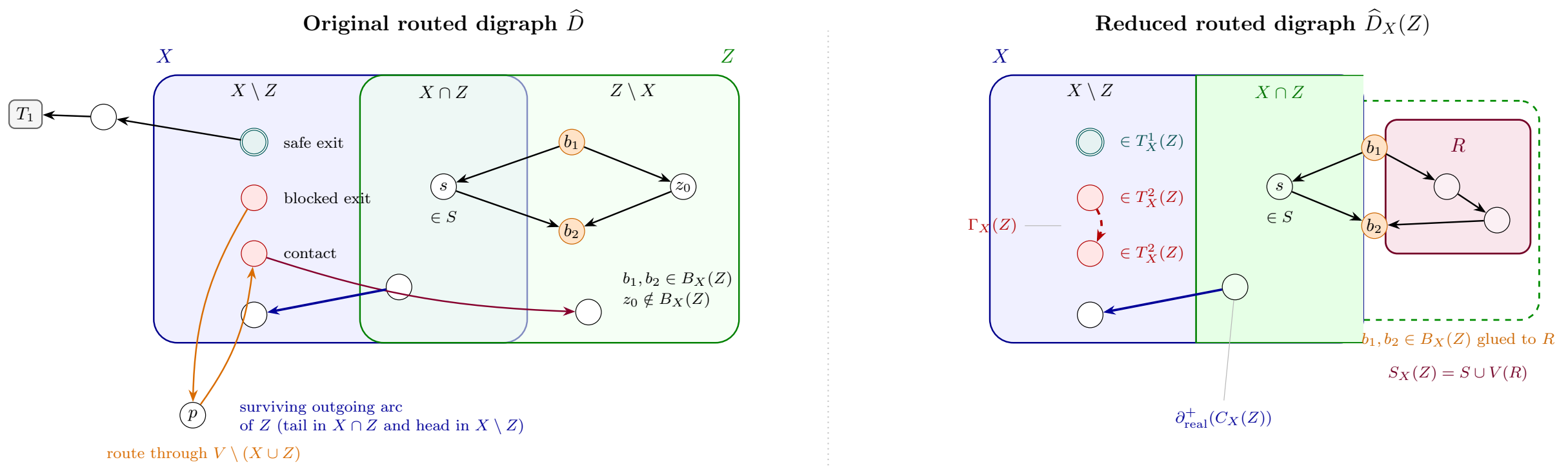
\begin{figure}[t]
\centering
\resizebox{\textwidth}{!}{%
\begin{tikzpicture}[
  >={Stealth[length=2.2mm]},
  vtx/.style={circle, draw, fill=white, inner sep=0.4pt, minimum size=4.6mm, font=\footnotesize},
  bvtx/.style={circle, draw=orange!80!black, fill=orange!22, inner sep=0.4pt, minimum size=4.6mm, font=\footnotesize},
  svtx/.style={circle, draw=teal!70!black, double, fill=teal!10, inner sep=0.4pt, minimum size=4.6mm, font=\footnotesize},
  fvtx/.style={circle, draw=red!70!black, fill=red!10, inner sep=0.4pt, minimum size=4.6mm, font=\footnotesize},
  Xset/.style={rounded corners=12pt, draw=blue!55!black, fill=blue!6, line width=0.8pt},
  Zset/.style={rounded corners=12pt, draw=green!50!black, fill=green!7, fill opacity=0.55, line width=0.8pt},
  Capz/.style={draw=green!50!black, fill=green!10, line width=0.7pt},
  Rbox/.style={rounded corners=6pt, draw=purple!60!black, fill=purple!10, line width=0.9pt},
  tbox/.style={rounded corners=3pt, draw=black!60, fill=black!4, line width=0.7pt, font=\footnotesize},
  real/.style={->, thick, black},
  outc/.style={->, very thick, blue!60!black},
  aux/.style={->, very thick, red!70!black, dashed},
  ext/.style={->, thick, orange!85!black},
  cont/.style={->, thick, purple!70!black},
  lead/.style={gray!55, line width=0.4pt},
  reg/.style={font=\footnotesize},
  setn/.style={font=\small\bfseries},
  tag/.style={font=\scriptsize},
  title/.style={font=\bfseries},
]

\begin{scope}
\draw[Xset] (0.5,-0.6) rectangle (7.2,4.2);
\draw[Zset] (4.2,-0.6) rectangle (11.0,4.2);
\node[setn, blue!55!black]  at (0.7,4.55) {$X$};
\node[setn, green!50!black] at (10.8,4.55) {$Z$};
\node[reg] at (2.3,3.9) {$X\setminus Z$};
\node[reg] at (5.7,3.9) {$X\cap Z$};
\node[reg] at (9.1,3.9) {$Z\setminus X$};
\node[title] at (5.7,5.15) {Original routed digraph $\widehat D$};

\node[vtx]  (s)  at (5.7,2.2) {$s$};
\node[tag]  at (5.7,1.65) {$\in S$};
\node[bvtx] (b1) at (8.0,3.0) {$b_1$};
\node[bvtx] (b2) at (8.0,1.4) {$b_2$};
\node[vtx]  (z0) at (10.0,2.2) {$z_0$};
\node[svtx] (es) at (2.3,3.0) {};
\node[fvtx] (eb) at (2.3,2.0) {};
\node[fvtx] (ec) at (2.3,1.0) {};
\node[vtx]  (o1) at (2.3,-0.1) {};
\node[vtx]  (cp) at (8.3,-0.05) {};
\node[vtx]  (p)  at (1.2,-1.9) {$p$};

\node[vtx]  (m1) at (4.9,0.4) {};

\node[tag, anchor=west] at (2.7,3.0) {safe exit};
\node[tag, anchor=west] at (2.7,2.0) {blocked exit};
\node[tag, anchor=west] at (2.7,1.0) {contact};

\node[tbox] (T1) at (-1.8,3.5) {$T_1$};
\node[vtx]  (hs) at (-0.4,3.45) {};

\draw[real] (b1) -- (s);
\draw[real] (s)  -- (b2);
\draw[real] (b1) -- (z0);
\draw[real] (z0) -- (b2);

\draw[real] (es) -- (hs);
\draw[real] (hs) -- (T1);

\draw[ext] (eb) to[bend right=15] (p);
\draw[ext] (p)  to[bend right=15] (ec);

\draw[cont] (ec) to[bend right=8] (cp);
\node[tag, orange!85!black] at (1.4,-2.6) {route through $V\setminus(X\cup Z)$};

\draw[outc] (m1) -- (o1);
\node[tag, blue!60!black, align=left] (Olab) at (4.6,-1.95)
  {surviving outgoing arc\\ of $Z$ (tail in $X\cap Z$ and head in $X\setminus Z$)};

\node[tag, align=left] at (9.9,0.35)
  {$b_1,b_2\in B_X(Z)$\\[1pt] $z_0\notin B_X(Z)$};
\end{scope}

\draw[gray!45, dotted, line width=0.7pt] (12.6,-2.8) -- (12.6,5.0);

\begin{scope}[xshift=15.0cm]
\draw[Xset] (0.5,-0.6) rectangle (7.2,4.2);
\begin{scope}
  \clip (0.5,-0.6) rectangle (7.2,4.2);
  \fill[Capz] (4.2,-0.6) rectangle (7.4,4.2);
\end{scope}
\draw[green!50!black, line width=0.7pt] (4.2,-0.6) -- (4.2,4.2);

\node[setn, blue!55!black] at (0.7,4.55) {$X$};
\node[reg] at (2.3,3.9) {$X\setminus Z$};
\node[reg, green!45!black] at (5.7,3.9) {$X\cap Z$};
\node[title] at (5.4,5.15) {Reduced routed digraph $\widehat D_X(Z)$};

\node[Rbox, minimum width=2.6cm, minimum height=2.4cm] (Rnode) at (8.9,2.2) {};
\node[setn, purple!60!black] at (8.9,2.95) {$R$};

\node[vtx, fill=green!6]   (Rs)  at (5.7,2.2) {$s$};
\node[tag] at (5.7,1.65) {$\in S$};
\node[bvtx] (Rb1) at (7.4,2.9) {$b_1$};
\node[bvtx] (Rb2) at (7.4,1.5) {$b_2$};
\node[vtx, fill=purple!6] (Rr1) at (8.7,2.2) {};
\node[vtx, fill=purple!6] (Rr2) at (9.6,1.6) {};
\node[svtx] (Res) at (2.3,3.0) {};
\node[fvtx] (Reb) at (2.3,2.0) {};
\node[fvtx] (Rec) at (2.3,1.0) {};
\node[vtx]  (Ro1) at (2.3,-0.1) {};

\node[vtx, fill=green!6] (Rm1) at (4.9,0.4) {};

\node[tag, teal!60!black, anchor=west] at (2.7,3.0) {$\in T_X^1(Z)$};
\node[tag, red!70!black, anchor=west]  at (2.7,2.0) {$\in T_X^2(Z)$};
\node[tag, red!70!black, anchor=west]  at (2.7,1.0) {$\in T_X^2(Z)$};

\draw[real] (Rb1) -- (Rs);
\draw[real] (Rs)  -- (Rb2);
\draw[real] (Rb1) -- (Rr1);
\draw[real] (Rr1) -- (Rr2);
\draw[real] (Rr2) -- (Rb2);

\draw[aux] (Reb) to[bend left=30] (Rec);
\node[tag, red!70!black] (Glab) at (0.55,1.5) {$\Gamma_X(Z)$};
\draw[lead] (Glab.east) -- (1.78,1.5);

\coordinate (cz1) at (4.45,3.6);
\coordinate (cz2) at (4.45,-0.05);
\begin{scope}[on background layer]
  \node[draw=green!55!black, dashed, line width=0.9pt, rounded corners=10pt,
        fit=(cz1)(cz2)(Rs)(Rb1)(Rb2)(Rm1)(Rnode), inner sep=4pt] (Cbox) {};
\end{scope}

\draw[outc] (Rm1) -- (Ro1);
\node[tag, blue!60!black, align=left] (ROlab) at (4.7,-1.95)
  {$\partial^+_{\mathrm{real}}(C_X(Z))$};
\draw[lead] (ROlab.north) -- (Rm1);

\node[tag, orange!80!black] at (8.9,-0.55) {$b_1,b_2\in B_X(Z)$ glued to $R$};
\node[tag, purple!60!black] at (8.9,-1.15) {$S_X(Z)=S\cup V(R)$};
\end{scope}

\end{tikzpicture}%
}
\caption{The crossing reduction of \Cref{def:crossingReducedGraph}. In the
original routed digraph (left), the kept part $X\cap Z$ together with the
attachment boundary $B_X(Z)\subseteq Z\setminus X$ forms the visible side
$L_X(Z)=(X\cap Z)\cup B_X(Z)$, while the rest of $Z\setminus X$ is replaced.
The safe exit reaches $T_1$ outside $X\cup Z$, the blocked exit leaves $X$ with
no such continuation, and the contact vertex is adjacent to $Z\setminus X$.
The exterior route through $p\in V\setminus(X\cup Z)$ becomes a shortcut in
$\Gamma_X(Z)$. The thick blue arc from $X\cap Z$ into $X\setminus Z$ is an
outgoing arc of $Z$ with both endpoints in $X$, and it survives as part of the
outgoing boundary of the reduced chip. In the reduced routed digraph
$\widehat D_X(Z)$ (right), $Z\setminus X$ is replaced by a bounded
representative $R$ glued along $B_X(Z)$. The reduced chip
$C_X(Z)=(X\cap Z)\cup V(R)$ is the dashed region, and the reduced source is
$S_X(Z)=S\cup V(R)$. Safe exits form the local target $T_X^1(Z)$, while
blocked exits, contact vertices, and auxiliary tails lie in the forbidden set
$T_X^2(Z)$.}
\label{fig:crossingReduction}
\end{figure}

\Cref{fig:crossingReduction} illustrates \Cref{def:crossingReducedGraph}.
The construction depends on the unknown chip $Z$. In
\Cref{lem:preinductionStep} we generate all possible reduced instances by
guessing its bounded data, namely the attachment boundary, representative,
shortcut relation, and relevant target, forbidden, and request vertices.

\paragraph{Immediate properties of the reduced instance.}
The next two observations check the reduced data and the condition
$\pi$. Their proofs are in \Cref{app:attachment-proofs}.

\begin{observation}[Basic properties of the reduced chip]
\label{obs:basicReducedData}
Assume the notation of \Cref{def:crossingReducedGraph} and that the
instance is guarded. Then 
the following hold.
\begin{enumerate}
\item $S_X(Z)\subseteq C_X(Z)$.
\item $C_X(Z)$ is disjoint from $T_X^1(Z)\cup T_X^2(Z)$.
\item $Q_X(Z)\subseteq X\setminus Z$.
\item $C_X(Z)$ is auxiliary-respecting in $\widehat D_X(Z)$.
\end{enumerate}
\end{observation}

Arcs with both endpoints in $B_X(Z)$ are taken from the representative.
Removing them from the kept side gives the gluing identity below. The
underlying decomposition is given in \Cref{app:crossingReductionDetails}.

\begin{observation}[The real graph induced by the reduced chip]
\label{obs:reducedChipRealSubgraph}
With the notation of \Cref{def:crossingReducedGraph}, including the set $A_B$ defined there, we have
\[
\widehat D_X(Z)_{\mathrm{real}}[C_X(Z)]
=
\bigl(\widehat D_{\mathrm{real}}[L_X(Z)]-A_B\bigr)\oplus R.
\]
Consequently, if $\widehat D_{\mathrm{real}}[Z]\models \pi$, then
$\widehat D_X(Z)_{\mathrm{real}}[C_X(Z)]\models \pi$.
\end{observation}

\begin{observation}[Surviving real boundary]
\label{obs:survivingRealBoundary}
With the notation of \Cref{def:crossingReducedGraph}, we have
\[
\partial_{\mathrm{real}}^+(C_X(Z))
=
\{(u,v)\in \partial_{\mathrm{real}}^+(Z)\mid
u\in X\cap Z,\ v\in X\setminus Z\}.
\]
In particular,
\[
\emptyset\neq \partial_{\mathrm{real}}^+(C_X(Z))
\subsetneq
\partial_{\mathrm{real}}^+(Z).
\]
\end{observation}

\begin{proof}
All representative vertices lie in $C_X(Z)$, so no arc contributed by $R$
leaves it. Any remaining outgoing arc has head in $X\setminus Z$ and
tail in $(X\cap Z)\cup B_X(Z)$. A tail in $B_X(Z)\subseteq Z\setminus X$
would give an arc in
$\partial_{\mathrm{real}}^+(Z)\cap\partial_{\mathrm{real}}^-(X)$,
contrary to the hypothesis. Conversely, every original arc from $X\cap Z$
to $X\setminus Z$ is retained and leaves $C_X(Z)$. This proves the equality.

For non-emptiness, take an $S$-$x$ path in $\widehat D[X]$ with
$x\in X\setminus Z$. Its first arc leaving $Z$ is real, since $Z$ is
auxiliary-respecting, and belongs to the displayed surviving boundary.
For proper containment, \Cref{obs:routedCrossingBudgetDrop} supplies a
real outgoing arc of $Z$ with tail in $Z\setminus X$, which does not survive.
\end{proof}

The non-empty proper boundary inclusion will let us use budget $k-1$ and recover an original boundary arc from the recursive call.

\paragraph{Path replacement and bounded data.}
The shortcuts preserve paths between retained vertices even after an
additional set $W\subseteq X\setminus Z$ is removed. This is the path
replacement needed when lifting a larger chip.

\begin{observation}[Path replacement by auxiliary arcs]
\label{obs:pathReplacementAux}
With the notation of \Cref{def:crossingReducedGraph}, let
$W\subseteq X\setminus Z$ and let
$u,v\in X\setminus (Z\cup W)$. Then the following hold.
\begin{enumerate}
\item If there is a $u$-$v$ path in $\widehat D-(Z\cup W)$, then there is
a $u$-$v$ path in $\widehat D_X(Z)-(C_X(Z)\cup W)$.
\item If there is a $u$-$v$ path in
$\widehat D_X(Z)-(C_X(Z)\cup W)$, then there is a $u$-$v$ path in
$\widehat D-(Z\cup W)$.
\end{enumerate}
\end{observation}

\begin{proof}
For the first direction, replace each maximal excursion of a path through
$V(\widehat D)\setminus(X\cup Z)$ by its shortcut in $\Gamma_X(Z)$.
Every arc with both endpoints in $X\setminus Z$ is retained. If an old
auxiliary arc or shortcut is already real, use that real arc. The resulting
walk avoids $C_X(Z)\cup W$ because all its vertices lie in
$X\setminus(Z\cup W)$.

Conversely, a path avoiding $C_X(Z)\cup W$ uses no representative or
boundary vertex. Its real arcs are original arcs inside $X\setminus Z$.
Keep old auxiliary arcs and expand each remaining shortcut into a defining
path whose internal vertices lie outside $X\cup Z$. These vertices avoid
$W\subseteq X$, so the expanded walk avoids $Z\cup W$. In both directions,
removing cycles from the resulting walk gives the required path.
\end{proof}

The reduced instance is useful only because the information it depends on is
bounded. The next observation identifies the relevant domain. Every shortcut
endpoint, every exit or contact vertex, and every entry vertex that appears
in the reduced target, forbidden set, or request set lies in the routed
interface of $X$. Thus these objects can later be guessed by branching over
bounded subsets of this interface.

\begin{observation}[Shortcut endpoints are routed-interface vertices]
\label{obs:shortcutEndpointsRoutedInterface}
With the notation of \Cref{def:crossingReducedGraph}, suppose moreover that
$\widehat D$ is $\eta$-guarded. Then
\[
\Gamma_X(Z)\subseteq
\rintf_{\widehat D}(X)\times \rintf_{\widehat D}(X).
\]
Moreover,
\[
\safeExit{Z}{X}\cup \blockedExit{Z}{X}\cup \contact{Z}{X}
\subseteq \rintf_{\widehat D}(X),
\]
and
\[
\entry{Z}{X}{T_1}{Q\setminus X}
\cup
\entry{Z}{X}{T_1}{\head(\partial_{\mathrm{real}}^+(Z))\setminus X}
\subseteq \rintf_{\widehat D}(X).
\]
\end{observation}

The proof is a direct verification from the definitions of the routed
interface, exits, contacts, shortcut arcs, and entry vertices, and is given
in \Cref{app:attachment-proofs}.

\subsubsection{Reduced chips and lifting}

We now prove the two main correctness statements for the crossing
reduction. First, $C_X(Z)$ is a valid chip in the reduced routed instance.
Second, if $C_X(Z)$ could be enlarged in the reduced instance without
increasing its real outgoing boundary, then this enlargement could be lifted
back to an enlargement of the original chip $Z$. This would contradict the
importance of $Z$.

\begin{observation}[Strictness in routed digraphs]
\label{obs:routedStrictnessCriterion}
Let $\widehat D$ be a routed digraph, and let $Y$ be an $S$-$T$
reachability set that is auxiliary-respecting. Then $Y$ is strict
if and only if, for every vertex $v$ that is the head of a real arc in
$\partial_{\mathrm{real}}^+(Y)$, there is a $v$-$T$ path in
$\widehat D-Y$.
\end{observation}

\begin{proof}
The proof is the same as the proof of \Cref{lem:basicStrictFacts}. The only
difference is that cuts and boundaries use only real arcs, while the
witnessing paths may use all arcs of $\widehat D$. Since $Y$ is auxiliary-respecting, every path from $S\subseteq Y$ to $T$ must leave $Y$
through a real arc. Thus inclusion-wise minimality of the real cut
$\partial_{\mathrm{real}}^+(Y)$ is equivalent to the stated reachability
condition for the heads of real outgoing arcs.
\end{proof}

The next lemma verifies the chip conditions for the reduced instance.
Reachability uses the representative included in the new source, strictness
uses the shortcut arcs, and the request condition uses the entry points in
$Q_X(Z)$.

\begin{lemma}[Reduced chip in the crossing case]
\label{lem:crossingReducedChip}
Let $(\widehat D,S,T_1,T_2,Q)$ be an $\eta$-guarded routed instance. Let $X$ be a compact auxiliary-respecting $S$-$T_1$ reachability set in
$\widehat D$ and let $Z$ be an $\sttqpk$-chip in $\widehat D$ that
crosses $X$. Suppose that
$\partial_{\mathrm{real}}^+(Z)\cap
\partial_{\mathrm{real}}(X)=\emptyset$.
With the notation of \Cref{def:crossingReducedGraph}, the set $C_X(Z)$ is an
$
(S_X(Z),\allowbreak T_X^1(Z),\allowbreak T_X^2(Z),\allowbreak Q_X(Z),\allowbreak \pi,\allowbreak k-1)
 $-chip in $\widehat D_X(Z)$.
\end{lemma}

\begin{proof}
Write $C=C_X(Z)$, $S'=S_X(Z)$, $T_1'=T_X^1(Z)$, $T_2'=T_X^2(Z)$, and
$Q'=Q_X(Z)$. By \Cref{obs:basicReducedData}, $S'\subseteq C$, the set $C$
avoids $T_1'\cup T_2'$ and is auxiliary-respecting, and $Q'\subseteq X\setminus Z$.
By \Cref{obs:survivingRealBoundary,obs:reducedChipRealSubgraph}, its real
boundary has size at most $k-1$ and its induced real graph satisfies $\pi$.

To check source reachability, take $c\in X\cap Z$ and an $S$-$c$ path in
$\widehat D[Z]$. Guardedness makes this path entirely real. If it stays in
$X$, it survives unchanged. Otherwise, its suffix from the last return
arc $(b,x)$ to $X$ survives, with $b\in B_X(Z)\subseteq V(R)\subseteq S'$.
Every vertex of $R$ is itself a source, so all of $C$ is reachable from $S'$.

We use one path argument for both strictness and requests. If
$v\in X\setminus Z$ reaches $T_1$ in $\widehat D-Z$, take the last exit
$(t,h)$ from $X$ on such a path. This arc is real because $X$ is
auxiliary-respecting, and its suffix avoids $X\cup Z$. Hence
$t\in\safeExit{Z}{X}=T_1'$. By \Cref{obs:pathReplacementAux}, the prefix
from $v$ to $t$ projects to a path in $\widehat D_X(Z)-C$.

Every head of an arc leaving $C$ belongs to $X\setminus Z$ and reaches
$T_1$ in $\widehat D-Z$, by \Cref{obs:survivingRealBoundary} and the
strictness of $Z$. The preceding argument and
\Cref{obs:routedStrictnessCriterion} prove strictness of $C$. Each
$q'\in Q'$ likewise reaches $T_1$ in $\widehat D-Z$: it is either an old
request or an entry point on a chosen request or strictness path.
Applying the same argument proves all reduced requests.
\end{proof}

The more technical direction is lifting. A potential dominator of $C_X(Z)$ in the
reduced graph adds vertices only from $X\setminus Z$. The forbidden sets were
chosen so that these added vertices cannot use the forgotten part
$Z\setminus X$ in a new way. This is what lets the enlargement transfer back
to the original graph.

\begin{lemma}[Lifting a dominating reduced chip]
\label{lem:crossingLifting}
Let $(\widehat D,S,T_1,T_2,Q)$ be an $\eta$-guarded routed instance. Let
$X$ be a compact $S$-$T_1$ reachability set in $\widehat D$ that is auxiliary-respecting, and let $Z$ be an important $\sttqpk$-chip in
$\widehat D$ that crosses $X$. Suppose that
$\partial_{\mathrm{real}}^+(Z)\cap
\partial_{\mathrm{real}}(X)=\emptyset$. 
With the notation of \Cref{def:crossingReducedGraph}, let $C=C_X(Z)$.
There is no
$
(S_X(Z),\allowbreak T_X^1(Z),\allowbreak T_X^2(Z),\allowbreak Q_X(Z),\allowbreak \pi,\allowbreak k-1)
$-chip $C'$ in $\widehat D_X(Z)$ such that
\[
C\subset C'
\qquad\text{and}\qquad
|\partial_{\mathrm{real}}^+(C')|
\le
|\partial_{\mathrm{real}}^+(C)|.
\]
\end{lemma}

\begin{proof}
Suppose, for a contradiction, that such a chip $C'$ exists. Let
$C\coloneqq C_X(Z)$ and 
$W\coloneqq C'\setminus C$.
Since $C$ contains $X\cap Z$ and all vertices of $R$, we have
$W\subseteq X\setminus Z$.
Define
\[
Z'\coloneqq Z\cup W.
\]
Since $C'\supset C$, the set $W$ is non-empty, and hence $Z'\supset Z$.

We show that $Z'$ is an original chip with no larger real outgoing boundary than $Z$.

We summarize the consequences of the forbidden sets in the reduced instance
that will be used repeatedly. Since $C'$ is disjoint from
$T_X^1(Z)\cup T_X^2(Z)$, since $(T_2\cap X)\subseteq T_X^2(Z)$, and since
$W\subseteq X\setminus Z$, we have
\[
W\cap
\bigl(
\safeExit{Z}{X}\cup
\blockedExit{Z}{X}\cup
\contact{Z}{X}\cup
T_2
\bigr)=\emptyset .
\label{eq:WavoidsForbidden}
\]
Moreover, since $\tail(A_{\mathrm{aux}}')\subseteq T_X^2(Z)$ and $C'$ is
disjoint from $T_X^2(Z)$, no auxiliary arc of the reduced instance has its
tail in $C'$.

We shall use the following consequences: 
\begin{enumerate}[label=(F\arabic*)]
\item\label{F:no-exit-from-W}
No vertex of $W$ is the tail of a real arc leaving $X$ whose head avoids
$Z$.
\item\label{F:no-contact-from-W}
There is no real arc in either direction between $W$ and $Z\setminus X$.
\end{enumerate}

\begin{claim}\label{claim:lifting-elementary-chip-conditions}
The set $Z'$ is reachable from $S$ in $\widehat D[Z']$, is disjoint from
$T_1\cup T_2$, and is auxiliary-respecting.
\end{claim}

\begin{claimproof}
For $w\in W$, take an $S_X(Z)$-$w$ path in the reduced chip $C'$.
It uses only real arcs, since auxiliary tails are forbidden. Its last
departure from $C$ has tail in $(X\cap Z)\cup B_X(Z)\subseteq Z$;
the remaining suffix lies in $W$ and consists of original arcs.
Prepending a source path inside $Z$ therefore reaches $w$ in $\widehat D[Z']$.

The set $Z'$ avoids $T_1$ because both $X$ and $Z$ do, and avoids $T_2$
by \eqref{eq:WavoidsForbidden}. Guardedness then excludes all auxiliary tails from $Z'$, so it is
auxiliary-respecting.
\end{claimproof}

\begin{claim}\label{claim:lifting-pi-condition}
We have $\widehat D_{\mathrm{real}}[Z']\models\pi$.
\end{claim}

\begin{claimproof}
By \ref{F:no-contact-from-W}, the attachment boundary is still
$B=B_X(Z)$. With $A_B$ as in \Cref{def:crossingReducedGraph}, put
\[
D_1'=\widehat D_{\mathrm{real}}[(Z'\cap X)\cup B]-A_B,
\qquad D_2=\widehat D_{\mathrm{real}}[Z\setminus X].
\]
Then the reduced real graph on $C'$ is $D_1'\oplus R$, which satisfies
$\pi$. Since $R\equiv_\pi D_2$, so does $D_1'\oplus D_2$.
This is exactly $\widehat D_{\mathrm{real}}[Z']$: the same attachment
separates the two sides, and the arcs removed from $D_1'$ are supplied by
$D_2$.
\end{claimproof}

\begin{claim}[Safe-exit lifting]\label{claim:safeExitLifting}
Let $a\in X$. If
$a \leadsto T_X^1(Z)$
 in 
$\widehat D_X(Z)-C'$,
then
$a\in X\setminus (Z\cup W)$ and  $a \leadsto T_1$ in 
$\widehat D-Z'$.
\end{claim}

\begin{claimproof}
Let $t\in T_X^1(Z)$ be the endpoint of the reduced path. Its endpoints
$a,t$ lie in $X\setminus(Z\cup W)$. By \Cref{obs:pathReplacementAux},
the path lifts to an $a$-$t$ path avoiding $Z'=Z\cup W$.
The safe exit $t$ has a real arc to the exterior followed by a path to
$T_1$ outside $X\cup Z$. This continuation also avoids $W\subseteq X$,
giving the required path.
\end{claimproof}

Thus, in the strictness and request arguments below, it will be enough to
show that the relevant vertex of $X$ reaches the reduced target
$T_X^1(Z)$ in $\widehat D_X(Z)-C'$.

\begin{claim}\label{claim:lifting-strictness}
$Z'$ is a strict $S$-$T_1$ reachability set in $\widehat D$.
\end{claim}

\begin{claimproof}
By \Cref{obs:routedStrictnessCriterion}, it suffices to show that every
head of a real arc leaving $Z'$ reaches $T_1$ in $\widehat D-Z'$.
Let $(u,y)\in \partial_{\mathrm{real}}^+(Z')$.

First suppose that $y\in X$. The tail $u$ cannot lie in $Z\setminus X$,
since then
$(u,y)\in
\partial_{\mathrm{real}}^+(Z)\cap
\partial_{\mathrm{real}}^-(X)$, 
contrary to $\partial_{\mathrm{real}}^+(Z)\cap\partial_{\mathrm{real}}(X)=\emptyset$.

Thus $u\in (X\cap Z)\cup W\subseteq C'$, and $(u,y)$ is a real arc of the
reduced graph leaving $C'$. Since $C'$ is strict in the reduced instance,
\Cref{obs:routedStrictnessCriterion} gives a $y$-$T_X^1(Z)$ path in
$\widehat D_X(Z)-C'$.
By \Cref{claim:safeExitLifting}, $y\leadsto T_1$ in $\widehat D-Z'$.

Now suppose that $y\notin X$. By \ref{F:no-exit-from-W} and \ref{F:no-contact-from-W}, the tail $u$ cannot lie in $W$. Hence $u\in Z$. Since $(u,y)$ leaves $Z'$ and $Z\subseteq Z'$, we also have $y\notin Z$. Therefore the arc $(u,y)$ must belong to $\partial_{\mathrm{real}}^+(Z)$. Therefore
$y\in \head(\partial_{\mathrm{real}}^+(Z))\setminus X$. 
If there is a $y$-$T_1$ path in $\widehat D-(X\cup Z)$, then this path
avoids $Z'$ because $W\subseteq X$, and we are done.

Otherwise, consider the chosen path $\avoidpath{Z}{X}{T_1}{y}$. Since $Z$
is strict in $\widehat D$, such a path exists in $\widehat D-Z$. Since no
$y$-$T_1$ path exists in $\widehat D-(X\cup Z)$, the chosen path enters
$X$. Let
$r\in \entry{Z}{X}{T_1}{y}$.
Then
\[
r\in
\entry{Z}{X}{T_1}{\head(\partial_{\mathrm{real}}^+(Z))\setminus X}
\subseteq Q_X(Z).
\]
The prefix of $\avoidpath{Z}{X}{T_1}{y}$ from $y$ to $r$ avoids $Z$ and
avoids $X\setminus\{r\}$ by the first-entry property.

Since $C'$ is a reduced chip and $r\in Q_X(Z)$, we have that  $r\leadsto T_X^1(Z)$ in $\widehat D_X(Z)-C'$. By \Cref{claim:safeExitLifting},
$r\in X\setminus (Z\cup W)$ and $r\leadsto T_1$ in $\widehat D-Z'$.
Hence the prefix from $y$ to $r$ also avoids $W$, and therefore avoids
$Z'$. Concatenating this prefix with the lifted $r$-$T_1$ path gives a
$y$-$T_1$ path in $\widehat D-Z'$.

Thus every head of a real arc leaving $Z'$ reaches $T_1$ in
$\widehat D-Z'$. By \Cref{obs:routedStrictnessCriterion}, $Z'$ is strict.
\end{claimproof}

\begin{claim}\label{claim:lifting-requests}
For every $q\in Q$, there is a $q$-$T_1$ path in $\widehat D-Z'$.
\end{claim}

\begin{claimproof}
If $q\in Q\cap X$, its reduced request and \Cref{claim:safeExitLifting}
give the required path. If $q\notin X$, use the same first-entry argument
as in \Cref{claim:lifting-strictness}. The chosen path
$\avoidpath{Z}{X}{T_1}{q}$ either avoids $X$ and hence $Z'$, or first enters
at $r\in Q_X(Z)$. Its prefix avoids $X\setminus\{r\}$, while the reduced
request and \Cref{claim:safeExitLifting} give $r\notin Z'$ and an
$r$-$T_1$ path avoiding $Z'$. Concatenation supplies the path for $q$.
\end{claimproof}

\begin{claim}\label{claim:lifting-boundary-size}
$|\partial_{\mathrm{real}}^+(Z')|
\le 
|\partial_{\mathrm{real}}^+(Z)|\leq k$.
\end{claim}

\begin{claimproof}
Put
\[
O=\{(u,v)\in\partial_{\mathrm{real}}^+(Z):
u\in Z\setminus X,\ v\notin X\cup Z\}.
\]
Boundary disjointness excludes outgoing arcs of $Z$ crossing $X$ in
either direction. Thus \Cref{obs:survivingRealBoundary} gives
\[
\partial_{\mathrm{real}}^+(Z)
=O\mathbin{\dot\cup}\partial_{\mathrm{real}}^+(C).
\]
An outgoing arc of $Z'$ whose tail lies in $Z\setminus X$ belongs to
$O$. Every other such arc has tail in $(X\cap Z)\cup W$ and head in
$X\setminus(Z\cup W)$: boundary disjointness excludes an exterior head
for a tail in $X\cap Z$, and \ref{F:no-exit-from-W} excludes it for a
tail in $W$. This arc is retained and leaves $C'$. Hence
\[
\begin{aligned}
|\partial_{\mathrm{real}}^+(Z')|
&\le |O|+|\partial_{\mathrm{real}}^+(C')|\\
&\le |O|+|\partial_{\mathrm{real}}^+(C)|
=|\partial_{\mathrm{real}}^+(Z)|\le k.
\end{aligned}
\]
\end{claimproof}

By
\Cref{claim:lifting-elementary-chip-conditions,claim:lifting-pi-condition,claim:lifting-strictness,claim:lifting-requests,claim:lifting-boundary-size},
 the set $Z'$ is an $(S,T_1,T_2,Q,\pi,k)$-chip in $\widehat D$ satisfying
\[
|\partial_{\mathrm{real}}^+(Z')|
\le
|\partial_{\mathrm{real}}^+(Z)|.
\]
Moreover, $Z'\supset Z$ because $W\ne\emptyset$. This contradicts the
importance of $Z$. Therefore no such reduced chip $C'$ exists.
\end{proof}

Combining the reduced-chip and lifting lemmas gives the fixed-chip reduction.

\begin{lemma}[Crossing reduction for a fixed chip]\label{lem:representativeGluing}
Let $(\widehat D,S,T_1,T_2,Q)$ be an $\eta$-guarded routed instance. Let
$X$ be a compact $S$-$T_1$ reachability set in $\widehat D$ that is auxiliary-respecting, and let $Z$ be an important $\sttqpk$-chip in
$\widehat D$ that crosses $X$. Suppose that
$\partial_{\mathrm{real}}^+(Z)\cap
\partial_{\mathrm{real}}(X)=\emptyset$.
With the notation of \Cref{def:crossingReducedGraph}, the set $C_X(Z)$ is
an important
$
(S_X(Z),\allowbreak T_X^1(Z),\allowbreak T_X^2(Z),\allowbreak Q_X(Z),\allowbreak \pi,\allowbreak k-1)
$-chip in $\widehat D_X(Z)$. Moreover,
$\emptyset\neq \partial_{\mathrm{real}}^+(C_X(Z))
\subsetneq
\partial_{\mathrm{real}}^+(Z)$.
\end{lemma}

\begin{proof}
The chip conditions and importance follow from
\Cref{lem:crossingReducedChip,lem:crossingLifting}, respectively.
The boundary inclusion is \Cref{obs:survivingRealBoundary}.
\end{proof}

\subsubsection{The bounded reduced family}

The previous subsection shows that a fixed crossing chip $Z$ gives an
important chip in a smaller reduced instance. We now remove the dependence
on knowing $Z$. The point is that all information used to build the reduced
instance comes from bounded sets. The attachment boundary is bounded by
balance, representatives are chosen from a finite set, and shortcut
endpoints, exits, contacts, and request entry points all lie in the bounded
routed interface of $X$. Hence only boundedly many reduced instances have
to be considered.

\begin{lemma}[Crossing reduction family]\label{lem:preinductionStep}
Let $(\widehat D,S,T_1,T_2,Q)$ be an $\eta$-guarded routed instance whose
real part is $\alpha$-balanced. Let $X$ be a compact auxiliary-respecting
$S$-$T_1$ reachability set in $\widehat D$. There is a family
\[
\Pcal=
\{(\widehat D_i,S_i,T_i^1,T_i^2,Q_i)\mid i\in[p]\}
\]
of routed instances, where
$p\le p_{\pi,k,\alpha,\eta}$ 
for a constant depending only on $\pi,k,\alpha,\eta$, with the following
property.

For every important $\sttqpk$-chip $Z$ in $\widehat D$ that crosses $X$ and
satisfies
$\partial_{\mathrm{real}}^+(Z)\cap\partial_{\mathrm{real}}(X)=\emptyset$,
there is an index $i\in[p]$ and an important
$
(S_i,T_i^1,T_i^2,Q_i,\pi,k-1)
$-chip $C_i$ in $\widehat D_i$ such that
\[
\emptyset \neq \partial_{\mathrm{real}}^+(C_i)
\subsetneq
\partial_{\mathrm{real}}^+(Z).
\]
Moreover, each routed instance in $\Pcal$ is
$\eta'_{\pi,k,\alpha,\eta}$-guarded and has
$\alpha'_{\pi,k,\alpha,\eta}$-balanced real part, for constants
$\eta'_{\pi,k,\alpha,\eta}$ and $\alpha'_{\pi,k,\alpha,\eta}$ depending
only on $\pi,k,\alpha,\eta$.
\end{lemma}

\begin{proof}
If $|\partial_{\mathrm{real}}^+(X)|>k$, then there is no $\sttqpk$-chip in
$\widehat D$. Indeed, every $\sttqpk$-chip is an $S$-$T_1$ reachability set
of real cut size at most $k$, while $X$ is compact. Hence in this case we
take $\Pcal=\emptyset$. We may therefore assume that
$|\partial_{\mathrm{real}}^+(X)|\le k$.

Let
\[
b\coloneqq 2k+\alpha+2\eta.
\]
Since the real part of $\widehat D$ is $\alpha$-balanced, we have
\[
|\partial_{\mathrm{real}}^-(X)|
\le
|\partial_{\mathrm{real}}^+(X)|+\alpha
\le
k+\alpha.
\]
Thus
\[
|\intf_{\widehat D_{\mathrm{real}}}(X)|\le 2k+\alpha
\qquad\text{and}\qquad
|\outf_{\widehat D_{\mathrm{real}}}(X)|\le 2k+\alpha.
\]
Since $|\auxv(\widehat D)|\le 2\eta$, it follows that
\[
|\rintf_{\widehat D}(X)|\le b.
\]

We construct the family by guessing all bounded data that can arise from a
crossing chip. For every set
$B\subseteq \outf_{\widehat D_{\mathrm{real}}}(X)$, 
fix an injective labeling of $B$ by a label set $I_B\subseteq [b]$ with
$|I_B|=|B|$. For every representative
$R\in \Rcal_{\equiv_\pi}[\Dcal_{\subseteq [b]}]$ 
whose label set is $I_B$, and for every irreflexive relation

$\Gamma\subseteq \rintf_{\widehat D}(X)\times \rintf_{\widehat D}(X)$,   
build the routed graph $\widehat D'$ as follows. Let
\[
A_B\coloneqq
\{(u,v)\in A_{\mathrm{real}}(\widehat D)\mid u,v\in B\}.
\]
Its real part is
\[
D_{\mathrm{real}}'\coloneqq
\bigl(\widehat D_{\mathrm{real}}[X\cup B]-A_B\bigr)\oplus R.
\]

Let
\[
A_{\mathrm{old}}^X
\coloneqq
\{(u,v)\in A_{\mathrm{aux}}(\widehat D)\mid u,v\in X\}.
\]
Set
\[
A_{\mathrm{aux}}'\coloneqq
(A_{\mathrm{old}}^X\cup \Gamma)\setminus A(D_{\mathrm{real}}').
\]
Then
\[
\widehat D'\coloneqq
\bigl(V(D_{\mathrm{real}}'),A(D_{\mathrm{real}}'),A_{\mathrm{aux}}'\bigr).
\]

For every choice of sets
\[
T^1,T^\circ,Q^\circ\subseteq \rintf_{\widehat D}(X),
\]
we add the tuple
\[
\bigl(
\widehat D',
S\cup V(R),
T^1,
(T_2\cap X)\cup T^\circ\cup \tail(A_{\mathrm{aux}}'),
(Q\cap X)\cup Q^\circ
\bigr)
\]
to $\Pcal$.

The number of tuples is bounded by a function of $\pi,k,\alpha,\eta$ only.
Indeed, both
$|\outf_{\widehat D_{\mathrm{real}}}(X)|$
 and 
$|\rintf_{\widehat D}(X)|$ 
are bounded by $b$, and the number and size of the representatives in
$\Rcal_{\equiv_\pi}[\Dcal_{\subseteq [b]}]$ 
are bounded by $r_\pi(b)$. We call the resulting bound
$p_{\pi,k,\alpha,\eta}$.

The number of auxiliary arcs in each reduced instance is also bounded.
Indeed,
\[
|A_{\mathrm{aux}}'|
\le
|A_{\mathrm{aux}}(\widehat D)|+|\rintf_{\widehat D}(X)|^2
\le
\eta+b^2.
\]
By construction, the forbidden set contains every auxiliary tail. Together
with the preceding bound, this shows that each produced instance is
$\eta'_{\pi,k,\alpha,\eta}$-guarded, where
we may take
$\eta'_{\pi,k,\alpha,\eta}\coloneqq \eta+b^2$.

For balance, \Cref{lem:balanceStability} bounds the balance of
$\widehat D_{\mathrm{real}}[X]$ by $2\alpha+2k$. Add $B$ as isolated
vertices and then the arcs between $B$ and $X$, of which there are at most
$2k+\alpha$. This gives
$\widehat D_{\mathrm{real}}[X\cup B]-A_B$ with balance at most
$3\alpha+4k$. Gluing $R$ increases it by at most $|A(R)|\le r_\pi(b)$.
Thus we may take $\alpha'_{\pi,k,\alpha,\eta}=3\alpha+4k+r_\pi(b)$.

Let $Z$ be an important
$\sttqpk$-chip crossing $X$ and satisfying
\[
\partial_{\mathrm{real}}^+(Z)\cap\partial_{\mathrm{real}}(X)=\emptyset.
\]
We show that the guesses include the reduced instance for $Z$. 

Set
\[
B=B_X(Z).
\]
By \Cref{obs:boundedAttachmentBoundary}, applied to the real part of
$\widehat D$, we have
\[
B\subseteq \outf_{\widehat D_{\mathrm{real}}}(X)
\qquad\text{and}\qquad
|B|\le 2k+\alpha.
\]
Equip $B$ with the fixed labeling chosen in the construction of $\Pcal$.
Let $R$ be the representative of the boundaried digraph
$\widehat D_{\mathrm{real}}[Z\setminus X]$ with boundary $B$ and this
labeling.

By \Cref{obs:shortcutEndpointsRoutedInterface}, all shortcut endpoints,
exits, contacts, and entry vertices lie in $\rintf_{\widehat D}(X)$.
Thus the following data are among the guesses:
\[
\Gamma=\Gamma_X(Z),
\]
\[
T^1=\safeExit{Z}{X},
\]
\[
T^\circ=\blockedExit{Z}{X}\cup \contact{Z}{X},
\]
and
\[
Q^\circ=
\entry{Z}{X}{T_1}{Q\setminus X}
\cup
\entry{Z}{X}{T_1}{\head(\partial_{\mathrm{real}}^+(Z))\setminus X}.
\]
For these choices, the corresponding tuple in $\Pcal$ is exactly
\[
\bigl(
\widehat D_X(Z),
S_X(Z),
T_X^1(Z),
T_X^2(Z),
Q_X(Z)
\bigr).
\]

By \Cref{lem:representativeGluing}, the set
$C_X(Z)=(X\cap Z)\cup V(R)$ 
is an important
$
(S_X(Z),\allowbreak T_X^1(Z),\allowbreak T_X^2(Z),\allowbreak Q_X(Z),\allowbreak \pi,\allowbreak k-1)
$-chip in $\widehat D_X(Z).$ Moreover,
$\emptyset\neq \partial_{\mathrm{real}}^+(C_X(Z))
\subsetneq
\partial_{\mathrm{real}}^+(Z)$. 
Thus the required index $i$ exists, with $C_i=C_X(Z)$.
\end{proof}

Before combining the nested and crossing reductions, we prove the two
boundary-band analogues of \Cref{lem:nestedWithin}. The main nested lemma
handles chips that lie between two consecutive members of the minimum cut
sequence. The initial band, between $S$ and $X_1$, keeps the source as $S$
and uses $T_{X_1}^\star$ as the local target. The final band, above
$X_q$, keeps the target as $T_1$. We do not treat the final band by setting
$X'=R(S,T_1)$ in \Cref{lem:nestedWithin}. In a routed instance,
$R(S,T_1)$ need not be auxiliary-respecting, since (in principle) an auxiliary arc may
leave $R(S,T_1)$ directly into $T_1$.

\begin{lemma}[Initial nested localization]
\label{lem:routedInitialNestedWithin}
Suppose that there is an $S$-$T_1$ path in the routed digraph $\widehat D$. Let
$\mathcal X=(X_1,\ldots,X_q)$ be a minimum $S$-$T_1$ cut sequence in
$\widehat D$. Let $Z$ be an important
$
(S,T_1,T_2,Q,\pi,k)
$-chip in $\widehat D$ such that
$S\subseteq Z\subset X_1$ and 
$\partial_{\mathrm{real}}^+(Z)\cap
\partial_{\mathrm{real}}^+(X_1)=\emptyset$.

Set
\[
\Delta_Z\coloneqq \entry{Z}{X_1}{T_1}{Q\setminus X_1}
\qquad\text{and}\qquad
Q^\star\coloneqq (Q\cap X_1)\cup \Delta_Z .
\]
Then $Z$ is an important
$
(S,T_{X_1}^\star,T_2,Q^\star,\pi,k)
$-chip in $\widehat D$ and  
$
\lambda_{\widehat D}(S,T_{X_1}^\star)>
\lambda_{\widehat D}(S,T_1).
$
\end{lemma}

\begin{lemma}[Final nested localization]\label{lem:routedFinalNestedWithin}
Suppose that there is an $S$-$T_1$ path in the routed digraph $\widehat D$. Let
$\mathcal X=(X_1,\ldots,X_q)$ be a minimum $S$-$T_1$ cut sequence in
$\widehat D$, and set
$R\coloneqq R(S,T_1)$. 
Let $Z$ be an important
$
(S,T_1,T_2,Q,\pi,k)
$-chip in $\widehat D$ such that
$X_q\subset Z\subseteq R$ and 
$\partial_{\mathrm{real}}^+(Z)\cap
\partial_{\mathrm{real}}^+(X_q)=\emptyset$.
Then $Z$ is an important
$
(S_{X_q}^\star,T_1,T_2,Q,\pi,k)
$-chip in $\widehat D$ and $\lambda_{\widehat D}(S_{X_q}^\star,T_1)>
\lambda_{\widehat D}(S,T_1)$.
\end{lemma}

\subsubsection{The inductive bound}

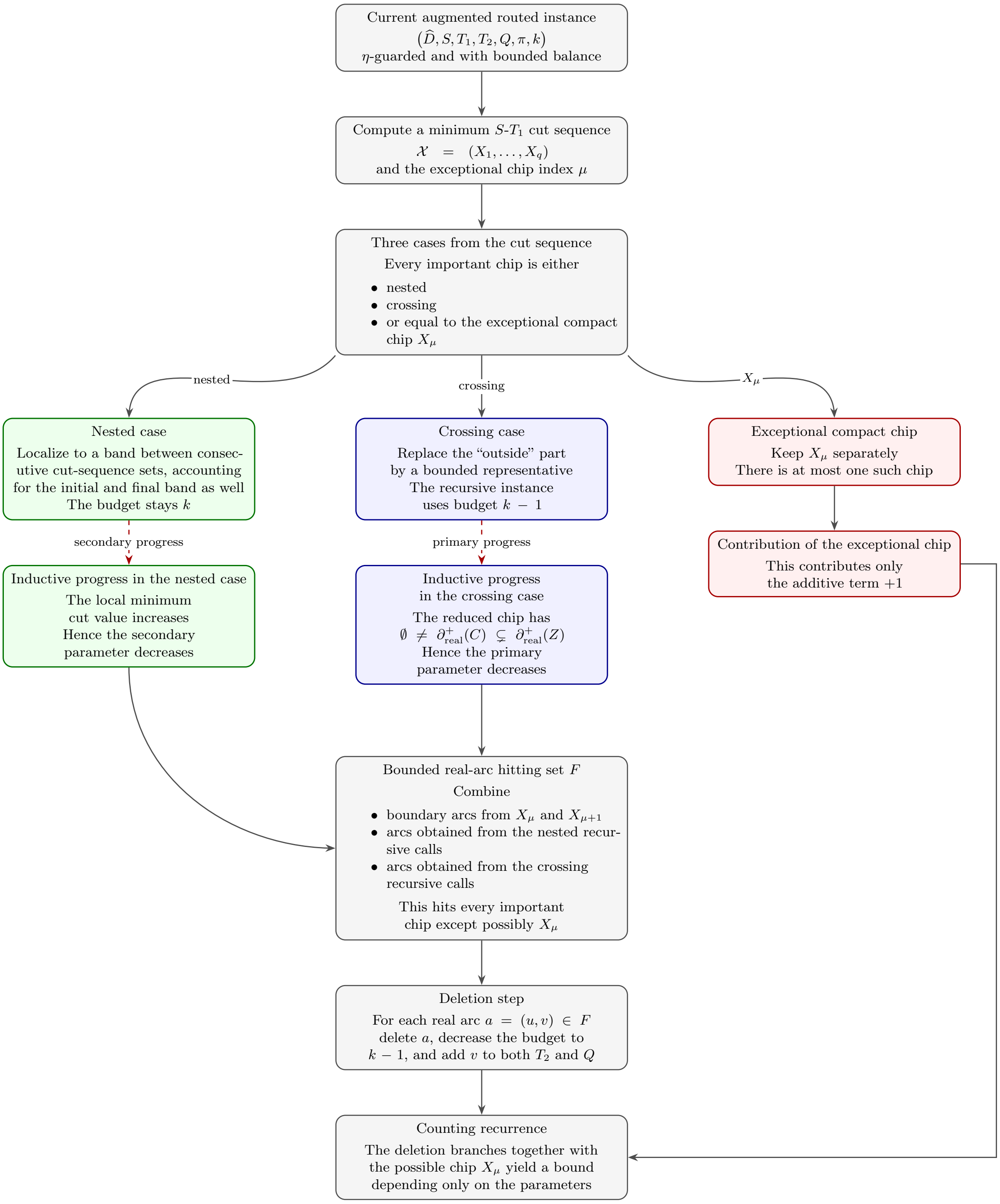
\begin{figure}[p]
\centering
\resizebox{0.97\textwidth}{!}{%
\begin{tikzpicture}[
  >={Stealth[length=2.2mm]},
  node distance=10mm and 12mm,
  box/.style={
    rounded corners=6pt,
    draw=black!70,
    fill=black!4,
    line width=0.7pt,
    align=center,
    inner sep=5pt,
    text width=6.2cm,
    font=\footnotesize
  },
  smallbox/.style={
    rounded corners=6pt,
    draw=black!70,
    fill=black!4,
    line width=0.7pt,
    align=center,
    inner sep=5pt,
    text width=5.2cm,
    font=\footnotesize
  },
  greenbox/.style={
    rounded corners=6pt,
    draw=green!45!black,
    fill=green!7,
    line width=0.8pt,
    align=center,
    inner sep=5pt,
    text width=5.3cm,
    font=\footnotesize
  },
  bluebox/.style={
    rounded corners=6pt,
    draw=blue!55!black,
    fill=blue!6,
    line width=0.8pt,
    align=center,
    inner sep=5pt,
    text width=5.3cm,
    font=\footnotesize
  },
  redbox/.style={
    rounded corners=6pt,
    draw=red!65!black,
    fill=red!6,
    line width=0.8pt,
    align=center,
    inner sep=5pt,
    text width=5.3cm,
    font=\footnotesize
  },
  lab/.style={
    font=\scriptsize,
    fill=white,
    inner sep=1.2pt
  },
  flow/.style={
    ->,
    >=Stealth,
    line width=0.75pt,
    draw=black!70
  },
  prog/.style={
    ->,
    >=Stealth,
    line width=0.75pt,
    draw=red!65!black,
    dashed
  }
]

\node[box] (inst) {
Current augmented routed instance\\[1mm]
$\bigl(\widehat D,S,T_1,T_2,Q,\pi,k\bigr)$\\
$\eta$-guarded and with bounded balance
};

\node[box, below=of inst] (seq) {
Compute a minimum $S$-$T_1$ cut sequence\\[1mm]
$\mathcal X=(X_1,\ldots,X_q)$\\
and the exceptional chip index $\mu$
};

\node[box, below=of seq] (dich) {
Three cases from the cut sequence\\[1mm]
Every important chip is either
\begin{itemize}
\item nested
\item crossing
\item or equal to the exceptional compact chip $X_\mu$
\end{itemize}
};

\draw[flow] (inst) -- (seq);
\draw[flow] (seq) -- (dich);

\node[greenbox, below left=14mm and 18mm of dich] (nested) {
Nested case\\[1mm]
Localize to a band between consecutive cut-sequence sets, accounting for the initial and final band as well\\
The budget stays $k$
};

\node[bluebox, below=14mm of dich] (crossing) {
Crossing case\\[1mm]
Replace the ``outside'' part by a bounded representative\\
The recursive instance uses budget $k-1$
};

\node[redbox, below right=14mm and 18mm of dich] (exception) {
Exceptional compact chip\\[1mm]
Keep $X_\mu$ separately\\
There is at most one such chip
};

\draw[flow] (dich.south west) to[out=-130,in=90] node[lab] {nested} (nested.north);
\draw[flow] (dich) -- node[lab] {crossing} (crossing);
\draw[flow] (dich.south east) to[out=-50,in=90] node[lab] {$X_\mu$} (exception.north);

\node[greenbox, below=of nested] (nestedprog) {
Inductive progress in the nested case\\[1mm]
The local minimum cut value increases\\
Hence the secondary parameter decreases
};

\node[bluebox, below=of crossing] (crossprog) {
Inductive progress in the crossing case\\[1mm]
The reduced chip has
$\emptyset\neq \partial^+_{\mathrm{real}}(C)\subsetneq \partial^+_{\mathrm{real}}(Z)$\\
Hence the primary parameter decreases
};

\node[redbox, below=of exception] (exceptionprog) {
Contribution of the exceptional chip\\[1mm]
This contributes only the additive term $+1$
};

\draw[prog] (nested) -- node[lab] {secondary progress} (nestedprog);
\draw[prog] (crossing) -- node[lab] {primary progress} (crossprog);
\draw[flow] (exception) -- (exceptionprog);

\node[box, below=16mm of crossprog] (hit) {
Bounded real-arc hitting set $F$\\[1mm]
Combine
\begin{itemize}
\item boundary arcs from $X_\mu$ and $X_{\mu+1}$
\item arcs obtained from the nested recursive calls
\item arcs obtained from the crossing recursive calls
\end{itemize}
This hits every important chip except possibly $X_\mu$
};

\draw[flow] (nestedprog.south) to[out=-90,in=180] (hit.west);
\draw[flow] (crossprog) -- (hit);

\node[box, below=of hit] (delete) {
Deletion step\\[1mm]
For each real arc $a=(u,v)\in F$\\
delete $a$, decrease the budget to $k-1$, and add $v$ to both $T_2$ and $Q$
};

\node[box, below=of delete] (recur) {
Counting recurrence\\[1mm]
The deletion branches together with the possible chip $X_\mu$ yield a bound
depending only on the parameters
};

\draw[flow] (hit) -- (delete);
\draw[flow] (delete) -- (recur);
\draw[flow] (exceptionprog.east) -- ++(8mm,0) |- (recur.east);

\end{tikzpicture}%
}
\caption{A representation of the full induction. In the bottom blue box, $Z$ is the original crossing chip and $C$ is its reduced chip.
From the current augmented routed instance, we compute a minimum
$S$-$T_1$ cut sequence and the index $\mu$. Nested chips are handled by
local instances with the same budget but larger minimum cut value, so the
secondary induction parameter decreases. Crossing chips are handled by
boundedly many reduced routed instances with budget $k-1$, so the primary
induction parameter decreases. The possible compact chip $X_\mu$
contributes only one additional chip. The recursive calls and the
cut-sequence boundary arcs produce a bounded real-arc hitting set $F$.
Finally, for each $a=(u,v)\in F$, the deletion step removes $a$, lowers
the budget, and adds $v$ to both $T_2$ and $Q$.}
\label{fig:inductive-assembly}
\end{figure}

The purpose of the
induction is to build a bounded set $F$ of real arcs that hits the outgoing
boundary of every important chip except possibly one compact chip from the
minimum cut sequence. Once such an $F$ is available, the deletion step
bounds the number of chips hit by each arc of $F$ using instances with budget $k-1$. \Cref{fig:inductive-assembly} illustrates the full induction step. 

\begin{lemma}[Bounded number of important augmented chips]
\label{lem:boundedAugmentedChips}
For every $k,\alpha,\eta$ and hereditary CMSO sentence $\pi$, there is a
constant $c_{k,\pi,\alpha,\eta}$ such that the following holds. Let
$\widehat D$ be a routed digraph whose real part is $\alpha$-balanced, and
let $S,T_1,T_2,Q\subseteq V(\widehat D)$ satisfy
$S\cap (T_1\cup T_2\cup Q)=\emptyset$.
Suppose moreover that $(\widehat D,S,T_1,T_2,Q)$ is $\eta$-guarded. Then
the number of important
$
(S,T_1,T_2,Q,\pi,k)
$-chips in $\widehat D$ is at most $c_{k,\pi,\alpha,\eta}$.
\end{lemma}

\begin{proof}
We prove a stronger statement by induction. For every $k$, every balance
bound $\alpha$, every bound $\eta$ on the number of auxiliary arcs, every
hereditary CMSO sentence $\pi$, and every $\rho\in\{0,\ldots,k\}$, there is
a constant $c_{k,\pi,\alpha,\eta,\rho}$ with the following property. Let
$\widehat D$ be a routed digraph whose real part is $\alpha$-balanced, and
let $S,T_1,T_2,Q\subseteq V(\widehat D)$ satisfy
$S\cap (T_1\cup T_2\cup Q)=\emptyset$.
Suppose that $(\widehat D,S,T_1,T_2,Q)$ is $\eta$-guarded and that either
$\lambda_{\widehat D}(S,T_1)>k$,
or
$\lambda_{\widehat D}(S,T_1)\le k$
and
$k-\lambda_{\widehat D}(S,T_1)\le \rho$.
Then the number of important
$
(S,T_1,T_2,Q,\pi,k)
$-chips in $\widehat D$ is at most $c_{k,\pi,\alpha,\eta,\rho}$.

The role of $\rho$ is only to make nested recursion measurable. A nested
recursive call may keep the same budget, so the induction needs a second
quantity that decreases when the minimum cut value increases. Since every
chip has real outgoing boundary of size at most $k$, every recursive
instance that can contain a chip has minimum cut value at most $k$.

Call an instance \emph{invalid} if its source meets the target, forbidden
set, or request set, and discard it. An invalid instance has no chip, since
every chip contains the source and avoids these three sets. Every valid instance
of budget $k-1$ admits secondary bound $k-1$, including the no-chip case
when its minimum cut exceeds the budget.

The induction is on the lexicographic order of the pair $(k,\rho)$, and it is simultaneous over all balance bounds and all bounds on the number of auxiliary arcs. If $\lambda_{\widehat D}(S,T_1)>k$, then there are no chips. If there is no $S$-$T_1$ path, then \Cref{ex:trivialChip} gives at most one important chip. Thus assume that $ 1\le \lambda_{\widehat D}(S,T_1)\le k$. In particular, $k\ge 1$.

Set
\[
\ell\coloneqq \lambda_{\widehat D}(S,T_1).
\]
By \Cref{prop:routedCutSequenceExists}, let
\[
\mathcal X=(X_1,\ldots,X_q)
\]
be a minimum $S$-$T_1$ cut sequence in $\widehat D$. Let $\mu$ be the largest index in
$[q]$ such that $X_\mu$ is an
$
(S,T_1,T_2,Q,\pi,k)
$-chip, and set $\mu=0$ if no such index exists. Set 
\[
X_0=S
\qquad\text{and}\qquad
X_{q+1}=R(S,T_1).
\]

The rest of the proof has two phases. First we generate a bounded temporary
hitting set $F$ using the nested and crossing reductions. Then we count chips
by fixing one arc of $F$ and applying the deletion step.

Let us start with the first phase. 
We add some arcs to $F$ right away as follows. If $\mu\ge 1$, add all arcs of
$\partial_{\mathrm{real}}^+(X_\mu)$. If $\mu<q$, add all arcs of
$\partial_{\mathrm{real}}^+(X_{\mu+1})$. In particular, this adds
$\partial_{\mathrm{real}}^+(X_1)$ when $\mu=0$. Each of these compact sets has outgoing
real boundary of size $\ell\le k$, so this contributes at most $2k$ arcs.

\paragraph{Case 1: Nested chips.} We next handle the nested chips whose outgoing boundary avoids the arcs
just added. 

\paragraph{Case 1.a: $\mu\ge 1$ and $\mu<q$.} For every set
\[
\Delta\subseteq \rintf_{\widehat D}(X_{\mu+1}),
\]
consider the local instance
\[
(S_{X_\mu}^\star,T_{X_{\mu+1}}^\star,T_2,
(Q\cap X_{\mu+1})\cup \Delta,\pi,k)
\]
in $\widehat D$. 
Let
\[
\ell^\star
\coloneqq
\lambda_{\widehat D}(S_{X_\mu}^\star,T_{X_{\mu+1}}^\star).
\]
If $\ell^\star\le \ell$, then we add nothing in this subcase. This is safe:
if there were an important nested chip $Z$ with
$X_\mu\subset Z\subset X_{\mu+1}$ whose outgoing boundary avoids the arcs
already added to $F$, then \Cref{lem:nestedWithin} would imply
$\ell^\star>\ell$.

Assume therefore that $\ell^\star>\ell$. If $\ell^\star>k$, then none of
these local instances has a chip of budget at most $k$. Otherwise
$\ell^\star\le k$, and hence
\[
k-\ell^\star\le k-(\ell+1)\le \rho-1.
\]
For every choice of $\Delta$, if the local instance is invalid, then it
contributes nothing, since we discard it.
Otherwise the induction hypothesis bounds the number of important chips in
that local instance by
$c_{k,\pi,\alpha,\eta,\rho-1}$.

For each such important local chip $Y$, add all arcs of
$\partial_{\mathrm{real}}^+(Y)$ to $F$.

This adds only boundedly many arcs. Indeed, since $X_{\mu+1}$ is compact
and the real part of $\widehat D$ is $\alpha$-balanced, we have
\[
|\intf_{\widehat D_{\mathrm{real}}}(X_{\mu+1})|\le 2k+\alpha .
\]
Since the instance is $\eta$-guarded, we have
$|\auxv(\widehat D)|\le 2\eta$. Therefore
\[
|\rintf_{\widehat D}(X_{\mu+1})|\le 2k+\alpha+2\eta .
\]
Hence there are only boundedly many choices of $\Delta$, and each
important local chip has real outgoing boundary of size at most $k$.

\paragraph{Case 1.b: $\mu=0$.} The initial band is the band between $S$
and $X_1$. Let
$\ell^\star
\coloneqq
\lambda_{\widehat D}(S,T_{X_1}^\star)$. 
If $\ell^\star\le \ell$, then we add nothing in this subcase. This is safe:
if there were an important nested chip $Z$ with $S\subseteq Z\subset X_1$
whose outgoing boundary avoids the arcs already added to $F$, then
\Cref{lem:routedInitialNestedWithin} would imply $\ell^\star>\ell$.

For $\ell<\ell^\star\le k$, use the same recursion as in Case 1.a on
\[
(S,T_{X_1}^\star,T_2,(Q\cap X_1)\cup\Delta,\pi,k),
\qquad \Delta\subseteq\rintf_{\widehat D}(X_1).
\]
The secondary bound drops to $\rho-1$. Discard invalid instances and add
the real outgoing boundaries of all important local chips to $F$.
If $\ell^\star>k$, there are no local chips.

\paragraph{Case 1.c: $\mu=q$.} Let
\[
\ell^\star
\coloneqq
\lambda_{\widehat D}(S_{X_q}^\star,T_1).
\]
If $\ell^\star\le \ell$, then we add nothing in this subcase. This is safe:
if there were an important chip nested between $X_q$ and $R(S,T_1)$ whose
outgoing boundary avoids $\partial_{\mathrm{real}}^+(X_q)$, then
\Cref{lem:routedFinalNestedWithin} would imply $\ell^\star>\ell$.

For $\ell<\ell^\star\le k$, apply the same recursion to the single local
instance $(S_{X_q}^\star,T_1,T_2,Q,\pi,k)$, again with secondary bound
$\rho-1$. If it is valid, add the outgoing boundaries of all its important
chips to $F$. If $\ell^\star>k$, it has no chips.

\paragraph{Case 2: Crossing chips.} Let $\mathcal Y$ be the following
collection of compact reachability sets:
\[
\mathcal Y=
\begin{cases}
\{X_1\}, & \text{if } \mu=0,\\
\{X_\mu,X_{\mu+1}\}, & \text{if } 1\le \mu<q,\\
\{X_q\}, & \text{if } \mu=q.
\end{cases}
\]
For every $X\in\mathcal Y$, add all arcs of $\partial_{\mathrm{real}}(X)$ to $F$. Since $X$ is compact,
$|\partial_{\mathrm{real}}^+(X)|=\ell\le k$. Since the real part of
$\widehat D$ is $\alpha$-balanced, we also have
\[
|\partial_{\mathrm{real}}^-(X)|\le k+\alpha .
\]
Thus this adds only boundedly many real arcs.

Fix $X\in\mathcal Y$. Apply \Cref{lem:preinductionStep} to the
$\eta$-guarded routed instance
$(\widehat D,S,T_1,T_2,Q)$ 
and the compact set $X$. We obtain a family
$\mathcal P_X=
\{(\widehat D_i,S_i,T_i^1,T_i^2,Q_i)\mid i\in[p]\}$
of routed instances, where
$p\le p_{\pi,k,\alpha,\eta}$. 

Each instance has an $\alpha'_{\pi,k,\alpha,\eta}$-balanced real part and is $\eta'_{\pi,k,\alpha,\eta}$-guarded. For every $i\in[p]$, consider the candidate recursive instance
\[
(\widehat D_i,S_i,T_i^1,T_i^2,Q_i).
\]
Discard invalid instances. For every remaining instance, induction applies
with budget and secondary bound $k-1$.
Thus the number of important $(S_i,T_i^1,T_i^2,Q_i,\pi,k-1)$-chips in
$\widehat D_i$ is bounded by
\[
c_{k-1,\pi,\alpha'_{\pi,k,\alpha,\eta},
\eta'_{\pi,k,\alpha,\eta},k-1}.
\]

 For every such important chip $Y$, add to $F$ all arcs of $\partial_{\mathrm{real}}^+(Y)$ that are real arcs of the original graph $\widehat D$.

This again adds only boundedly many arcs. The number of reduced instances is bounded by $p_{\pi,k,\alpha,\eta}$, the
number of important chips in each reduced instance is bounded by induction,
and each such chip has real outgoing boundary of size at most $k-1$.

At this stage $F$ contains arcs obtained from every local or reduced instance
that can arise from the dichotomy. We now check that these guesses cover all
important chips.

Specifically, we claim that $F$ hits every important
$
(S,T_1,T_2,Q,\pi,k)
$-chip, with the sole possible exception of $X_\mu$ when $\mu\ge1$.
Let $Z$ be an important chip other than this possible exception; when
$\mu=0$, this includes $Z=S$.
By \Cref{lem:possibleBranchings}, the chip $Z$ is either nested in one of
the cases above, or crosses one of the compact sets in $\mathcal Y$.

Suppose first that $Z$ is nested. If
$\partial_{\mathrm{real}}^+(Z)$ meets one of the boundary sets added at the
start, then $Z$ is hit by $F$. Otherwise, the appropriate localization
argument applies. By \Cref{obs:routedPrefPath}, the entry set used in that local instance is
contained in the relevant routed interface. Hence this local instance was
one of the instances considered above. Since $Z$ is important in that
local instance, we added all arcs of $\partial_{\mathrm{real}}^+(Z)$ to
$F$. Thus $Z$ is hit by $F$.

Now suppose that $Z$ crosses some $X\in\mathcal Y$. If
\[
\partial_{\mathrm{real}}^+(Z)\cap\partial_{\mathrm{real}}(X)\ne\emptyset,
\]
then $Z$ is hit by the real arcs of $\partial_{\mathrm{real}}(X)$ that we
added to $F$. Otherwise,
\[
\partial_{\mathrm{real}}^+(Z)\cap\partial_{\mathrm{real}}(X)=\emptyset .
\]
By \Cref{lem:preinductionStep}, one of
the reduced instances associated with $X$ contains an important reduced
chip $C$ such that
\[
\emptyset\ne \partial_{\mathrm{real}}^+(C)
\subsetneq
\partial_{\mathrm{real}}^+(Z).
\]
This is an inclusion of real arcs of the original graph $\widehat D$. Hence every arc of
$\partial_{\mathrm{real}}^+(C)$ was added to $F$ when we processed the
important chips in the reduced instances. Therefore $F$ hits $Z$.

Thus every important chip, except possibly $X_\mu$ when $\mu\ge1$,
has an outgoing real arc in $F$. The size of $F$ is bounded by a constant depending only on
$k,\pi,\alpha,\eta,\rho$. Let this bound be
$h_{k,\pi,\alpha,\eta,\rho}$.

We now count the important chips. When $\mu\ge1$, we set aside the
single possible exception $X_\mu$. Every remaining important chip has
some arc of $F$ in its outgoing real boundary; when $\mu=0$, all important
chips are hit. Hence
\[
|\Ical_{\widehat D}(S,T_1,T_2,Q,\pi,k)|
\le
1+
\sum_{a\in F}
|\Ical_{\widehat D}^a(S,T_1,T_2,Q,\pi,k)|.
\]

Let $a=(u,v)\in F$. If $v\in S$, then
\[
\Ical_{\widehat D}^a(S,T_1,T_2,Q,\pi,k)=\emptyset,
\]
because every $\sttqpk$-chip contains $S$, while any chip whose outgoing
real boundary contains $a$ must exclude the head $v$ of $a$. Hence we may
assume that $v\notin S$.

Since $a$ is a real arc, the proof of \Cref{cor:inducedChips} applies in
the routed setting. Therefore
\[
|\Ical_{\widehat D}^a(S,T_1,T_2,Q,\pi,k)|
\le
|\Ical_{\widehat D-a}(S,T_1,T_2\cup\{v\},Q\cup\{v\},\pi,k-1)|.
\]

By \Cref{lem:balanceStability}, deleting one real arc increases the balance
by at most one. The instance
\[
(\widehat D-a,S,T_1,T_2\cup\{v\},Q\cup\{v\})
\]
is still $\eta$-guarded, since the auxiliary arcs are unchanged and the
forbidden set only grows. Discarding invalid instances as before, induction applies with budget and
secondary bound $k-1$.
Therefore
\[
|\Ical_{\widehat D-a}(S,T_1,T_2\cup\{v\},Q\cup\{v\},\pi,k-1)|
\le
c_{k-1,\pi,\alpha+1,\eta,k-1}.
\]

It follows that
\[
|\Ical_{\widehat D}(S,T_1,T_2,Q,\pi,k)|
\le
1+
h_{k,\pi,\alpha,\eta,\rho}\cdot
c_{k-1,\pi,\alpha+1,\eta,k-1}.
\]
Define
\[
c_{k,\pi,\alpha,\eta,\rho}
\coloneqq
1+
h_{k,\pi,\alpha,\eta,\rho}\cdot
c_{k-1,\pi,\alpha+1,\eta,k-1}.
\]
For budget $0$, \Cref{ex:trivialChip} shows that there is at most one chip, so
we set $c_{0,\pi,\alpha,\eta,0}\coloneqq 1$.
This completes the induction.

Finally, take
\[
c_{k,\pi,\alpha,\eta}
\coloneqq
\max_{\rho\in\{0,\ldots,k\}} c_{k,\pi,\alpha,\eta,\rho}.
\]
This proves the lemma.

\end{proof}

We are now ready to complete the proof of our first main theorem.

\begin{proof}[Proof of \Cref{thm:boundImpChips}] View the ordinary digraph $D$ as the routed digraph 
$\widehat D=(V(D),A(D),\emptyset)$.  
Its real part is $D$, it has no auxiliary arcs, and it is $0$-guarded. Apply \Cref{lem:boundedAugmentedChips} with  $T_1=T, \qquad T_2=\emptyset, \qquad Q=\emptyset, \qquad \eta=0$.   
For this choice, an $(S,T,\emptyset,\emptyset,\pi,k)$-chip in $\widehat D$ is exactly an ordinary $(S,T,\pi,k)$-chip in $D$. The forbidden and request conditions disappear, and the auxiliary-respecting condition is vacuous. 
The domination relation defining importance is also the same, since real outgoing boundaries in $\widehat D$ are precisely ordinary outgoing boundaries in $D$. Therefore the number of important $(S,T,\pi,k)$-chips in $D$ is bounded by the constant supplied by \Cref{lem:boundedAugmentedChips}. This proves the theorem. \end{proof}

\subsection{Enumerating important chips}
\begin{proof}[Proof of \Cref{thm:enumerateBoundImpChips}]
We describe the algorithm for the augmented routed instances used in the
proof of \Cref{lem:boundedAugmentedChips}. The theorem follows by applying
this algorithm to the ordinary routed instance
$(V(D),A(D),\emptyset)$ with $T_1=T$, $T_2=Q=\emptyset$, and $\eta=0$.

The algorithmic proof is close to the counting proof. The
counting proof builds a bounded search tree whose nodes are local or reduced
instances. For enumeration, we build the same tree explicitly and test the
chip conditions whenever a candidate set is produced.

Fix $k,\alpha,\pi$. The algorithm is non-uniform in these parameters. More
generally, recursive calls are made on augmented routed instances with a
current balance bound and a current bound on the number of auxiliary arcs.
In the proof of \Cref{lem:boundedAugmentedChips}, both bounds remain
bounded by functions of the original parameters. Thus all boundary sizes
that arise in crossing reductions are bounded in terms of $k,\alpha,\pi$.
Whenever the crossing reduction uses representatives of the finite
equivalence classes of $\equiv_\pi$, these representatives are treated as
part of the fixed advice for the algorithm.

Let us follow the proof of \Cref{lem:boundedAugmentedChips}. Given
an augmented routed instance
$(\widehat D,\allowbreak S,\allowbreak T_1,\allowbreak T_2,\allowbreak Q,\allowbreak \pi,\allowbreak k)$,
it first handles the trivial cases. If there is no $S$-$T_1$ path, then by
\Cref{ex:trivialChip} there is at most one chip, namely the set reachable
from $S$, and this set can be checked directly. If
$\lambda_{\widehat D}(S,T_1)>k$, then there is no chip.

Otherwise, compute a minimum $S$-$T_1$ cut sequence
$\mathcal X=(X_1,\ldots,X_q)$ 
using \Cref{prop:routedCutSequenceExists}. We compute the largest index
$\mu$ for which $X_\mu$ is an augmented chip, setting $\mu=0$ if no such
index exists. These indices form an initial segment: all $X_i$ are compact
auxiliary-respecting reachability sets, while heredity, avoidance of $T_2$,
and the request paths are preserved when passing to an earlier $X_i$.
Hence binary search finds $\mu$ using at most
$\lceil\log_2(q+1)\rceil$ chip tests. All chip conditions except
$\widehat D_{\mathrm{real}}[X_i]\models\pi$ are checkable in polynomial
time. The latter is decided using the assumed test for $\pi$. Since $X_i$
is compact, $|\partial^+_{\mathrm{real}}(X_i)|\le k$, and
\Cref{lem:balanceStability} bounds the balance of
$\widehat D_{\mathrm{real}}[X_i]$ in terms of $k$ and the current balance
bound.

Having found $\mu$, the algorithm constructs the bounded real-arc set $F$
as in the proof of \Cref{lem:boundedAugmentedChips}. First it adds the
bounded boundary arcs used there. In the nested cases, it forms the same
bounded list of local instances. For each such instance, it first computes
the relevant local minimum cut value. If this value does not give the
increase required in the proof of \Cref{lem:boundedAugmentedChips}, the
local instance is ignored. Otherwise, the algorithm recursively enumerates
the important chips of the local instance and adds their outgoing real
boundaries to $F$.

In the crossing case, the algorithm constructs the bounded family of
reduced routed instances from \Cref{lem:preinductionStep}. This construction
is effective once the representatives for $\equiv_\pi$ are fixed as advice.
It then recursively enumerates the important chips of these reduced
instances and adds to $F$ those outgoing real arcs that belong to the
original routed graph.

By the proof of \Cref{lem:boundedAugmentedChips}, the resulting set $F$ has
size bounded by a function of $k,\alpha,\pi$ only, and it intersects the
outgoing real boundary of every important chip, except possibly the single
compact chip $X_\mu$ when $\mu\ge1$. The algorithm therefore forms a candidate family as follows. If $\mu\ge 1$,
it includes $X_\mu$. For every real arc $a=(u,v)\in F$, it recursively
enumerates the important
$(S,T_1,T_2\cup\{v\},Q\cup\{v\},\pi,k-1)$-chips in $\widehat D-a$ and adds the returned vertex sets to the candidate family.

Correctness follows from \Cref{cor:inducedChips}. If an important chip of
the current instance is not the possible exception $X_\mu$ with
$\mu\ge1$, then its outgoing boundary
contains some arc $a\in F$, and hence the same vertex set appears as an
important chip in the corresponding recursive instance with budget $k-1$.
Thus every important chip of the current instance is contained in the
candidate family.

It remains only to remove spurious candidates. For each candidate set, the
algorithm checks whether it is an augmented chip in the current instance,
checking the combinatorial conditions, including the boundary bound, before
testing $\pi$. Again, all conditions except the test for $\pi$ are polynomial-time
checkable, and the test for $\pi$ is valid because any candidate with real
outgoing boundary of size at most $k$ induces a real subdigraph whose
balance is bounded by \Cref{lem:balanceStability} as a function of $k$ and
the current balance bound. Finally, among
the surviving candidates, the algorithm keeps exactly those sets $Z$ for
which there is no surviving candidate $Z'\supset Z$ with
$|\partial^+_{\mathrm{real}}(Z')|
\le
|\partial^+_{\mathrm{real}}(Z)|$.
This removes precisely the non-important chips. Indeed, if a chip is not
important, then it is dominated by a maximal chip with no larger outgoing
boundary, and such a maximal chip is important, hence already appears in the
candidate family by the previous paragraph.

The recursion depth is bounded as in the proof of
\Cref{lem:boundedAugmentedChips}. Nested calls keep the budget fixed and
decrease the secondary measure $k-\lambda$. Deletion and crossing calls
decrease the budget. The branching factor and hence the number of nodes
are bounded by functions of the initial parameters. Apart from tests for
$\pi$, the work at each node is polynomial in the current graph size, with
an absolute exponent $d_0$.

For later use, we bound the cost of the tests for $\pi$ explicitly. Start
with an augmented instance on $n$ vertices and parameters
$\pi,k,\alpha,\eta$. Every crossing reduction adds only a parameter-bounded
number of representative vertices, and the other recursive operations add
none. Thus every graph encountered has at most $n$ plus a
parameter-bounded number of vertices. Each node makes logarithmically many
chip tests to find $\mu$, and only a parameter-bounded number of further
tests on candidate sets. By \Cref{lem:balanceStability}, the balance of
every graph submitted to the test for $\pi$ is also parameter-bounded.
Consequently, there is an integer $A=A(\pi,k,\alpha,\eta)\ge1$, independent
of $n$, such that the entire enumeration makes at most
$A\log_2(n+2)$ tests, each on a graph with at most $n+A$ vertices and balance
at most $A$, and does at most $A(n+1)^{d_0}$ other work. We use $A$ as the
balance bound in all these tests. If $\tau_\pi(N,\gamma)$ bounds the time for testing $\pi$ on graphs with at
most $N$ vertices and balance at most $\gamma$, the total running time is
therefore at most
\begin{equation}\label{eq:chipEnumerationOracleBound}
A\bigl((n+1)^{d_0}+\log_2(n+2)\,\tau_\pi(n+A,A)\bigr).
\end{equation}
The assumed test for $\pi$ has a polynomial exponent independent of the
balance. This gives the claimed running time
$c'_{\pi,k,\alpha}\cdot n^{\bigoh(1)}$ for $\eta=0$, with an exponent
independent of $k$ and $\alpha$, and the algorithm outputs exactly all
important $(S,T,\pi,k)$-chips.
\end{proof}

\section{Application: non-uniform FPT algorithm for \arcDoct{}}
\label{sec:arcDoctApplication}
In this section, we use important-chip enumeration to solve \arcDoct{} on balanced digraphs. 

\subsection{\arcDoct{}  preliminaries}

A directed cycle with an odd number of arcs is called a {\em directed odd cycle}. 
Given a digraph $D$ and a set $X\subseteq A(D)$, we say that $X$ is an {\em arc directed odd cycle transversal} (in short, {\em \arcDoct{}}) of $D$ if $X$ intersects every directed odd cycle of $D$.

\defprob{\arcDoct}{
    A digraph $D$ and an integer $k$.
}{
    Is there an arc subset $F \subseteq A(D)$ of size at most $k$ such that $D-F$ has no directed odd cycle?
}

   For an instance $(D,k)$ of {\sc Arc DOCT}, we use the
 term {\em solution}  to refer to any set of size at most $k$ that is an \arcDoct{}.

The version of {\arcDoct} where the goal is to delete vertices is called {\vertexDoct}.

We first note that \arcdoct{} remains NP-complete on Eulerian digraphs. See
\Cref{app:doct-hardness-proofs} for details.
Not every arc-deletion problem is NP-hard on Eulerian digraphs. For instance, the classic {\sc Directed Multiway Cut} problem, which asks
whether at most $k$ arcs can be deleted so that no directed path joins two
distinct specified terminals, is known to be polynomial-time solvable on
Eulerian digraphs. 
%
We next use important-chip enumeration to prove fixed-parameter tractability of \arcDoct{} on balanced digraphs.

%
%

Our proofs rely on the following well-known proposition.

\begin{proposition}[Folklore]\label{prop:underlyingBipartite}
    Let $D$ be a strongly connected directed graph. Then, $D$ does not contain a  directed odd cycle if and only if the underlying undirected graph of $D$ is a bipartite graph.
\end{proposition}

Let $\mk_\ell$ denote the CMSO sentence expressing the following: 
\begin{quote}There is an arc set $F$ of size at most $\ell$ such that for every odd-size arc set $X$ that forms a union of directed cycles (that is, every vertex $v$ is either not incident on any arc in $X$, or appears as the head of exactly one arc in $X$ and as the tail of exactly one arc in $X$), $F\cap X\ne\emptyset$.
	
\end{quote}

Any arc set $X\subseteq A(D)$ satisfying the
stated degree condition for every vertex forms a vertex-disjoint union of directed cycles. Moreover, if $|X|$ is odd, one of these cycles is odd. Conversely, every directed odd cycle itself defines such an odd arc set $X$.

So, for a digraph $D$, we have $D\models \mk_\ell$ if and only if $(D,\ell)$ is a yes-instance of {\arcDoct}.

 \subsection{Compression when the modulator lies in one strongly connected component}

The following subroutine follows from the compression algorithm of Lokshtanov et al.~\cite{LokshtanovRSZ20} by the reduction below. 
 
\begin{lemma}\label{lem:specialCaseDOCT}
	There is an algorithm that, given an instance $(D,k)$ of {\sc Arc DOCT} along with a set $W\subseteq V(D)$ such that $D-W$ has no directed odd cycles, runs in time $2^{(k+|W|)^{\bigoh(1)}}\cdot n^{\bigoh(1)}$ and either solves the instance or correctly concludes that there is no solution $S\subseteq A(D)$ such that $W$ intersects exactly one strongly connected component of $D-S$.
\end{lemma}

\begin{proof}
If $W=\emptyset$, then the empty arc set is a solution. If $k=0$, we
test in polynomial time whether $D$ has a directed odd cycle. Hence assume
that $W\ne\emptyset$ and $k\ge1$.

Suppose that a solution $S$ leaves all of $W$ in one strongly connected
component of $D-S$. By \Cref{prop:underlyingBipartite}, this component has
a bipartite underlying graph. We guess the two sides of its bipartition
on $W$, using at most $2^{|W|}$ guesses. For each guess, identify the
vertices in each non-empty part to one terminal, denoted by $w_L$ or
$w_R$, and let $W^*$ be the set of the resulting terminals. In this
intermediate quotient we retain every original arc as a distinct arc
occurrence, including any loops and parallel arcs created by identification.

Replace each original arc occurrence $e=(u,v)$ in the quotient by a
length-three path $u,x_e,y_e,v$, with fresh subdivision vertices for each
occurrence. If $|W^*|=2$, add the two direct arcs $(w_L,w_R)$ and
$(w_R,w_L)$ after this subdivision. Finally, replace every original
nonterminal vertex by $k+1$ independent twins with the same in-neighbors
and out-neighbors; leave the terminals and subdivision vertices unchanged.
Call the resulting simple digraph $H$. The set $W^*$ has size at most two,
$H[W^*]$ is strongly connected, and $H-W^*$ has no directed odd cycle.
Any such cycle would project, through the twins and subdivision paths,
to an odd closed walk in $D-W$.

We apply the algorithm for \emph{Restricted Colorful Walk Cover} from
\cite[Lemma~5.1 in the full version]{LokshtanovRSZ20}, with alphabet size
two and every arc labeled by the transposition. In this specialization,
colorful walk covers are precisely directed odd cycle transversals. The
algorithm finds a set $Y\subseteq V(H)\setminus W^*$ of size at most $k$
whose deletion destroys all directed odd cycles, or correctly concludes
that no such set exists. Its hypotheses hold because $W^*$ is a directed
odd cycle transversal, $H[W^*]$ is strongly connected, and
$|W^*|\le2\le2k+1$. In particular, the terminals are undeletable, so the
direct arcs between them survive every allowed solution.

For completeness, consider the guess induced by a solution $S$ as above, let
$C$ be the strongly connected component of $D-S$ that contains $W$, and delete
$x_e$ for each original arc $e\in S$. For all $u,v\in W$, there is a directed
$u$-$v$ path in $C$, of length zero if $u=v$, since $C$ is strongly connected.
By \Cref{prop:underlyingBipartite}, the underlying graph of $C$ is bipartite,
so every such path has even length if $u$ and $v$ lie in the same guessed part
and odd length otherwise. In particular, an original arc with both endpoints
in one part does not survive in $D-S$, since it would lie in $C$. Hence every
loop created by identification has its subdivision vertex $x_e$ deleted.

Suppose that the resulting graph contains a directed odd cycle. Projecting
twins and contracting subdivision paths gives an odd closed walk $\omega$ in the
quotient whose original arcs avoid $S$. A closed walk that uses only added
direct arcs alternates between $w_L$ and $w_R$ and has even length. Otherwise,
consider a maximal segment of $\omega$ that enters a terminal through an
original arc with head $u\in W$, possibly continues through added direct arcs,
and then leaves through an original arc with tail $v\in W$. The parity of the
number of added arcs on this segment equals the parity of a directed $u$-$v$
path in $C$. Replacing every such segment by that path gives a closed walk in
$D-S$ of the same odd parity. An odd closed walk contains a directed odd
cycle, a contradiction. Subdivision preserves parity, and projecting twins
preserves walk length, so the specified subdivision vertices form a solution
in $H$.

Conversely, normalize any returned solution $Y$ by discarding its twin
vertices. This preserves feasibility: each twin group has a surviving
vertex, and replacing any restored twin by a surviving twin would turn
a newly created odd cycle into an odd closed walk in $H-Y$. The
normalized solution therefore consists only of subdivision vertices.
Delete from $D$ each original arc whose subdivision path meets this
solution. At most $k$ arcs are deleted. Any surviving directed odd cycle
in $D$ would project under identification and lift through the intact
subdivision paths and surviving twins to an odd closed walk in $H-Y$,
a contradiction. Hence every returned solution gives a valid original
arc solution, regardless of the guess.

Each constructed graph has size $\bigoh(k(|V(D)|+|A(D)|))$, and each
call takes $2^{\bigoh(k^2)}\cdot |V(D)|^{\bigoh(1)}$ time. Including all
guesses gives the claimed running time. If every guess fails, completeness
shows that no solution leaves $W$ in one strongly connected component.
\end{proof}

\Cref{lem:specialCaseDOCT} holds for all digraphs and needs no balance assumption.

\subsection{The important-chip branching algorithm}
 
To solve an instance of \arcDoct{} with budget $k$, the important-chip enumeration algorithm needs the property of having a solution of strictly smaller budget to be efficiently testable. We therefore prove the theorem by induction on $k$. 
 
\thmgeneralBalancedDOCT*

\begin{proof}
We prove the theorem by induction on the budget $k$, simultaneously for
all balance bounds. Write $T_k(n,\alpha)$ for the worst-case running time
on inputs with at most $n$ vertices and balance at most $\alpha$, and set
$L(n)=\log_2(n+2)$. We establish the stronger bound
\begin{equation}\label{eq:doctStrongRuntime}
T_k(n,\alpha)\le F_{k,\alpha}(n+1)^dL(n)^k,
\end{equation}
where $d$ is an absolute constant large enough for the polynomial work in
\eqref{eq:chipEnumerationOracleBound}, the approximation routine, and
\Cref{lem:specialCaseDOCT}. The value of $k$ is fixed throughout one
induction step, and $h\le k$ denotes the remaining budget in recursive
calls.

For $k=0$, the algorithm checks whether $D$ contains a directed odd cycle.
By \Cref{prop:underlyingBipartite}, this is equivalent to checking whether
the underlying undirected graph of every strongly connected component of $D$
is bipartite. This also tests $\pi_0$ in polynomial time for every balance bound.

Fix $k\ge 1$, and assume the theorem has been proved for every pair
$(m,\gamma)\in\{0,\ldots,k-1\}\times\mathbb N_0$, where $m$ is the
deletion budget and $\gamma$ is the balance bound. Thus, for every $m<k$, the CMSO sentence
$\mk_m$ is efficiently testable on $\gamma$-balanced digraphs for every
$\gamma\in\mathbb N_0$.

We now prove the statement for this fixed budget $k$. The main work is the
following recursive claim.

\begin{claim}\label{claim:recursiveArcDoct}
There is an algorithm that takes as input a tuple $(D,h,\beta,W)$, where
$h\le k$, the digraph $D$ is $\beta$-balanced, the set
$W\subseteq V(D)$ satisfies $|W|\le 4k$, and $D-W$ has no directed odd
cycle. The algorithm decides whether $(D,h)$ is a yes-instance of
{\sc Arc DOCT}. Moreover, when invoked with balance bound $\beta$, all
recursive calls made by the algorithm have balance bounded by a function of
$k$ and $\beta$, and the running time is
$f(k,\beta)\cdot |V(D)|^{\bigoh(1)}$ for some function $f$.
\end{claim}

Assuming \Cref{claim:recursiveArcDoct}, the theorem is proved as follows. On
the original input $(D,k)$, whose balance is at most $\alpha$, we first
invoke the factor-2 FPT-approximation for {\vertexDoct} in
\cite{LokshtanovRSZ20} (see also \cite{LokshtanovMRSZ21} for an improved
running time). Applied to the instance $(D,2k)$ of {\vertexDoct}, this
algorithm either computes a set $W\subseteq V(D)$ of size at most $4k$ such
that $D-W$ has no directed odd cycles, or correctly concludes that there is
no such set of size at most $2k$. In the latter case, we return \no, since
the endpoints of the arcs in any solution to $(D,k)$ form a vertex set of
size at most $2k$ that intersects all directed odd cycles. In the former
case, if $W=\emptyset$, then $D$ itself has no directed odd cycle and we
return \yes. Otherwise we run the algorithm of
\Cref{claim:recursiveArcDoct} on $(D,k)$ with modulator $W$ and balance
bound $\alpha$.

It remains to prove \Cref{claim:recursiveArcDoct}. 
We prove it by lexicographic induction on
$(h,|W|)$. At a pair $(h,w)$, the claim is proved for every admissible input with $|W|=w$ and every balance bound $\beta\in\mathbb N_0$. The balance bound is not part of the induction measure.
Recursive calls may increase it, but only in a way that is later bounded in the running time analysis.

\medskip
\noindent
\textit{Step 1: Base case.} If $h=0$, we check whether $D$ has
a directed odd cycle. As in the case $k=0$, this takes linear time by \Cref{prop:underlyingBipartite}. 

Assume now
that $h\ge 1$ and that the claim holds for all
pairs $(h',w')\in\{0,\ldots,k\}\times\{0,\ldots,4k\}$ that are
lexicographically smaller than $(h,|W|)$, for every balance bound. Let the current graph have balance at most
$\beta$, and let the current modulator be $W\ne\emptyset$.

\medskip
\noindent
\textit{Step 2: Guessing the ordered partition of $W$.}
Let $F\subseteq A(D)$ denote a hypothetical solution of size at most $h$.
Consider the ordered partition of $W$ into groups $W_1,\dots,W_t$ such that
in $D-F$, each $W_i$ lies in exactly one strongly connected component, no
two distinct groups $W_i$ and $W_j$ belong to the same strongly connected
component, and there is no directed path in $D-F$ from $W_i$ to $W_j$ for
$i<j$. Since $|W|\le 4k$, the number of ordered partitions of $W$ into at
most $|W|$ non-empty groups is at most $(4k)^{4k}=k^{\bigoh(k)}$. For each
guess, we proceed as follows.

If $t=1$, then the current guess asserts that $W$ lies in a single
strongly connected component of $D-F$. We invoke
\Cref{lem:specialCaseDOCT}. If this call finds a solution, then we return
\yes. If it concludes that no solution has this property, then the current
ordered partition cannot correspond to a solution, and we continue with the
next ordered partition.

\medskip
\noindent
\textit{Step 3: Branching via important chips when $t > 1$.}

\paragraph{The empty-separator case.}
We first test whether there is a path from $W_1$ to $W\setminus W_1$ in
$D$. If no such path exists, let $U\coloneqq R(W_1,\emptyset)$
be the set of vertices reachable from $W_1$ in $D$.

Since no arc leaves $U$, every directed cycle of $D$ is either contained in
$D[U]$ or contained in $D-U$. Moreover, $D[U]-W_1$ has no directed odd
cycle, because $U\cap W=W_1$ and $D-W$ has no directed odd cycle.
Similarly, $(D-U)-(W\setminus W_1)$ has no directed odd cycle, since it is
an induced subdigraph of $D-W$.

We try all budget splits. For every $s\in\{0,\ldots,h\}$, we run
\Cref{lem:specialCaseDOCT} on $(D[U],s)$ (we call this the inside instance) with modulator $W_1$, and we
recursively solve $(D-U,h-s)$ (we call this the outside instance) with modulator $W\setminus W_1$. We accept if there is a choice of $s$ for which 
both calls accept.

The correctness of this split is as follows. Suppose first that there is a
solution $F$ compatible with the guessed ordered partition. Then
$F\cap A(D[U])$ is a solution for $D[U]$, while
$F\cap A(D-U)$ is a solution for $D-U$. The strongly connected component of
$D-F$ containing $W_1$ is contained in $U$, since no arc leaves $U$. Hence
\Cref{lem:specialCaseDOCT} applies to the call on $(D[U],s)$ with
modulator $W_1$. Conversely, if the inside and outside instances have
solutions
$F_{\mathrm{in}}\subseteq A(D[U])$ and
$F_{\mathrm{out}}\subseteq A(D-U)$ with
$|F_{\mathrm{in}}|+|F_{\mathrm{out}}|\le h$, then
$F_{\mathrm{in}}\cup F_{\mathrm{out}}$ hits all directed odd cycles of
$D$, because no directed cycle can use vertices both in $U$ and outside
$U$.

Since no arc leaves $U$, at most $\beta$ arcs enter it. Passing to $D-U$
therefore changes the imbalance of any remaining vertex set by at most
$\beta$, so $D-U$ is $2\beta$-balanced. The inside routine needs no balance
bound. The outside call has a strictly smaller modulator $W\setminus W_1$
and no larger budget, so its induction measure decreases.

If no budget split succeeds, then the current ordered partition cannot yield
a solution, and the algorithm continues with the next ordered partition.

\noindent
\paragraph{The non-empty-separator case.}
For the current ordered partition, assume now that $D$ contains a path
from $W_1$ to $W\setminus W_1$.

Let $F_0$ be an inclusion-wise minimal subset of $F$ that hits all paths
from $W_1$ to $W\setminus W_1$ in $D$, and let $Z'\coloneqq R(W_1,F_0)$.
Since the empty-separator case has already been handled, $F_0\ne\emptyset$.

Reachability gives $\partial^+(Z')\subseteq F_0$. This arc set is itself
a $W_1$-$(W\setminus W_1)$ cut, so minimality gives
$F_0=\partial^+(Z')$ and makes $Z'$ strict.
Put $F_{\mathrm{in}}=F\cap A(D[Z'])$ and $\ell_0=|F_{\mathrm{in}}|$.
The set $F_{\mathrm{in}}$ hits every directed odd cycle in $D[Z']$, and
$\ell_0+|F_0|\le h$. Since $F_0\ne\emptyset$, we have $\ell_0<h$, and $Z'$ is
a $(W_1,W\setminus W_1,\pi_{\ell_0},|F_0|)$-chip. We now push to an important
chip without increasing the solution size.

\begin{claim}\label{claim:imp_chip_exists}
There exists a solution $\widehat F$ for the instance $(D,h)$ of \arcDoct{}
and an important
$(W_1,W\setminus W_1,\pi_{\ell_0},|F_0|)$-chip $Z''$ such that
$\widehat F$ contains an arc from $\partial^+(Z'')$.
\end{claim}

\begin{claimproof}
Choose an important chip $Z''\supseteq Z'$ with
$|\partial^+(Z'')|\le |F_0|$, allowing $Z''=Z'$.
Let $F_{\mathrm{bd}}=\partial^+(Z'')$, and choose an arc directed odd
cycle transversal $F_{\mathrm{rep}}$ of $D[Z'']$ of size at most $\ell_0$.
Define
\[
\widehat F=
\bigl(F\setminus(F_0\cup F_{\mathrm{in}})\bigr)
\cup F_{\mathrm{bd}}\cup F_{\mathrm{rep}}.
\]
Since $F_0$ and $F_{\mathrm{in}}$ are disjoint,
\[
|\widehat F|\le |F|-|F_0|-|F_{\mathrm{in}}|
+|F_{\mathrm{bd}}|+|F_{\mathrm{rep}}|\le |F|\le h.
\]
Every restored arc lies inside $Z''$. This is immediate for
$F_{\mathrm{in}}$, and an arc of $F_0$ starts in $Z'\subseteq Z''$ and
either ends in $Z''$ or belongs to $F_{\mathrm{bd}}$.
An odd cycle surviving $\widehat F$ would have to use a restored arc,
since $F$ was a solution. It would then lie entirely in $Z''$, whose
outgoing arcs all belong to $F_{\mathrm{bd}}\subseteq\widehat F$, contradicting the choice of
$F_{\mathrm{rep}}$. Thus $\widehat F$ is a solution.
Finally, $F_{\mathrm{bd}}$ is non-empty because there is a
$W_1$-$(W\setminus W_1)$ path in $D$, and it is contained in $\widehat F$.
\end{claimproof}

\medskip
\noindent
\textit{Step 4: Enumeration and recursion.}
For every $\ell<h$, the property defined by $\pi_\ell$ is hereditary: if a
digraph has an arc directed odd cycle transversal of size at most $\ell$, then
every induced subdigraph has one, obtained by restricting the transversal.
Moreover, since $\ell<h\le k$, the outer induction hypothesis implies that
$\pi_\ell$ is efficiently testable on $\gamma$-balanced digraphs for every
balance bound $\gamma$. Thus the testability hypothesis of
\Cref{thm:enumerateBoundImpChips} is satisfied for each property $\pi_\ell$
that appears below.

For every pair $(\ell,p)$ with $\ell+p\le h$, $\ell<h$, and $p>0$, we enumerate the
set
\[
\CC_{\ell,p}=
\bigcup_{Z \in \Ical(W_1,W\setminus W_1,\pi_\ell,p)}
\partial^{+}(Z)
\]
of all outgoing arcs of important
$(W_1,W\setminus W_1,\pi_\ell,p)$-chips in time
$c'_{\pi_\ell,p,\beta}\cdot n^{\bigoh(1)}$.

By \Cref{claim:imp_chip_exists}, for the values $\ell=\ell_0$ and
$p=|F_0|$, some arc $a\in \CC_{\ell,p}$ belongs to a solution of size at most
$h$. The algorithm branches over all arcs in
\[
\bigcup_{\ell,p:\,\ell+p\le h,\,\ell<h,\,p>0}\CC_{\ell,p},
\]
and for each such arc $a$, recursively solves the instance $(D-a,h-1)$.
If the recursive call accepts, then $D-a$ has an arc directed odd cycle
transversal of size at most $h-1$, and adding $a$ to such a transversal
gives a solution of size at most $h$ for $D$. Conversely, by
\Cref{claim:imp_chip_exists}, whenever a solution compatible with the
current ordered partition exists in the non-empty-separator case, at least
one branch deletes an arc $a$ contained in some solution $\widehat F$ of
size at most $h$. Then $\widehat F\setminus\{a\}$ is a solution for
$(D-a,h-1)$, so that branch accepts by the inner induction hypothesis.

The recursive call is valid. By \Cref{lem:localChanges}, applied to the
single-arc set $\{a\}$, the digraph $D-a$ is $(\beta+1)$-balanced. Also,
$(D-a)-W$ has no directed odd cycle, since it is a subdigraph of $D-W$.
Thus the recursive call on $(D-a,h-1)$ uses the same modulator $W$ and a
valid balance bound $\beta+1$, and its induction measure is lexicographically smaller.

The algorithm returns \yes if any branch, over all ordered partitions and
all arcs selected in Step 4, succeeds, and returns \no otherwise.

\medskip
\noindent
\textit{Running time analysis.}
We verify that the running time is bounded by a function of the original
$k$ and $\alpha$ times a polynomial in $n$.

The approximation step is used only once, before the recursive routine is
called, and runs in time $2^{\bigoh(k\log k)}\cdot n^{\bigoh(1)}$.
Every modulator encountered by the recursion is a subset of the initial
modulator, and therefore has size at most $4k$.

At any recursive node, the number of ordered partitions of the current
modulator is at most $k^{\bigoh(k)}$. For each ordered partition, the
empty-separator case tries at most $k+1$ budget splits. The call to
\Cref{lem:specialCaseDOCT} runs in time
$2^{k^{\bigoh(1)}}\cdot n^{\bigoh(1)}$, since the current budget is at most
$k$ and the current modulator has size at most $4k$.

It remains to bound the important-chip branching. For a current budget
$h\le k$, the number of pairs $(\ell,p)$ with $\ell+p\le h$, $\ell<h$, and $p>0$ is
at most $k^2$. For each such pair, \Cref{thm:boundImpChips} bounds the
number of important chips by a constant depending only on $\ell,p$ and the
current balance bound. Each such chip has at most $p\le k$ outgoing arcs.

We now bound the balance values that can occur when the recursive algorithm
is started from the original $\alpha$-balanced graph. Although the inner
induction is uniform in the balance bound, the running time of the chip
enumeration depends on that bound, so we need a bound depending only on
$k$ and $\alpha$.

An arc-deletion branch decreases the budget and increases the balance by
at most one. An empty-separator split does not increase the budget, and the only
recursive continuation replaces the current modulator $W$ by the proper
subset $W\setminus W_1$, and increases the balance by at most a factor of
two. Along any root-to-leaf path there are at most $k$ arc-deletion
branches. Between two arc-deletion branches there are at most $4k$
empty-separator continuations, since each such continuation strictly
decreases the modulator. Hence all graphs encountered have balance bounded
by a function $B_{k,\alpha}$ depending only on $k$ and $\alpha$.

For fixed $\ell,p,\beta$, let $c_{\pi_\ell,p,\beta}$ be the constant from
\Cref{thm:boundImpChips}, and set
\[
\widehat c_{k,\alpha}
=
\max_{\substack{\beta\le B_{k,\alpha}\\ \ell<k,\ p\le k}}
c_{\pi_\ell,p,\beta}.
\]
The important-chip branching factor is at most
$k^3\widehat c_{k,\alpha}$. The number of recursive levels is bounded by a
function of $k$, since $(h,|W|)$ decreases lexicographically. Thus the
number of nodes of this recursive routine is bounded in terms of
$k,\alpha$.

We must also account for the lower-budget decision algorithms used to test
$\pi_\ell$ inside enumeration. Apply \eqref{eq:chipEnumerationOracleBound}
to the finitely many parameter choices $\ell<k$, $p\le k$, and
$\beta\le B_{k,\alpha}$. Together with the preceding node bound, this gives
an integer $A=A_{k,\alpha}\ge1$, independent of $n$, such that one full
budget-$k$ decision call makes at most $A L(n)$ lower-budget property tests,
each on at most $n+A$ vertices and with balance at most $A$. We use $A$ as
the balance bound in these lower-budget tests. All other work, including
the approximation and special-case calls, takes at most
$A(n+1)^d$ time. Each property test has budget $\ell<k$, so
\begin{equation}\label{eq:doctRuntimeRecurrence}
T_k(n,\alpha)
\le A(n+1)^d+A L(n)\max_{0\le \ell<k}T_\ell (n+A,A).
\end{equation}
The outer induction hypothesis applies simultaneously at all balance
bounds, in particular at $A$. Since
$n+A+1\le(A+1)(n+1)$ and $L(n+A)\le(A+1)L(n)$, it gives
\[
\max_{0\le \ell<k}T_\ell (n+A,A)
\le
(A+1)^{d+k-1}(n+1)^dL(n)^{k-1}
\max_{0\le \ell<k}F_{\ell,A}.
\]
As $L(n)\ge1$, \eqref{eq:doctStrongRuntime} follows by taking
\[
F_{k,\alpha}
=A\left(1+(A+1)^{d+k-1}\max_{0\le \ell<k}F_{\ell,A}\right).
\]
The case $k=0$ uses the polynomial test already described. Finally, for
every $k$ there is a constant $C_k$ such that
$L(n)^k\le C_k(n+1)$ for all $n\ge0$. Hence the running time is at most
$F_{k,\alpha}C_k(n+1)^{d+1}$, with a polynomial exponent independent of
$k$ and $\alpha$. This proves the theorem.
\end{proof}

Since Eulerian digraphs are $0$-balanced, the following is an immediate corollary.
 
\begin{corollary}
{\sc Arc DOCT} on Eulerian digraphs is non-uniformly fixed-parameter tractable parameterized by the solution size.
\end{corollary}

\section{Uniformity via  weak representatives}
\label{sec:weakRep}
Recall that the enumeration algorithm of \Cref{thm:enumerateBoundImpChips} is
non-uniform. For every fixed $k,\alpha,\pi$ it uses a fixed set of CMSO representatives. Their existence follows from the finite-state property of CMSO, but the proof gives no procedure that computes such a set from $\pi$. 
For the applications in this paper, we do not need the full strength of
CMSO representatives. In the crossing reduction, a representative
is used only to replace the part
$Z\setminus X$ of a chip by a bounded boundaried graph $R$. To make the
proof go through, we need two implications. First, the reduced chip must
still satisfy $\pi$. Second, if a larger chip appears in the reduced
instance, then it must lift back to a larger chip in the original instance.
This motivates a  weaker notion that we call weak representatives.

\subsection{Weak representatives}

\begin{definition}[Boundary-clean boundaried digraph]
Let $D$ be a boundaried digraph with boundary $B$. We say that $D$ is
\emph{boundary-clean} if $D$ has no arc with both endpoints in $B$.
\end{definition}

\begin{definition}[Computable weak representatives]
\label{def:computableWeakRepresentatives}
Let $\pi$ be a hereditary digraph property. We say that $\pi$ has
\emph{computable weak representatives} if there is a computable function
$f_\pi$ and an algorithm which, given $b\in\mathbb N_0$, outputs in time
$f_\pi(b)$ a finite set $\mathcal R_{\pi,b}$ of $b$-boundaried digraphs
such that every graph in $\mathcal R_{\pi,b}$ has at most $f_\pi(b)$
vertices and at most $f_\pi(b)$ arcs.

Moreover, for every pair of compatible $b$-boundaried digraphs $D_1,D_2$
such that $D_1$ is boundary-clean and
$D_1\oplus D_2\models \pi$, 
there is a graph $R\in\mathcal R_{\pi,b}$, compatible with $D_1$ and
$D_2$, satisfying the following two one-sided conditions.
\begin{enumerate}
\item \emph{Completeness for the fixed context:}
\[
D_1\oplus R\models \pi.
\]
\item \emph{No false positives in boundary-clean contexts.} For every
boundary-clean $b$-boundaried digraph $D_1'$ compatible with both $R$ and
$D_2$, 
\[
\text{if }D_1'\oplus R\models \pi,
\quad
\text{ then }
\quad
D_1'\oplus D_2\models \pi .
\]
\end{enumerate}
We call such a graph $R$ a \emph{weak representative of $D_2$ relative to
$D_1$}.
\end{definition}

The boundary-clean requirement makes
arcs with both endpoints in the boundary belong to the replaced side only, so
that they are not counted twice in an arc-deletion budget after gluing.

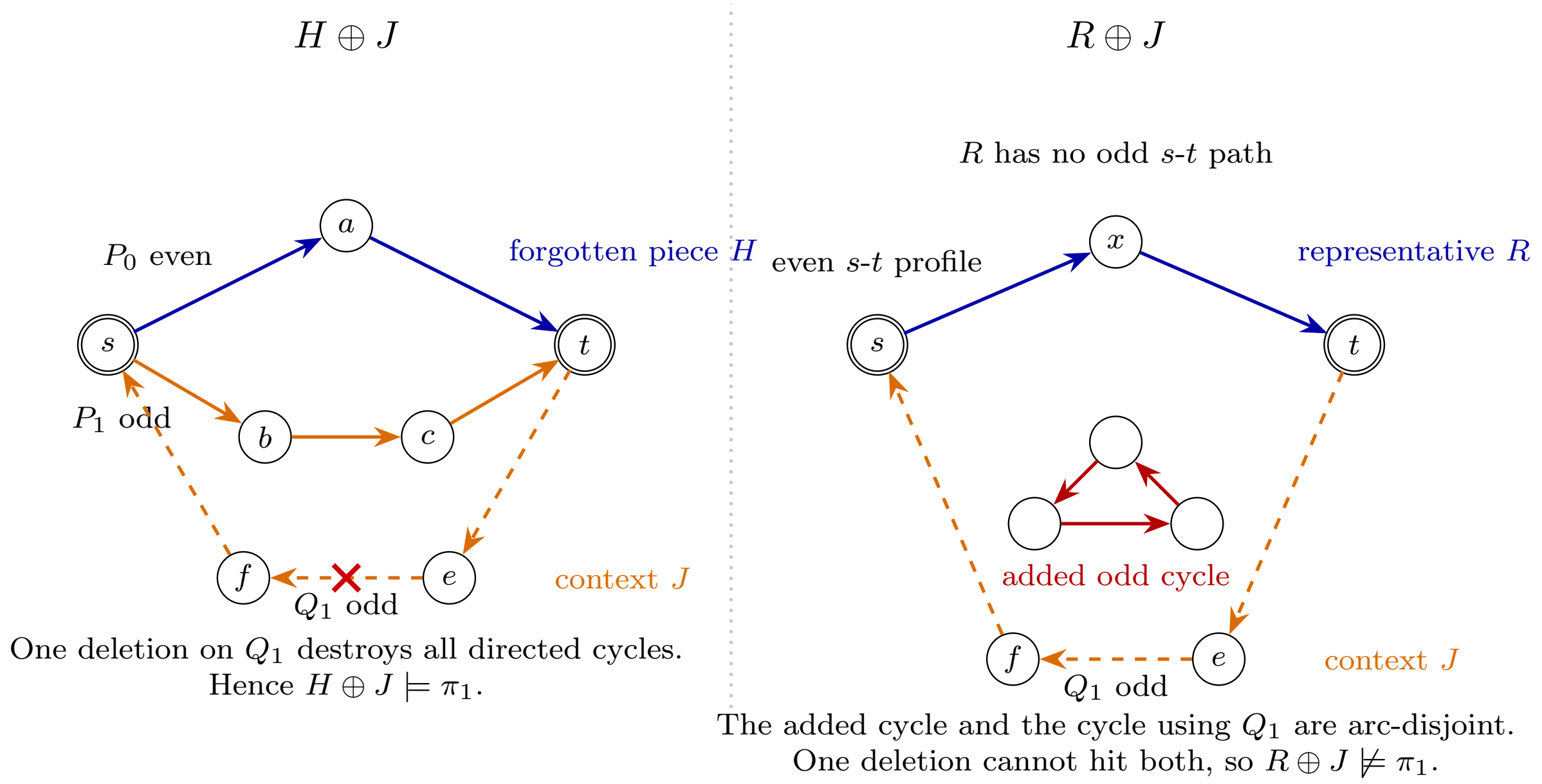
\begin{figure}[t]
\centering
\resizebox{\textwidth}{!}{%
\begin{tikzpicture}[
  >=Stealth,
  v/.style={
    circle,
    draw,
    fill=white,
    inner sep=1pt,
    minimum size=4.8mm,
    font=\scriptsize
  },
  bv/.style={
    circle,
    draw,
    double,
    fill=white,
    inner sep=1pt,
    minimum size=5.2mm,
    font=\scriptsize
  },
  even/.style={->, line width=0.9pt, draw=blue!65!black},
  odd/.style={->, line width=0.9pt, draw=orange!85!black},
  ret/.style={->, line width=0.9pt, draw=orange!85!black, dashed},
  marker/.style={->, line width=0.9pt, draw=red!70!black},
  xcut/.style={draw=red!80!black, line width=1.3pt},
  lab/.style={font=\scriptsize},
  title/.style={font=\small\bfseries},
  note/.style={font=\scriptsize, align=center}
]

\node[title] at (2.2,2.85) {$H\oplus J$};

\node[bv] (sL) at (0,0) {$s$};
\node[bv] (tL) at (4.4,0) {$t$};

\node[v] (aL) at (2.2,1.1) {$a$};
\node[v] (bL) at (1.45,-0.85) {$b$};
\node[v] (cL) at (2.95,-0.85) {$c$};

\node[v] (eL) at (3.15,-2.15) {$e$};
\node[v] (fL) at (1.25,-2.15) {$f$};

\draw[even] (sL) -- node[lab, above left] {$P_0$ even} (aL);
\draw[even] (aL) -- (tL);

\draw[odd] (sL) -- node[lab, below left] {$P_1$ odd} (bL);
\draw[odd] (bL) -- (cL);
\draw[odd] (cL) -- (tL);

\draw[ret] (tL) -- (eL);
\draw[ret] (eL) -- node[lab, below] {$Q_1$ odd} (fL);
\draw[ret] (fL) -- (sL);

\coordinate (xL) at ($(eL)!0.5!(fL)$);
\draw[xcut] ($(xL)+(-0.12,-0.12)$) -- ($(xL)+(0.12,0.12)$);
\draw[xcut] ($(xL)+(-0.12,0.12)$) -- ($(xL)+(0.12,-0.12)$);

\node[note] at (2.2,-3.0)
  {One deletion on $Q_1$ destroys all directed cycles.\\
   Hence $H\oplus J\models\pi_1$.};

\node[lab, blue!65!black] at (4.85,0.85) {forgotten piece $H$};
\node[lab, orange!85!black] at (4.75,-2.15) {context $J$};

\draw[gray!45, dotted, line width=0.8pt] (5.75,-3.35) -- (5.75,3.15);

\begin{scope}[xshift=7.1cm]
\node[title] at (2.2,2.85) {$R\oplus J$};

\node[bv] (sR) at (0,0) {$s$};
\node[bv] (tR) at (4.4,0) {$t$};

\node[v] (xR) at (2.2,0.95) {$x$};

\node[v] (m1) at (2.2,-0.9) {};
\node[v] (m2) at (1.45,-1.65) {};
\node[v] (m3) at (2.95,-1.65) {};

\node[v] (eR) at (3.15,-2.9) {$e$};
\node[v] (fR) at (1.25,-2.9) {$f$};

\draw[even] (sR) -- node[lab, above left] {even $s$-$t$ profile} (xR);
\draw[even] (xR) -- (tR);

\draw[marker] (m1) -- (m2);
\draw[marker] (m2) -- (m3);
\draw[marker] (m3) -- (m1);

\draw[ret] (tR) -- (eR);
\draw[ret] (eR) -- node[lab, below] {$Q_1$ odd} (fR);
\draw[ret] (fR) -- (sR);

\node[note] at (2.2,1.75)
  {$R$ has no odd $s$-$t$ path};
\node[note, red!70!black] at (2.2,-2.15)
  {added odd cycle};

\node[note] at (2.2,-3.7)
  {The added cycle and the cycle using $Q_1$ are arc-disjoint.\\
   One deletion cannot hit both, so $R\oplus J\not\models\pi_1$.};

\node[lab, blue!65!black] at (4.95,0.85) {representative $R$};
\node[lab, orange!85!black] at (4.75,-2.9) {context $J$};
\end{scope}

\end{tikzpicture}%
}
\caption{An example distinguishing weak representatives from CMSO
representatives. The full piece $H$ and the representative $R$ need
not be CMSO-equivalent for $\pi_1$. A context $J$ may
distinguish them in a direction allowed by weak representativeness.
Here $H\oplus J$ satisfies $\pi_1$, while $R\oplus J$ does not. Such a context
$J$ is not the fixed relative context with respect to which $R$ is chosen as
a weak representative. Here $Q_1$ is an odd path in $J$.}
\label{fig:weak-not-cmso}
\end{figure}

\begin{remark}[Weak representatives versus CMSO representatives]
\label{rem:weakVsMyhillNerode}
{\em A CMSO representative preserves the truth of $\pi$ in every compatible
context and in both directions. A weak representative is relative and
one-sided. 

For \arcDoct{}, this one-sidedness is useful. Once the arc deletions inside
the replaced part $D_2$ are fixed, the representative only needs to
remember which walk parities between boundary vertices remain and how many
arcs were deleted inside $D_2$. It need not be CMSO-equivalent to $D_2$.
\Cref{fig:weak-not-cmso} illustrates this. An arbitrary fixed context may
distinguish the original piece $H$ from a representative $R$ in the
direction $H\oplus J\models\pi_1$ but $R\oplus J\not\models\pi_1$. This is
compatible with weak representativeness because the first condition concerns
only the context $D_1$ relative to which $R$ was chosen, and $J$ cannot be
$D_1$ since $D_1\oplus R\models\pi_1$.
}
\end{remark}

The next lemma shows that the two conditions in
\Cref{def:computableWeakRepresentatives} suffice for the crossing reduction.

\begin{lemma}[Crossing reduction with weak representatives]
\label{lem:weakRepresentativeGluing}
Let $\pi$ be a hereditary property with computable weak representatives.
Let $(\widehat D,S,T_1,T_2,Q)$ be an $\eta$-guarded routed instance.
Let $X$ be a compact auxiliary-respecting
$S$-$T_1$ reachability set, and let $Z$ be an important
$(S,T_1,T_2,Q,\pi,k)$-chip in $\widehat D$ that crosses $X$. Suppose that
$\partial_{\mathrm{real}}^+(Z)\cap \partial_{\mathrm{real}}(X)=\emptyset$. Let 
$B\coloneqq B_X(Z)$ 
 and 
$A_B\coloneqq
\{(u,v)\in A_{\mathrm{real}}(\widehat D)\mid u,v\in B\}$.

Fix a bijective labeling of $B$ by $[|B|]$, and view
\[
D_1\coloneqq \widehat D_{\mathrm{real}}[L_X(Z)]-A_B
\qquad\text{and}\qquad
D_2\coloneqq \widehat D_{\mathrm{real}}[Z\setminus X]
\]
as compatible $|B|$-boundaried digraphs with boundary $B$. Let
$R\in\mathcal R_{\pi,|B|}$ be a weak representative of $D_2$ relative to
$D_1$.
Construct the reduced routed instance exactly as in
\Cref{def:crossingReducedGraph}, except that this graph $R$ is used in
place of the CMSO representative of $D_2$. 

Then the reduced chip $C_X(Z)=(X\cap Z)\cup V(R)$ is an important
$(S_X(Z),\allowbreak T_X^1(Z),\allowbreak T_X^2(Z),\allowbreak Q_X(Z),\allowbreak \pi,\allowbreak k-1)$-chip in the reduced routed instance. Moreover,
$\emptyset\neq
\partial_{\mathrm{real}}^+(C_X(Z))
\subsetneq
\partial_{\mathrm{real}}^+(Z)$.
\end{lemma}
\begin{proof}
In \Cref{lem:representativeGluing}, the representative is used only to
check $\pi$ for the reduced chip and for a lifted dominator. All path,
boundary, and forbidden-set arguments are unchanged.

For the reduced chip, the attachment decomposition gives
$\widehat D_{\mathrm{real}}[Z]=D_1\oplus D_2$ and
$\widehat D_X(Z)_{\mathrm{real}}[C_X(Z)]=D_1\oplus R$.
The first weak-representative condition therefore gives $\pi$.

For lifting, suppose $C'\supset C_X(Z)$ is a reduced chip with no larger
boundary, and put $W=C'\setminus C_X(Z)$ and $Z'=Z\cup W$.
As in \Cref{claim:lifting-pi-condition}, the forbidden contacts ensure
that the attachment boundary is still $B$. The context
\[
D_1'=\widehat D_{\mathrm{real}}[(Z'\cap X)\cup B]-A_B
\]
is boundary-clean, and the reduced graph on $C'$ is $D_1'\oplus R$.
The second weak-representative condition gives
$D_1'\oplus D_2=\widehat D_{\mathrm{real}}[Z']\models\pi$.
Here the boundary-internal arcs removed from each context are supplied by
the replaced side. The remaining claims of \Cref{lem:crossingLifting}
make $Z'$ an original dominator, a contradiction. The boundary drop remains
\Cref{obs:survivingRealBoundary}.
\end{proof}
The preceding lemma allows us to make the enumeration theorem uniform for
properties admitting computable weak representatives.

\thmenumerateChipsWeakRepresentatives*


\begin{proof}
The proof follows the enumeration proof of
\Cref{thm:enumerateBoundImpChips}. We indicate the only changes needed to
make the algorithm uniform.
The construction of the minimum cut sequence, the nested reductions, the
deletion step, and the verification of all chip conditions except the
condition $\widehat D_{\mathrm{real}}[Z]\models\pi$ are algorithmic. The
latter condition is checked using the assumed uniform test for $\pi$ on
graphs of bounded balance. This is the only point where efficient
testability parameterized by the balance is needed.
In the crossing case, the non-uniform algorithm enumerates the fixed set of
CMSO representatives. We replace this step as follows. Whenever
the crossing reduction is applied to a compact set $X$, the attachment
boundary has size bounded by a function of $k,\alpha$, and the current
auxiliary bound. For each possible boundary $B$ of this bounded size, the
algorithm computes the finite set $\mathcal R_{\pi,|B|}$ supplied by
\Cref{def:computableWeakRepresentatives} and branches over all
$R\in\mathcal R_{\pi,|B|}$. Here each guessed set $B$ is labeled bijectively
by $[|B|]$, as in \Cref{lem:weakRepresentativeGluing}, in place of the
labeling by $I_B\subseteq[b]$ used in the proof of \Cref{lem:preinductionStep}.
Let $Z$ be a crossing chip and let $D_1,D_2$ be as in \Cref{lem:weakRepresentativeGluing}. By \Cref{def:computableWeakRepresentatives}, some $R\in\mathcal R_{\pi,|B|}$ is a weak representative of $D_2$ relative to $D_1$, and by \Cref{lem:weakRepresentativeGluing} the branch using this $R$ gives the same conclusion for $Z$ as the CMSO representative did. The rest of the proof of
\Cref{thm:enumerateBoundImpChips} is unchanged. The number of reduced instances remains bounded by a
computable function of the parameters, because the number and size of the
graphs in each set $\mathcal R_{\pi,b}$ are bounded by the computable
function $f_\pi(b)$.
Thus every branching step of the same recursive enumeration algorithm
is computable. Its recursion depth is bounded as before, and
the balance, auxiliary width, representative sizes, and number of nodes
are bounded by computable functions of the initial parameters. In
particular, \eqref{eq:chipEnumerationOracleBound} holds with a computable
bound $A$. Combining it with the assumed uniform test for $\pi$ gives the
running time $g_\pi(k,\alpha)\cdot n^{\bigoh(1)}$ for a computable function
$g_\pi$, and the algorithm outputs exactly the important
$(S,T,\pi,k)$-chips.
\end{proof}

It remains to compute weak representatives for the property needed in the
application. For \arcDoct{}, it suffices to remember which parities of walks connect each
ordered pair of boundary vertices after a bounded number of arc deletions
inside the replaced part. The next observations make this precise.

\begin{observation}[Parity-profile replacement for \textsc{Arc DOCT}]
\label{obs:doctParityProfileReplacement}
Let $G$ and $R$ be boundaried digraphs with the same boundary $B$, and let
$r\in\mathbb N_0$. For an arc set $F$, define
\[
  \Pi(G-F)
  =
  \{(x,y,\epsilon)\in B\times B\times \{0,1\}
    \mid
    \text{$G-F$ contains an $x$-$y$ walk of length $\epsilon$ modulo $2$}
  \}.
\]
Suppose that for every arc set $F_R\subseteq A(R)$ with $|F_R|\le r$ such
that $R-F_R$ has no directed odd cycle, there is an arc set
$F_G\subseteq A(G)$ with $|F_G|\le |F_R|$ such that $G-F_G$ has no directed
odd cycle and
\[
  \Pi(G-F_G)\subseteq \Pi(R-F_R).
\]
Then, for every boundary-clean boundaried digraph $H$ compatible with $G$ and $R$, if
$H\oplus R$ has an arc directed odd cycle transversal of size at most $r$,
then $H\oplus G$ has an arc directed odd cycle transversal of size at most
$r$.
\end{observation}

\begin{proof}
Let $Y$ be an arc directed odd cycle transversal of $H\oplus R$ with
$|Y|\le r$. Write
\[
  Y_H=Y\cap A(H)
  \qquad\text{and}\qquad
  Y_R=Y\cap A(R).
\]
Since $H$ is boundary-clean, the arc sets contributed by $H$ and by $R$
are disjoint in the gluing. Hence
$|Y_H|+|Y_R|=|Y|$.

Every directed odd cycle of $R-Y_R$ is a directed odd cycle of $H\oplus R-Y$, so $R-Y_R$ has no directed odd cycle. Since also $|Y_R|\le r$, by assumption there is an arc set $Y_G\subseteq A(G)$ with
$|Y_G|\le |Y_R|$ such that $G-Y_G$ has no directed odd cycle and
\[
  \Pi(G-Y_G)\subseteq \Pi(R-Y_R).
\]
We claim that
$Y' = Y_H\cup Y_G$ 
is an arc directed odd cycle transversal of $H\oplus G$. Moreover
$|Y'|\le |Y_H|+|Y_G|\le |Y_H|+|Y_R|=|Y|$.

Suppose for a contradiction that $H\oplus G-Y'$ contains a directed odd
cycle $C$. Since $G-Y_G$ has no directed odd cycle, the cycle $C$ is not
contained entirely in $G$. Decompose $C$ into maximal subpaths whose internal
vertices lie outside the boundary and whose endpoints lie in $B$, together
with the remaining parts contained in $H-Y_H$. Every subpath that lies in
$G-Y_G$ has endpoints $x,y\in B$ and some parity $\epsilon$. Since
  $\Pi(G-Y_G)\subseteq \Pi(R-Y_R)$,
there is an $x$-$y$ walk of the same parity in $R-Y_R$. Replacing all
$G$-subpaths of $C$ by such walks gives a directed closed walk of odd length
in $H\oplus R-Y$. Every directed closed walk of odd length contains a
directed odd cycle, contradicting that $Y$ is an arc directed odd cycle
transversal of $H\oplus R$. Hence $H\oplus G-Y'$ has no directed odd cycle,
as required.
\end{proof}

\begin{definition}[Parity profiles]
Let $B$ be a finite boundary set. A \emph{parity profile} on $B$ is a set $\mathcal R\subseteq B\times B\times \{0,1\}$. We call it \emph{reflexive} if $(b,b,0)\in\mathcal R$ for every $b\in B$. Walks of length zero are allowed, so every profile $\Pi(G-F)$ is reflexive. We say that $\mathcal R$ is \emph{closed under parity composition} if, whenever $(x,y,p)\in \mathcal R$ and $(y,z,q)\in \mathcal R$, we also have $(x,z,p+q \bmod 2)\in \mathcal R$.

We say that $\mathcal R$ contains an \emph{odd closed boundary walk} if there are triples \[(b_0,b_1,p_1),\ldots,(b_{m-1},b_m,p_m)\in \mathcal R\] with $b_m=b_0$ and $p_1+\cdots+p_m\equiv 1 \pmod 2$.
\end{definition}

\begin{observation}[Realizing a residual parity profile]
\label{obs:realizeResidualParityProfile}
Let $B$ be a boundary set, let $\ell,r\in\mathbb N_0$, and let
$\mathcal R\subseteq B\times B\times\{0,1\}$ be reflexive, closed under parity
composition, and contain no odd closed boundary walk. Then one can construct
a boundaried digraph $R_{\mathcal R,r}$ with boundary $B$, of size bounded
by a function of $|B|,\ell$, and $r$, such that the following hold.
\begin{enumerate}
\item $R_{\mathcal R,r}$ contains $r$ pairwise arc-disjoint directed odd
cycles whose vertices are disjoint from $B$.
\item After deleting one arc from each of these $r$ odd cycles, the remaining
graph has no directed odd cycle.
\item For every $(b_1,b_2,p)\in\mathcal R$ with $b_1\ne b_2$, the graph $R_{\mathcal R,r}$
contains $\ell+1$ pairwise arc-disjoint directed $b_1$-$b_2$ paths of
parity $p$ whose internal vertices are outside $B$. The triples $(b,b,0)$ are realized by length-zero walks and require no added arcs.
\item For every $b_1,b_2\in B$ and $p\in\{0,1\}$, if the part of
$R_{\mathcal R,r}$ obtained by deleting the $r$ added odd cycles contains a
directed $b_1$-$b_2$ walk of parity $p$, then
$(b_1,b_2,p)\in\mathcal R$.
\end{enumerate}
\end{observation}

\begin{proof}
First construct the profile-realizing part. For every triple
$(b_1,b_2,p)\in\mathcal R$ with $b_1\ne b_2$, add $\ell+1$ directed
$b_1$-$b_2$ paths of length two when $p=0$ and length three when $p=1$,
using fresh internal vertices and fresh arcs for every path and every triple.
Thus these paths are pairwise arc-disjoint across all
copies and all triples, and they share vertices only at their boundary endpoints.

Every positive-length boundary-to-boundary directed walk in the profile-realizing part decomposes into a
concatenation of the boundary-to-boundary paths added for triples of
$\mathcal R$. Hence such a walk induces a boundary walk in $\mathcal R$.
Since $\mathcal R$ is closed under parity composition, every
boundary-to-boundary walk in the profile-realizing part realizes a triple
already in $\mathcal R$. Length-zero walks give the reflexive triples. Since $\mathcal R$ contains no odd closed boundary
walk, the profile-realizing part has no directed odd cycle.

Now add $r$ pairwise arc-disjoint directed odd cycles, all using fresh
vertices and fresh arcs, and with no vertices in common with the
profile-realizing part. This gives the graph $R_{\mathcal R,r}$.

Deleting one arc from each of the $r$ added odd cycles leaves exactly the
profile-realizing part together with directed paths, and hence leaves an
odd-cycle-free graph. The third item follows because
there are $\ell+1$ arc-disjoint copies of each required positive-length boundary path. The fourth item follows from the same closure argument for
boundary-to-boundary walks in the profile-realizing part. 
\end{proof}

\begin{lemma}[Weak representatives for \textsc{Arc DOCT}]
\label{lem:arcDoctWeakRepresentatives}
For every $\ell\in\mathbb{N}_0$, let $\pi_\ell$ be the property that a
digraph has an arc directed odd cycle transversal of size at most $\ell$.
Then $\pi_\ell$ has computable weak representatives. Moreover, the
representative sets are computable uniformly in $\ell$ and in the boundary
size.

\end{lemma}

\begin{proof}
We define $\mathcal R_{\pi_\ell,b}$ by taking the union over every
label set $I\subseteq[b]$. For each such $I$, let $B$ be a boundary with
label set $I$. For every $r\in[\ell]_0$, and for every parity profile $\mathcal R\subseteq B\times B\times\{0,1\}$
that is reflexive, closed under parity composition, and contains no odd closed boundary
walk, include the graph $R_{\mathcal R,r}$ given by
\Cref{obs:realizeResidualParityProfile} for the current value of $\ell$. There are only finitely many such
profiles, and each graph has size bounded by a computable function of
$b$ and $\ell$. Hence the family is computable uniformly in $b$ and $\ell$.

We verify the two conditions from
\Cref{def:computableWeakRepresentatives}. Let $D_1,D_2$ be compatible
$b$-boundaried digraphs such that $D_1$ is boundary-clean and $D_1\oplus D_2\models \pi_\ell$.
Write $B$ for their common boundary after gluing, with its inherited labels.
Let $F$ be an arc directed odd cycle transversal of $D_1\oplus D_2$ of
size at most $\ell$. Let 
\[
F_1=F\cap A(D_1)
\qquad\text{and}\qquad
F_2=F\cap A(D_2),
\]
and let $r=|F_2|$. Since $D_1$ is boundary-clean, the arc sets of $D_1$ and
$D_2$ are disjoint in the gluing. Hence
\[
|F_1|+|F_2|=|F|\le \ell .
\]

 Since $F$ hits every directed odd cycle in
$D_1\oplus D_2$, the graph $D_2-F_2$ has no directed odd cycle. Let
\[
\mathcal R=\Pi(D_2-F_2).
\]
The profile is reflexive by length-zero walks and closed under parity
composition by concatenation. An odd closed boundary walk would expand to
an odd closed walk, hence an odd cycle, in $D_2-F_2$. Thus the profile has
all properties required by \Cref{obs:realizeResidualParityProfile}.

Therefore $R=R_{\mathcal R,r}$ belongs to $\mathcal R_{\pi_\ell,b}$. We first show that $D_1\oplus R\models \pi_\ell$. Delete the arcs of $F_1$ from $D_1$, and delete one arc from each of the $r$ added odd cycles in $R$. The total number of deleted arcs is $|F_1|+r=|F_1|+|F_2|\le \ell$.

Suppose the remaining graph contains a directed odd cycle. This cycle cannot
be contained entirely in the profile-realizing part of $R$, by
\Cref{obs:realizeResidualParityProfile}, and it cannot be contained in one
of the added odd cycles, since each of those has been hit. Decompose the
cycle into maximal subpaths that lie in the profile-realizing part of $R$
and have endpoints in $B$, together with the remaining parts lying in
$D_1-F_1$. By \Cref{obs:realizeResidualParityProfile}, each such
$R$-subpath corresponds to a triple of $\mathcal R=\Pi(D_2-F_2)$, and hence
can be replaced by a walk of the same parity in $D_2-F_2$. This gives a
directed closed walk of odd length in
$D_1\oplus D_2-F$,
and therefore a directed odd cycle, contradicting the choice of $F$. Hence
$D_1\oplus R\models \pi_\ell$.

For lifting, apply \Cref{obs:doctParityProfileReplacement} with $G=D_2$
and budget $\ell$. Let $Y_R\subseteq A(R)$ have size at most $\ell$ and
make $R-Y_R$ odd-cycle-free. The $r$ added odd cycles are pairwise arc-disjoint, so they force
$|Y_R|\ge r=|F_2|$. Every triple $(b_1,b_2,p)\in\mathcal R=\Pi(D_2-F_2)$ with $b_1\neq b_2$ has
$\ell+1$ arc-disjoint realizing paths in $R$, so at least one of them avoids
$Y_R$. Every triple $(b,b,p)\in\mathcal R$ has $p=0$, since $\mathcal R$
contains no odd closed boundary walk, and is realized by a length-zero walk.
Consequently
\[
\Pi(D_2-F_2)\subseteq\Pi(R-Y_R).
\]
Hence choosing $F_G=F_2$ verifies the hypotheses of \Cref{obs:doctParityProfileReplacement}. It follows
that every boundary-clean context $D_1'$ with
$D_1'\oplus R\models\pi_\ell$ also satisfies
$D_1'\oplus D_2\models\pi_\ell$.

Thus $R$ is a weak representative of $D_2$ relative to $D_1$. This proves
that $\pi_\ell$ has computable weak representatives, uniformly in $\ell$
and in the boundary size.
\end{proof}

\subsection{Uniform FPT algorithm for {\arcDoct} on balanced digraphs}

Computable weak representatives make the algorithm for \arcDoct{} uniform
on $\alpha$-balanced digraphs, with $\alpha$ treated as a parameter. The Eulerian case is obtained by setting $\alpha=0$. Recall that
the non-uniformity in \Cref{thm:generalBalancedDOCT} is only due to the
enumeration of important chips for the properties
\[
\pi_\ell:
\quad
\text{at most $\ell$ arcs can be deleted to remove all directed odd cycles}.
\]
For fixed $\ell$, the sentence $\pi_\ell$ is hereditary. By
\Cref{lem:arcDoctWeakRepresentatives}, these properties have computable weak
representatives uniformly in $\ell$ and in the boundary size. The remaining
issue is uniform testability of $\pi_\ell$ on balanced digraphs. We obtain
this testability by induction on the arc-deletion budget. When the current
budget is $k$, the proof of \Cref{thm:generalBalancedDOCT} only enumerates
chips for properties $\pi_\ell$ with $\ell<k$.

We prove the following stronger statement.

\thmuniformBalancedArcDoct*

\begin{proof}[Proof sketch]
We repeat the budget induction from \Cref{thm:generalBalancedDOCT},
including the strengthened running-time bound
\eqref{eq:doctStrongRuntime}. The base case is the same polynomial test.
In a call with current budget $h\le k$, only properties $\pi_\ell$ with
$\ell<h$ are tested, so their uniform tests are available by induction.

By \Cref{lem:arcDoctWeakRepresentatives} and
\Cref{thm:enumerateChipsWeakRepresentatives}, the enumeration uses
computable weak representatives and retains
\eqref{eq:chipEnumerationOracleBound} with computable parameter bounds.
Thus the constant $A_{k,\alpha}$ in \eqref{eq:doctRuntimeRecurrence} can be
chosen computably. The displayed recurrence for $F_{k,\alpha}$ then defines
computable bounds, since every term on its right has smaller budget, even
though its balance bound may increase. Consequently the running time is
$F_{k,\alpha}(n+1)^dL(n)^k$, and absorbing the logarithmic factor as before
gives $f(k,\alpha)\cdot n^{\bigoh(1)}$ for a computable function $f$, with a
polynomial exponent independent of $k$ and $\alpha$.
\end{proof}

Since Eulerian digraphs are $0$-balanced, we obtain the desired uniform
Eulerian statement.

\begin{corollary}[Uniform FPT algorithm for Eulerian \arcDoct]
\label{cor:uniformEulerianArcDoct}
\arcDoct{} on Eulerian digraphs is fixed-parameter tractable parameterized by the solution size.
\end{corollary}

\section{Additional consequences: undirected edge and vertex chips}
\label{sec:undirectedPort}

In this section we prove undirected versions of the important-chip results. For edge cuts, the result follows formally by replacing each
undirected edge by two opposite arcs and applying the directed theorem to the
resulting Eulerian digraph. For vertex separators, we retain the counting
scheme from \Cref{sec:genImpCuts} and supply the charging and lifting
arguments needed to preserve vertex-separator witnesses.

\subsection{Edge cuts in undirected graphs}
\label{subsec:undirectedEdgeCuts}

Throughout this subsection, undirected graphs are simple and loopless. Let
$G$ be such a graph. For $Z\subseteq V(G)$, let
$\partial_G(Z)
\coloneqq
\{\{u,v\}\in E(G)\mid u\in Z,\ v\notin Z\}
$
be the edge boundary of $Z$.

\begin{definition}[Undirected edge reachability sets]
\label{def:undirectedEdgeReachability}
Let $G$ be a graph and let $S,T\subseteq V(G)$ be disjoint non-empty vertex
sets. A set $Z\supseteq S$ is an \emph{$S$-$T$ reachability set} if
$Z\cap T=\emptyset$ and every vertex of $Z$ is reachable from $S$ in
$G[Z]$. We say that $Z$ is \emph{strict} if $\partial_G(Z)$ is an
inclusion-wise minimal $S$-$T$ edge cut in $G$.
\end{definition}

\begin{definition}[Undirected edge chips]
\label{def:undirectedEdgeChips}
Let $\pi$ be a CMSO sentence over undirected graphs. An
\emph{$(S,T,\pi,k)$-edge chip} in $G$ is a vertex set $Z$ such that
\begin{enumerate}
\item $Z$ is a strict $S$-$T$ reachability set in $G$.
\item $|\partial_G(Z)|\le k$.
\item $G[Z]\models \pi$.
\end{enumerate}
\end{definition}

\begin{definition}[Important undirected edge chips]
\label{def:undirectedImportantEdgeChips}
An $(S,T,\pi,k)$-edge chip $Z$ in $G$ is \emph{important} if there is no
$(S,T,\pi,k)$-edge chip $Z'$ in $G$ such that $Z'\supset Z$ and
$|\partial_G(Z')|\le |\partial_G(Z)|$.
\end{definition}

We now reduce the edge-cut version to the directed theorem.

\begin{definition}[Bidirected graph]
\label{def:bidirectedGraph}
For an undirected graph $G$, let $D_G$ be the digraph obtained from $G$ by
replacing every edge $\{u,v\}$ by the two arcs $(u,v)$ and $(v,u)$, while
keeping the same vertex set.
\end{definition}

The digraph $D_G$ is Eulerian. The next observation shows that bidirection preserves strict reachability sets and boundary size.

\begin{observation}
\label{obs:strictUndirectedDirected}
Let $G$ be an undirected graph, let $S,T\subseteq V(G)$ be disjoint, and let
$Z\subseteq V(G)$. Then $Z$ is a strict $S$-$T$ reachability set in $G$ if
and only if $Z$ is a strict $S$-$T$ reachability set in $D_G$. Moreover,
$|\partial_G(Z)|=
|\partial^+_{D_G}(Z)|$.
\end{observation}

\begin{proof}
Paths in $G$ and directed paths in $D_G$ correspond by orienting each edge
along the traversal, so reachability is preserved. The map sending a
boundary edge $\{u,v\}$, with $u\in Z$, to $(u,v)$ is a bijection onto
$\partial^+_{D_G}(Z)$ and preserves boundary size.

For a reachability set $Z$, its edge boundary is minimal exactly when
every outside boundary endpoint $v$ reaches $T$ in $G-Z$. Necessity follows
from a minimal-cut witness using just the edge $\{u,v\}$ of the boundary. After crossing
it, the witness cannot return to $Z$ without using another boundary edge. For sufficiency, concatenate an
$S$-$u$ path inside $Z$, that edge, and a $v$-$T$ path outside $Z$.
This is the undirected version of \Cref{lem:basicStrictFacts}. Since the
outside paths also correspond under bidirection, strictness is preserved.
\end{proof}

It remains to translate the logical condition. We use the standard
interpretation of undirected graphs inside digraphs by forgetting arc
orientations. Let $U(D)$ denote the simple undirected graph with vertex set $V(D)$ and
with an edge $\{u,v\}$ whenever $D$ contains at least one of the arcs
$(u,v)$ and $(v,u)$.

\begin{lemma}[Forgetting orientations in CMSO]
\label{lem:forgetOrientationsCMSO}
For every CMSO sentence $\pi$ over undirected graphs, there is a CMSO
sentence $\overrightarrow{\pi}$ over digraphs such that, for every digraph
$D$,
$D\models \overrightarrow{\pi}$
if and only if 
$U(D)\models \pi$.
Moreover, if $\pi$ is hereditary on undirected graphs, then
$\overrightarrow{\pi}$ is hereditary on digraphs.
\end{lemma}

\begin{proof}
In a digraph $D$, an arc set $M$ is an
orientation selector if it contains exactly one arc from each unordered pair
of adjacent vertices of $U(D)$. This is CMSO-definable. Namely, two arcs are
declared equivalent if they have the same two endpoints, possibly in opposite
orders. The formula for an orientation selector says that for every arc $a$ there
is exactly one arc $m\in M$ equivalent to $a$.

Given such a selector $M$, we translate an undirected CMSO formula to a
directed CMSO formula relative to $M$. Vertex variables are unchanged. Edge
variables become arc variables restricted to $M$. Edge-set variables become
arc-set variables restricted to $M$. 

The incidence predicate
$\mathbf{inc}(e,v)$ is replaced by
$\mathbf{head}(e,v)\vee \mathbf{tail}(e,v)$.

Membership and equality for edge variables are interpreted as membership and
equality of the selected arcs. Cardinality predicates for edge sets are
preserved because each undirected edge is represented by exactly one selected
arc.

For a sentence $\pi$, let $\overrightarrow{\pi}$ assert that there exists an
orientation selector $M$ such that the translation of $\pi$ relative to $M$
holds. A selector always exists, and for every selector $M$ the vertices of $D$ together with $M$ and the translated incidence predicate form a copy of $U(D)$. Hence $D\models\overrightarrow{\pi}$ exactly when $U(D)\models\pi$.

If $\pi$ is hereditary and $D'=D[X]$ is an induced subdigraph of $D$ with
$D\models\overrightarrow{\pi}$, then
\[
U(D')=U(D)[X].
\]
Hence $U(D')\models\pi$, and therefore
$D'\models\overrightarrow{\pi}$.
\end{proof}

\begin{corollary}
\label{cor:undirectedDirectedInduced}
Let $G$ be an undirected graph, let $\pi$ be a CMSO sentence over undirected
graphs, and let $Z\subseteq V(G)$. Then
\[
G[Z]\models \pi
\qquad\text{if and only if}\qquad
D_G[Z]\models \overrightarrow{\pi}.
\]
\end{corollary}

\begin{proof}
Since $G$ is simple, $U(D_G[Z])=G[Z]$, so the claim follows from \Cref{lem:forgetOrientationsCMSO}.
\end{proof}

\begin{lemma}[Edge chips are preserved by bidirection]
\label{lem:undirectedEdgeChipCorrespondence}
Let $G$ be an undirected graph, let $S,T\subseteq V(G)$ be disjoint
non-empty vertex sets, let $k\in\mathbb{N}_0$, and let $\pi$ be a hereditary
CMSO sentence over undirected graphs. Then a set $Z\subseteq V(G)$ is an
important $(S,T,\pi,k)$-edge chip in $G$ if and only if it is an important
$(S,T,\overrightarrow{\pi},k)$-chip in $D_G$.
\end{lemma}

\begin{proof}
By \Cref{obs:strictUndirectedDirected} and \Cref{cor:undirectedDirectedInduced},
bidirection preserves strict reachability, boundary size, and the logical
condition. Thus it preserves chips and their domination relation, and
hence importance.
\end{proof}

\corundirectedEdgeChipBound*

\begin{proof}
Let $\overrightarrow{\pi}$ be the directed CMSO sentence from
\Cref{lem:forgetOrientationsCMSO}. Since $D_G$ is Eulerian,
\Cref{thm:boundImpChips} applied with $\alpha=0$ gives a constant depending
only on $k$ and $\overrightarrow{\pi}$ that bounds the number of important
$(S,T,\overrightarrow{\pi},k)$-chips in $D_G$. By
\Cref{lem:undirectedEdgeChipCorrespondence}, the important
$(S,T,\pi,k)$-edge chips in $G$ are exactly these chips. This gives the
required constant.
\end{proof}


\begin{restatable}{corollary}{corundirectedEdgeChipEnumerationintro}
\label{cor:undirectedEdgeChipEnumerationintro}
Let $k\in\mathbb{N}_0$, and let $\pi$ be a hereditary CMSO sentence over
undirected graphs such that, for fixed $\pi$, deciding $H\models\pi$ on an
undirected graph $H$ takes polynomial time. Then all important
$(S,T,\pi,k)$-edge chips in an undirected graph $G$ can be enumerated in time
$c^e_{\pi,k}\cdot n^{\bigoh(1)}$.

\end{restatable}

\begin{proof}
By \Cref{lem:forgetOrientationsCMSO}, $D\models\overrightarrow{\pi}$ if and
only if $U(D)\models\pi$, and $U(D)$ can be computed in linear time. Hence
$\overrightarrow{\pi}$ is efficiently testable parameterized by the balance.
Construct $D_G$. Enumerate all important
$(S,T,\overrightarrow{\pi},k)$-chips in $D_G$ using
\Cref{thm:enumerateBoundImpChips} with $\alpha=0$. By
\Cref{lem:undirectedEdgeChipCorrespondence}, these are exactly the important
$(S,T,\pi,k)$-edge chips of $G$.
\end{proof}

\subsection{Vertex separators in undirected graphs}
\label{subsec:undirectedVertexSeparators}

We now prove the vertex-separator version using the counting scheme of
\Cref{thm:boundImpChips}. Here the open neighborhood of a set, defined below, plays the role of the set of outgoing arcs. To preserve strictness, we keep a path requirement for each separator vertex that has already been counted against the budget.

For a vertex set $Z\subseteq V(G)$, let
\[
N_G(Z)\coloneqq
\{v\in V(G)\setminus Z\mid \text{there is } u\in Z
\text{ with } \{u,v\}\in E(G)\}
\]
be the open neighborhood of $Z$. We write $\lambda_G^v(S,T)$ for the
minimum size of an $S$-$T$ vertex separator disjoint from $S\cup T$, with
the convention that $\lambda_G^v(S,T)=\infty$ if no such separator exists.

\begin{definition}[Undirected vertex reachability sets]
\label{def:undirectedVertexReachability}
Let $G$ be an undirected graph, let $S,T\subseteq V(G)$ be disjoint
non-empty vertex sets, and let $Z$ be an $S$-$T$ reachability set as in \Cref{def:undirectedEdgeReachability}. We say that $Z$ is \emph{vertex-strict} if $N_G(Z)$ is an
inclusion-wise minimal $S$-$T$ vertex separator disjoint from $S\cup T$.
We say that $Z$ is \emph{compact} if $N_G(Z)\cap T=\emptyset$ and
$|N_G(Z)|=\lambda_G^v(S,T)$.
\end{definition}

\begin{definition}[Undirected vertex chips]
\label{def:undirectedVertexChips}
Let $\pi$ be a CMSO sentence over undirected graphs. An
\emph{$(S,T,\pi,k)$-vertex chip} in $G$ is a vertex set $Z$ such that
\begin{enumerate}
\item $Z$ is a vertex-strict $S$-$T$ reachability set in $G$.
\item $|N_G(Z)|\le k$.
\item $G[Z]\models \pi$.
\end{enumerate}
\end{definition}

\begin{definition}[Important undirected vertex chips]
\label{def:undirectedImportantVertexChips}
An $(S,T,\pi,k)$-vertex chip $Z$ in $G$ is \emph{important} if there is no
$(S,T,\pi,k)$-vertex chip $Z'$ in $G$ such that $Z'\supset Z$ and
$|N_G(Z')|\le |N_G(Z)|$.
\end{definition}

The vertex proof uses the same induction on the boundary budget and the
minimum separator value as the directed proof. It needs two additions. First, each vertex already counted against the budget carries a path requirement that keeps its strictness witness. Second, when a chip crosses a minimum separator, the reduced instance keeps one vertex for each type of connected component outside the chip, together with the target sets that the component meets. This ensures that witness paths avoid the entire vertex
separator.

\paragraph{Augmented vertex instances.}
We use graphs $G\subseteq H$ on the same vertex set. The graph $G$ is seen by
$\pi$, while $H$ also contains auxiliary edges used by witness paths. An
instance has a source $S$, a separator target $T$, a forbidden set $F$, and
a charged set $Q\subseteq F$. Every endpoint of an auxiliary edge belongs
to $F$. We require $S\cap(T\cup F)=\emptyset$ and $T\cap Q=\emptyset$.
For each $q\in Q$, specify a request target set $U_q$. These sets may differ
from $T$, because the nested reduction changes the separator
target. For $Z\subseteq V(H)$, write
\[
 C_Z\coloneqq N_H(Z)\setminus Q,
\]
and call $C_Z$ the \emph{paid boundary} of $Z$.
An augmented chip contains $S$, avoids $T\cup F$, is reachable from $S$ in
$H[Z]$, and satisfies the following four conditions. The set $C_Z$ is an inclusion-wise
minimal $S$-$T$ separator in $H-Q$, disjoint from $S\cup T$. Its size satisfies $|C_Z|\le k$. The graph satisfies
$G[Z]\models\pi$. Finally, every $q\in Q$ has a $q$-$U_q$ path in
\[
 H-\bigl(Z\cup C_Z\cup(Q\setminus\{q\})\bigr).
\]
An augmented chip $Z$ is \emph{important} if there is no augmented chip $Z'\supsetneq Z$ with $|C_{Z'}|\le|C_Z|$. Auxiliary
endpoints are forbidden, so $H[Z]=G[Z]$ and $N_H(Z)=N_G(Z)$ for every
augmented chip. A charged vertex need not currently belong to $N_H(Z)$;
its request is imposed in either case. This permits the crossing reduction
to replace charged vertices by copies that have only auxiliary edges.
Ordinary vertex chips correspond to $G=H$ and
$F=Q=\emptyset$.
For any reachable $Z$ as above with $C_Z\cap T=\emptyset$, minimality of $C_Z$
is equivalent to every $c\in C_Z$ having a $c$-$T$ path in
$(H-Q)-(Z\cup(C_Z\setminus\{c\}))$. Indeed, adjoining a source path
inside $Z$ and its edge to $c$ gives the usual minimal-separator witness.

\thmundirectedVertexChipBound*

\begin{proof}
We prove the stronger augmented statement for every fixed upper bound $K$
on $k+|Q|$. The bound is independent of the forbidden set, the request target
sets, and the number of auxiliary edges. Put $J=H-Q$ and
$\lambda=\lambda_J^v(S,T)$. If $\lambda>k$, there are no chips. If
$\lambda=0$, minimality forces $C_Z=\emptyset$, and the only possible chip
is the set reachable from $S$ in $J$. This also handles an empty recursive
target. Hence assume $1\le\lambda\le k$.

\paragraph{Charging and monotonicity.}
If $x\in C_Z$, charge $x$ by replacing $Q,F,k$ with
$Q\cup\{x\},F\cup\{x\},k-1$, and give $x$ the request target $U_x=T$.
All other data remain unchanged. The old strictness witness for $x$ supplies
this request, while every other boundary or request witness already avoids
$x$. Thus $Z$ remains a chip.
Moreover, if a larger chip $Y$ dominates $Z$ after charging, then $x$ remains
adjacent to $Y$ and is forbidden from entering it. Its old paid boundary is
therefore its new paid boundary together with $x$. The new request for $x$
restores its old strictness witness, so $Y$ was also an old dominator.
Consequently every important chip whose paid boundary contains $x$ is still an important chip after charging $x$, and charging preserves $k+|Q|$.

Requests are monotone in the direction needed by the induction. For
$q\in Q$, their deleted set can be written as
\[
 Z\cup C_Z\cup(Q\setminus\{q\})
 =\bigl(N_H[Z]\cup Q\bigr)\setminus\{q\}.
\]
Since closed neighborhoods are monotone under inclusion, a request valid for
a larger set is valid for a smaller one. Together with heredity and the
forbidden-set condition, this gives the compact-chip monotonicity used in
\Cref{lem:monotonicityOfChips}.

\paragraph{Minimum separators and nested chips.}
Apply \Cref{prop:routedCutSequenceExists} to the usual vertex-splitting
network of $J$: nonterminal split arcs are real, while adjacency arcs
and terminal split arcs are auxiliary and hence undeletable. Projecting the
reachable source sides in the resulting minimum cut sequence and deleting
repetitions gives
\[
 X_1\subsetneq\cdots\subsetneq X_r,
 \qquad I_i=N_J(X_i),\qquad |I_i|=\lambda,
\]
where every minimum $S$-$T$ vertex separator is contained in
$\bigcup_i I_i$. Indeed, the finite cut arcs are precisely the split arcs
of the separator vertices, and a split vertex lies on such a cut exactly
when the original vertex belongs to $N_J(X_i)$. The nested source sides and
the cover of all minimum cuts therefore project to the stated properties.

Each $X_i$ already satisfies reachability, strictness, and the budget. Any
failure of its remaining chip conditions persists under enlargement, by the
preceding monotonicity. Hence the cut-sequence dichotomy of
\Cref{lem:possibleBranchings} applies. Apart from the last compact chip
$X_\mu$, important chips lie in one adjacent band or cross one of at most
two specified compact sets. Set $\mu=0$ when no compact chip exists.
In that case there is no exceptional chip. The initial and final artificial endpoints
are $S$ and the vertices reachable from $S$ in $J-T$.

For a band between consecutive $X\subsetneq X'$, add
$I=N_J(X)$ and $I'=N_J(X')$ to the temporary hitting set. An unhit chip
$X\subsetneq Z\subsetneq X'$ satisfies
\[
 X\cup I\subseteq Z,\qquad C_Z\subseteq X',\qquad Z\cap I'=\emptyset.
\]
Keep both graphs, $Q$, and every request target $U_q$ unchanged. Use source
$S^*=X\cup I$, separator target $T^*=I'$, and forbidden set
$F^*=F\cup(V(H)\setminus X')$. Original strictness witnesses, truncated at
their first visit to $I'$, show that $Z$ is a local chip.
Conversely, a local dominating chip $Y$ lies in $X'$ and has paid boundary
inside $X'$. Append to each local boundary witness ending at $e\in I'$ the
$e$-$T$ witness for the minimum separator $I'$ in $J$. This suffix avoids
$X'$ and $I'\setminus\{e\}$, so it avoids $Y$ and its other boundary
vertices. The old requests are unchanged. Since $S^*\subseteq Z$, local
source reachability also lifts. Thus $Y$ would be an original dominator,
and $Z$ is locally important.

Every minimal $S^*$-$I'$ separator in $J$ lies in
$X'\setminus(X\cup I)$ and is also an $S$-$T$ separator. No vertex of this
layer belongs to any $I_i$: earlier boundaries lie in $X\cup I$, and later
ones lie outside $X'$. The cover above therefore implies
\[
 \lambda_J^v(S^*,I')>\lambda.
\]
For the initial band, including $Z=S$ when it is a chip, use $S^*=S$,
$T^*=I_1$, and forbid the exterior of $X_1$. For the final band use $S^*=X_r\cup I_r$ and retain $T,F$.
The same arguments apply, using the corresponding endpoint layers.
Discard malformed local instances and those whose minimum cut does not
exceed $\lambda$ or exceeds the budget. The localization argument above guarantees
that no unhit chip is lost.

\paragraph{Crossing chips.}
Fix a compact $X$ and put $I=N_J(X)$. Add $I$ to the hitting set, and
consider a crossing chip $Z$ with $C_Z\cap I=\emptyset$. Set
\[
 B=I\cap Z,\qquad E=I\setminus Z,\qquad
 P=C_Z\setminus X,\qquad d=|P|.
\]
No vertex of $P$ has a neighbor in $X$, since $P\cap I=\emptyset$.
No vertex of $E$ has a neighbor in $Z$, since $E\cap C_Z=\emptyset$.
Furthermore, $C_Z\cap X\ne\emptyset$: take the first exit from $Z$ on a
path in $J[X]$ from $S$ to $X\setminus Z$.
Also $d\ge1$. A path in $J[Z]$ to $Z\setminus X$ first leaves $X$ at some
$b\in B$. Minimality of $I$ gives a $b$-$T$ path in
$J-(X\cup(I\setminus\{b\}))$. Its first exit from $Z$ belongs to $P$.

Choose a finite CMSO representative $R$ of $G[Z\setminus X]$ with labeled
boundary $B$. Keep $G[X\cup I\cup Q]$, remove its edges with both endpoints
in $B$, and glue in $R$. This is the real part of the reduced graph.
Take $S'=S\cup V(R)$. All remaining additions described below are auxiliary,
and their endpoints are forbidden.

To preserve exterior connectivity, consider the connected components of
\[
 H-\bigl(X\cup Z\cup P\cup Q\cup E\bigr).
\]
The signature of a component records its neighbors in
$L=E\cup Q\cup P$ and which of the sets $T$ and $U_q$, for $q\in Q$, it
meets. Retain one fresh vertex for each signature occurring, with auxiliary
edges to the recorded neighbors. Replace every $p\in P$ by a fresh copy
$p^*$. For every edge $\{p,y\}$ of $H$ with $y\in E\cup Q\cup P$, add the
auxiliary edge $\{p^*,y^*\}$, where $y^*=y$ if $y\in E\cup Q$ and $y^*$ is the
copy of $y$ if $y\in P$. Edges between $P$ and $B$ are not retained. No
witness uses them, since every reduced chip contains $B\subseteq V(R)\subseteq S'$.
Keep the old auxiliary edges on $X\cup I\cup Q$. They have no endpoint in
$(X\cap Z)\cup B$, since $Z\cap F=\emptyset$.
Component vertices and $P$-copies have no real edges.

For each old target set, keep its membership on $X\cup I\cup Q$ and mark
the component vertices whose signatures meet it. Membership in the
forgotten part of $Z$ is discarded, since that part is always inside the
chip. Denote the resulting target sets by $T'$ and $U'_q$. Give each $p^*$
the request target $U_{p^*}=T'$ and no memberships in the old request target
sets. Set
\[
 Q'=Q\cup\{p^*:p\in P\},\qquad k'=k-d.
\]
The new forbidden set $F'$ is the union of the retained old forbidden
vertices, $E$, $Q'$, and all component vertices. Let $G'\subseteq H'$ be the resulting
graphs and let
\[
 C=(X\cap Z)\cup V(R).
\]

This construction preserves exactly the exterior connectivity needed by
witnesses, including after deleting any subset of $E$. A path projects by
contracting its maximal subpaths in exterior components. Conversely, a
traversal through a component vertex expands inside one original connected
component with that signature. Its target marks give the correct possible
destinations. Merging equal signatures is valid because we require
individual path existence, never disjoint paths. All of $P,Q,E$ were removed
before forming components, so none is hidden inside a component vertex.

The reduced paid boundary is exactly
\[
 N_{H'}(C)\setminus Q'=C_Z\cap X.
\]
Indeed, $E$ has no neighbor in $Z$, the representative is entirely inside
$C$, and all new auxiliary endpoints are forbidden. Its size is at most
$k-d$. The CMSO condition follows by gluing. To see source reachability,
split original paths in $Z$ at their visits to $Z\setminus X$. Each retained
suffix starts in $S$ or in $B\subseteq V(R)$, and every representative
vertex is itself a source. Original strictness witnesses for $C_Z\cap X$
project to reduced strictness witnesses. The witness for $p\in P$ becomes
the request for $p^*$, while old requests project to their corresponding
$U'_q$. Hence $C$ is a reduced augmented chip.

Suppose a reduced chip $C'\supsetneq C$ dominates $C$. Every new or retained
vertex outside $X\cup V(R)$ is forbidden, so
$W=C'\setminus C\subseteq X\setminus Z$. Put $Y=Z\cup W$. Identifying
retained original vertices, we have the exact disjoint union
\[
 C_Y=P\mathbin{\dot\cup}\bigl(N_{H'}(C')\setminus Q'\bigr).
\]
New exterior neighbors of $W$ can only belong to $I\cup Q$. Indeed, $B$ is inside,
$E$ is retained explicitly, and $P$ remains adjacent through its old
neighbors in $Z\setminus X$. This proves the identity and the budget bound.
Reduced boundary witnesses expand to witnesses avoiding every other vertex
of $C_Y$. The requests for $p^*$ supply the witnesses for $p\in P$, and old
requests expand to their original targets. Component interiors contain
neither $P$ nor $E$, so the required avoidance is preserved.
Source paths lift through their last departure from $V(R)$, whose boundary
vertex is reachable from $S$ inside $Z$. They use no auxiliary edge, since
its endpoints are forbidden. Finally, the gluing
\[
 G[Y]=\bigl(G[(Y\cap X)\cup B]-E(G[B])\bigr)\oplus G[Z\setminus X]
\]
transfers $\pi$ from $C'$ to $Y$. Thus $Y$ is an original dominator, a
contradiction. The reduced chip is important.

\paragraph{Finite choices and the induction.}
The reduction does not require guessing vertices of $P$ in the input.
Guess $B\subseteq I$, $1\le d\le k$, a finite representative $R$, $d$ fresh
labeled copies, their auxiliary edges to $E\cup Q$ and to each other, and the
occurring component signatures. The interface has size at most $2K$ and there are at most
$K+1$ target labels, giving at most $2^{3K+1}$ possible signatures and hence at most
$2^{2^{3K+1}}$ sets of occurring signatures. Thus the number
of reduced instances is bounded by a constant $M(K,\pi)$. All old target
memberships on retained vertices are known. Discard malformed instances,
including those with $S'\cap(T'\cup F')\ne\emptyset$ or
$T'\cap Q'\ne\emptyset$.

Now induct lexicographically on $(k,\rho)$, where
$0\le\rho\le k$ and $k-\lambda\le\rho$, simultaneously for all
$k+|Q|\le K$. Nested calls decrease $\rho$, and crossing calls decrease
$k$ while replacing $|Q|$ by $|Q|+d$. Add the paid boundaries of the
important nested chips, and the projections to $X$ of the important
crossing chips' paid boundaries, to the temporary hitting set. The preceding
proofs show that this bounded set hits every original important chip except
possibly $X_\mu$ when $\mu\ge1$. Charging each hit vertex finishes the
induction.
Explicitly, let $f(0,0)=1$, $f(k,-1)=0$, and
$A_{k-1}=\max_{j<k,\,0\le\sigma\le j}f(j,\sigma)$. A sufficient recurrence is
\[
 f(k,\rho)=1+
 \bigl(2K+Kf(k,\rho-1)+2KM(K,\pi)A_{k-1}\bigr)A_{k-1}.
\]
There is no nested contribution when $\rho=0$. Taking $K=k$ and
$G=H$, $F=Q=\emptyset$ proves the ordinary bound.
\end{proof}

\begin{claim}[Enumeration for undirected vertex chips]
\label{thm:undirectedVertexChipEnumeration}
Let $k\in\mathbb{N}_0$, and let $\pi$ be a hereditary CMSO sentence over
undirected graphs such that, for fixed $\pi$, deciding $H\models\pi$ on
an undirected graph $H$ takes polynomial time. Then all important
$(S,T,\pi,k)$-vertex chips in an undirected graph $G$ can be enumerated in
time $c^v_{\pi,k}\cdot n^{\bigoh(1)}$.
\end{claim}

\begin{proof}
Use the augmented recursion just proved. Minimum-separator sequences are
computed by vertex splitting. Graph searches check strictness and requests,
and the assumed test for $\pi$ identifies compact chips and validates
candidates. Enumerating all representatives and signatures gives the bounded
crossing family without identifying an unknown forgotten piece.

Recursively compute the temporary hitting set, enumerate the charged
instances, include $X_\mu$ when it exists, and discard outputs that fail the
original chip conditions or are dominated by another valid output. Every
important chip occurs by the charging and localization arguments. Every
valid nonimportant output has an important dominator, which also occurs,
so this filtering is exact. The zero-cut case has its single explicit
candidate. The recurrence bounds all branching and recursion in terms of
$k,\pi$, and recursive graphs have $n+f(k,\pi)$ vertices. All other work is
polynomial. As in the directed theorem, the finite CMSO representative lists
are supplied non-uniformly unless a computable representative mechanism is
assumed.
\end{proof}

\section{Concluding Remarks}\label{sec:conclusion}

In independent work, Galby, Sharma, and van der Steenhoven~\cite{GalbySS26}
also prove that \arcDoct{} is fixed-parameter tractable on Eulerian digraphs.
Their algorithm uses an edge-cut version of treewidth reduction and has
running time $2^{2^{\bigoh(k^2)}}\cdot n^{\bigoh(1)}$.
This gives a different approach to the Eulerian special case of our result.

The bound on important chips in $\alpha$-balanced digraphs suggests the
following research directions.

\begin{enumerate}
    \item \textbf{Tighter bounds for specific sentences.}
    Our bounds for hereditary CMSO properties $\pi$ are intentionally phrased in as general a form as possible,
    and it is natural to ask whether substantially better bounds can be obtained 
    when $\pi$ is a specific CMSO or first-order (FO) sentence, even on
    undirected graphs. For instance, what is the best achievable bound when
    $\pi$ expresses bounded maximum degree? The recent work of
    Kenig~\cite{Kenig25cpis} demonstrates the viability of this line of
    inquiry, notably even for a non-hereditary property. Pinning down the
    optimal dependence on $k$ for natural CMSO and FO properties is a rich
    area for further investigation.

    \item \textbf{FPT algorithms for unweighted $\mathrm{W}$-hard problems on
    Eulerian digraphs.}
    Our algorithm for \arcDoct{} suggests asking which other unweighted
    $\mathrm{W}$-hard problems on general digraphs become fixed-parameter
    tractable on Eulerian or, more generally, $\alpha$-balanced digraphs. The aforementioned paper \cite{GalbySS26} already makes progress in this direction. 

    \item \textbf{Weighted problems and flow-augmentation.}
    Weighted variants of cut problems on general digraphs have recently seen
    major algorithmic progress through flow-augmentation techniques
    \cite{KimKPW25,KimMPSW24}. Can these methods be adapted to balanced
    digraphs to yield FPT algorithms for weighted problems that remain
    $\mathrm{W}$-hard in general?

    \item \textbf{Important balanced subgraphs in digraphs.}
    Dabrowski et al.~\cite{DabrowskiJOOW25} introduced important
    balanced subgraphs for undirected graphs. Is there an analogous notion,
    with accompanying results, on Eulerian digraphs?

    \item \textbf{Multi-budgeted chips.}
    In the spirit of the multi-budgeted directed cuts of Kratsch et
    al.~\cite{KratschLMPW20}, one can ask whether our notion of chips extends
    to the multi-budgeted setting. We believe that our approach, combined with
    the branching step of \cite{KratschLMPW20} that guesses the colors of the
    solution on each path of a maximum flow, could yield an analogous bound
    and enumeration algorithm for multi-budgeted chips.
\end{enumerate}

In summary, we believe important chips may prove to be a foundational tool
for parameterized algorithms on both undirected graphs and balanced digraphs.

\appendix

\section{Deferred proofs}\label{app:deferred-proofs}

This appendix contains proofs deferred from the main text. The order follows
the order in which the corresponding statements appear in the paper.
\Cref{obs:splitAlongGlue} and \Cref{obs:attachmentDecomposition} are stated
and proved in \Cref{app:crossingReductionDetails}.

\begin{center}
\begin{tabularx}{\textwidth}{@{}p{0.38\textwidth}X@{}}
\textbf{Statement} & \textbf{Location in this appendix} \\
\hline
\Cref{lem:basicStrictFacts} & \Cref{app:basic-strictness-balance} \\
\Cref{lem:balanceStability} & \Cref{app:basic-strictness-balance} \\
\Cref{obs:InterfaceSeparator} & \Cref{app:basic-strictness-balance} \\
\Cref{lem:inducedChips}, \Cref{cor:inducedChips} & \Cref{app:deletion-monotonicity} \\
\Cref{lem:monotonicityOfChips} & \Cref{app:deletion-monotonicity} \\
\Cref{obs:exitFromXprime}, \Cref{obs:localCutIncrease}, \Cref{lem:nestedWithin} & \Cref{app:nested-localization-proofs} \\
\Cref{lem:routedInitialNestedWithin}, \Cref{lem:routedFinalNestedWithin} & \Cref{app:nested-localization-proofs} \\
\Cref{obs:splitAlongGlue}, \Cref{obs:attachmentDecomposition}, \Cref{obs:boundedAttachmentBoundary} & \Cref{app:crossingReductionDetails} \\
\Cref{obs:basicReducedData}, \Cref{obs:reducedChipRealSubgraph}, \Cref{obs:shortcutEndpointsRoutedInterface} & \Cref{app:attachment-proofs} \\
\Cref{lem:multiToSimpleDOCT}, \Cref{thm:arcDoctNPComplete} & \Cref{app:doct-hardness-proofs} \\
\end{tabularx}
\end{center}

\subsection{Basic strictness and balance facts}
\label{app:basic-strictness-balance}

\begin{proof}[Proof of \Cref{lem:basicStrictFacts}]
For the first item, let $Z$ be compact. Then $\partial^+(Z)$ is an
$S$-$T$ cut of minimum cardinality. Hence no proper subset of
$\partial^+(Z)$ is an $S$-$T$ cut. Thus $\partial^+(Z)$ is inclusion-wise
minimal, and $Z$ is strict.

We prove the second item. First suppose that $Z$ is strict. Let
$v\in N^+(Z)$, and choose an arc $a=(u,v)\in\partial^+(Z)$. Since
$\partial^+(Z)$ is an inclusion-wise minimal $S$-$T$ cut, the set
$\partial^+(Z)\setminus\{a\}$ is not an $S$-$T$ cut. Hence there is an
$S$-$T$ path in $D-(\partial^+(Z)\setminus\{a\})$. Such a path must use
$a$. After using $a$, it cannot visit $Z$ again, since otherwise it would
need to leave $Z$ through an arc of
$\partial^+(Z)\setminus\{a\}$. Therefore the suffix of this path starting
at $v$ is a $v$-$T$ path in $D-Z$.

Conversely, suppose that every vertex of $N^+(Z)$ has a path to $T$ in
$D-Z$. Let $a=(u,v)\in\partial^+(Z)$. Since $Z$ is reachable from $S$ in
$D[Z]$, there is an $S$-$u$ path contained in $D[Z]$. By assumption, there
is a $v$-$T$ path in $D-Z$. Hence there is an $S$-$T$ path using $a$ and
no other arc of $\partial^+(Z)$. Therefore
$\partial^+(Z)\setminus\{a\}$ is not an $S$-$T$ cut. Since this holds for
every $a\in\partial^+(Z)$, the cut $\partial^+(Z)$ is inclusion-wise
minimal. Thus $Z$ is strict.

For the third item, let $a=(u,v)\in\partial^+(Z)$. Since the head of $a$ lies outside $Z$, we have $(D-a)[Z]=D[Z]$. Hence $Z$ is still
reachable from $S$ in $(D-a)[Z]$, and
\[
\partial_{D-a}^+(Z)=\partial_D^+(Z)\setminus\{a\}.
\]
Let $b\in\partial_{D-a}^+(Z)$. Since $Z$ is strict in $D$, the set
$\partial_D^+(Z)\setminus\{b\}$ is not an $S$-$T$ cut in $D$. Hence there
is an $S$-$T$ path in $D-(\partial_D^+(Z)\setminus\{b\})$. This path uses
$b$ and no other arc of $\partial_D^+(Z)$, so in particular it does not
use $a$. Therefore the same path exists in
\[
(D-a)-(\partial_{D-a}^+(Z)\setminus\{b\}).
\]
Thus no proper subset of $\partial_{D-a}^+(Z)$ is an $S$-$T$ cut in
$D-a$, and $Z$ is strict in $D-a$.
\end{proof}

\begin{figure}[t]
\begin{center}
\begin{tikzpicture}[
  >={Stealth[length=2.8mm]},
  vtx/.style={circle, draw, fill=white, inner sep=0pt, minimum size=7.5mm, font=\small},
  ext/.style={circle, draw, fill=black!5, inner sep=0pt, minimum size=7.5mm, font=\small},
  kept/.style={->, thick, black},
  qarc/.style={->, very thick, red!75!black, dashed},
  lbl/.style={font=\footnotesize},
  qlbl/.style={font=\scriptsize, red!75!black},
]

\node[vtx] (u1) at (0,1.5)     {$u_1$};
\node[vtx] (u2) at (-0.7,-0.5) {$u_2$};
\node[vtx] (u3) at (1.4,-0.6)  {$u_3$};

\begin{scope}[on background layer]
  \node[draw=blue!55!black, fill=blue!8, line width=0.8pt,
        rounded corners=18pt, fit=(u1)(u2)(u3),
        inner sep=16pt] (Z) {};
\end{scope}
\node[blue!55!black, font=\bfseries\large] at ($(Z.north)+(0,0.4)$) {$Z$};

\node[ext] (w1) at (-4.3,2.0)  {$w_1$};
\node[ext] (w2) at (-4.3,-0.7) {$w_2$};
\node[ext] (x1) at (4.6,1.5)   {$x_1$};
\node[ext] (x2) at (4.6,-1.5)  {$x_2$};

\draw[kept] (w1) -- (u1);
\draw[qarc] (w2) -- (u2);

\draw[kept] (u1) -- (x1);
\draw[qarc] (u3) -- (x2);

\draw[qarc] (u2) to[bend right=18] (u3);

\node[qlbl, align=center] at (-2.9,-1.35) {removed, enters $Z$\\(counts in $r^-$)};
\node[qlbl, align=center] at (3.55,-1.95) {removed, leaves $Z$\\(counts in $r^+$)};
\node[qlbl, align=center] at (0.35,-1.55) {removed, interior\\(counts in neither)};

\node[lbl, blue!55!black] at (-2.55,1.0) {$\partial^-(Z)$};
\node[lbl, blue!55!black] at (3.15,1.08) {$\partial^+(Z)$};

\begin{scope}[shift={(-4.3,-3.2)}]
  \draw[kept] (0,0) -- (0.9,0);
  \node[lbl, anchor=west] at (1.05,0) {arc kept in $D'=D-F$};
  \draw[qarc] (0,-0.55) -- (0.9,-0.55);
  \node[lbl, anchor=west] at (1.05,-0.55) {arc of $F$ (deleted to form $D'$)};
\end{scope}

\end{tikzpicture}	
\end{center}
\caption{An illustration of the set $Z$ in the proof of the first item of
\Cref{lem:balanceStability}.}
\label{fig:localChanges}
\end{figure}
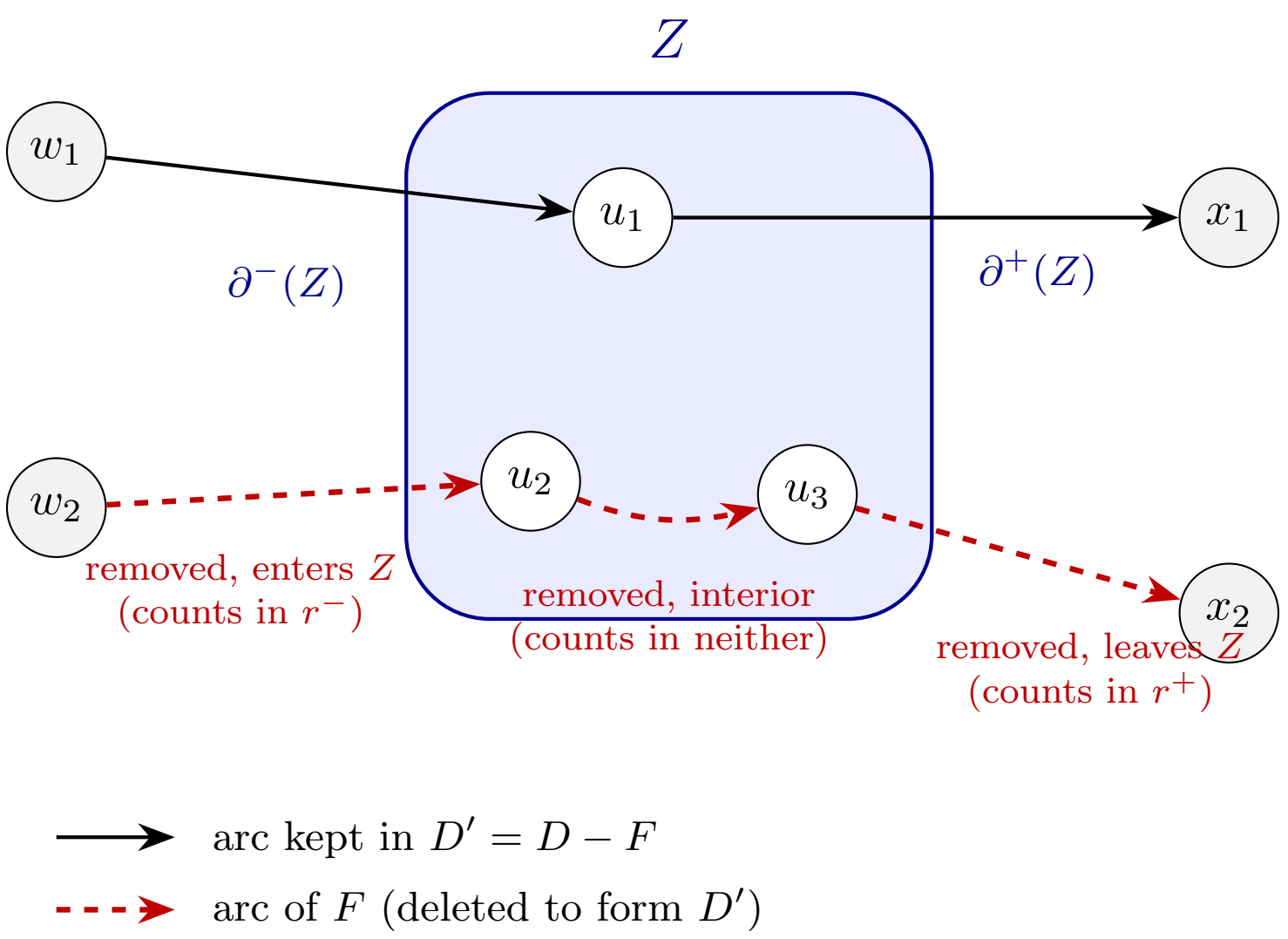

\begin{proof}[Proof of \Cref{lem:balanceStability}]
For the first item (see \Cref{fig:localChanges}), let $r^+$ and $r^-$ count the deleted arcs leaving and
entering $Z$, respectively. Their sum is at most $|F|$, and
\[
\begin{aligned}
\bigl||\partial_{D-F}^-(Z)|-|\partial_{D-F}^+(Z)|\bigr|
&=\bigl||\partial_D^-(Z)|-|\partial_D^+(Z)|-r^-+r^+\bigr|\\
&\le\alpha+r^-+r^+\le\alpha+|F|.
\end{aligned}
\]
For the second item, delete
$F=\partial_D^+(Z)\cup\partial_D^-(Z)$, of size at most $2r+\beta$.
By the first item, $W$ is $(\alpha+2r+\beta)$-balanced in $D-F$.
Since $W\subseteq Z$, its boundary there equals its boundary in $D[Z]$.

For the third item, start with $D_a$ and add the other vertices of $D_b$
as isolated vertices. Add the arcs contributed by $D_b$ that are not
already present, at most $|A(D_b)|$ arcs. The same imbalance calculation
as in the first item, with additions in place of deletions, increases the
balance by at most $|A(D_b)|$.
\end{proof}

\begin{proof}[Proof of \Cref{obs:InterfaceSeparator}]
Let $G$ be the undirected graph underlying $D$. Suppose, for a
contradiction, that there is a path in $G-X$ from a vertex of
$Z\setminus X$ to a vertex of $Y$. Since $Y$ is disjoint from $Z$, this
path contains an edge with one endpoint in $Z\setminus X$ and the other
endpoint in $V(D)\setminus Z$. The corresponding arc in $D$ has one
endpoint in $Z\setminus X$ and the other endpoint outside $Z$. Hence this
endpoint in $Z\setminus X$ belongs to $\intf(Z)=X$, a contradiction.
\end{proof}

\subsection{Deletion, boundary hitting, and monotonicity}
\label{app:deletion-monotonicity}

\begin{proof}[Proof of \Cref{lem:inducedChips}]
Write $\widehat D'=\widehat D-a$, $T_2'=T_2\cup\{v\}$, and
$Q'=Q\cup\{v\}$. Since $a$ leaves $Z$, its deletion preserves the induced
real graph, source reachability, and auxiliary-respecting property, and
lowers the boundary size by one. Every strictness or request witness in
$\widehat D-Z$ avoids $a$, whose tail lies in $Z$. This preserves the old
witnesses and supplies the new request at $v$, by
\Cref{obs:routedStrictnessCriterion}. Also $v\notin Z$, so $Z$ is a chip
in the smaller instance.

Suppose a larger chip $Z'$ dominates $Z$ there. It contains $u\in Z$ and
avoids $v\in T_2'$, so restoring $a$ adds one boundary arc to each of
$Z$ and $Z'$ and changes neither induced graph. The old strictness
witnesses persist, and the new boundary head $v$ has its witness from
$Q'$. All original request and forbidden conditions are inherited.
Thus $Z'$ is an original chip, with
\[
|\partial_{\mathrm{real},\widehat D}^+(Z')|
=|\partial_{\mathrm{real},\widehat D'}^+(Z')|+1
\le |\partial_{\mathrm{real},\widehat D'}^+(Z)|+1
=|\partial_{\mathrm{real},\widehat D}^+(Z)|,
\]
contradicting importance of $Z$.
\end{proof}

\begin{proof}[Proof of \Cref{cor:inducedChips}]
Let $Z \in \Ical_{\widehat D}^a\sttqpk$. Then $a \in \partial_{\mathrm{real}}^+(Z)$.
By \Cref{lem:inducedChips}, the same vertex set $Z$ is an important $(S, T_1, T_2 \cup \{v\}, Q \cup \{v\}, \pi, k-1)$-chip in $\widehat D - a$. The identity map on vertex sets gives an injection
from $\Ical_{\widehat D}^a\sttqpk$ into
  $\Ical_{\widehat D - a}(S, T_1, T_2 \cup \{v\}, Q \cup \{v\}, \pi, k-1)$,
which proves the bound.
\end{proof}

\begin{proof}[Proof of \Cref{lem:monotonicityOfChips}]
For the first item, reachability, strictness, the auxiliary-respecting property, and the
budget are assumed. The remaining conditions pass from $X$ to $Z\subseteq X$.
The condition on $\pi$ holds by heredity, avoidance of $T_2$ holds by inclusion, and the request paths
survive because $\widehat D-X\subseteq\widehat D-Z$.

For the second item, if $Z\subsetneq X$, the chip $X$ dominates $Z$, since
\[
|\partial_{\mathrm{real}}^+(X)|
=\lambda_{\widehat D}(S,T_1)
\le |\partial_{\mathrm{real}}^+(Z)|.
\]
Importance therefore forces $Z=X$.
\end{proof}

%
\subsection{Nested localization}
\label{app:nested-localization-proofs}

\begin{proof}[Proof of \Cref{obs:exitFromXprime}]
Let $t\in T_{X'}^\star$. By the definition of $T_{X'}^\star$, there is a
real arc $(t,h)\in\partial_{\mathrm{real}}^+(X')$. Since $X'$ is
auxiliary-respecting and strict, \Cref{obs:routedStrictnessCriterion} gives
an $h$-$T_1$ path in $\widehat D-X'$. Hence the arc $(t,h)$ followed by this
path gives the desired $t$-$T_1$ path.
\end{proof}

\begin{proof}[Proof of \Cref{obs:localCutIncrease}]
Since $X\subseteq Z$ and
$\partial^+_{\mathrm{real}}(Z)\cap\partial^+_{\mathrm{real}}(X)=\emptyset$,
every head of a real arc leaving $X$ belongs to $Z$. Since $X$ is
auxiliary-respecting, no auxiliary arc leaves $X$. Hence
\[
S_X^\star
=
X\cup \head(\partial^+_{\mathrm{real}}(X))
\subseteq Z\subseteq X'.
\]
Moreover $T_{X'}^\star\cap Z=\emptyset$, and $Z$ is
auxiliary-respecting. Therefore $\partial^+_{\mathrm{real}}(Z)$ is an
$S_X^\star$-$T_{X'}^\star$ cut, so
$\lambda_{\widehat D}(S_X^\star,T_{X'}^\star)$ is finite.

Let $F$ be a minimum $S_X^\star$-$T_{X'}^\star$ cut, and let $R$ be the set
of vertices reachable from $S_X^\star$ in $\widehat D-F$. By the minimality of $F$, we have 
$F=\partial^+_{\mathrm{real}}(R)$. 

We first show that $F$ is also an $S$-$T_1$ cut. Indeed, every
$S$-$T_1$ path leaves $X$ and later leaves $X'$. Since both $X$ and $X'$
are auxiliary-respecting, the first arc leaving each of these sets is real.
Thus the path contains an $S_X^\star$-$T_{X'}^\star$ subpath, and this
subpath meets $F$.

We next locate the reachable side $R$. By definition,
$S_X^\star\subseteq R$, and hence $X\subseteq R$. Also
$R\cap T_{X'}^\star=\emptyset$, since $F$ separates $S_X^\star$ from
$T_{X'}^\star$. We claim that $R\subseteq X'$. Otherwise, take a path in
$\widehat D-F$ from $S_X^\star$ to a vertex outside $X'$, and consider its
first arc leaving $X'$. This arc is real because $X'$ is
auxiliary-respecting, and its tail lies in $T_{X'}^\star$. This contradicts
$R\cap T_{X'}^\star=\emptyset$. Therefore, 
$X\subseteq R\subseteq X'$.

Furthermore, no real arc leaving $X$ belongs to $F$, because every head of
such an arc lies in $S_X^\star\subseteq R$. Also, no real arc leaving $X'$
belongs to $F$, because its tail lies in $T_{X'}^\star$, which is disjoint from
$R$. Since $X\subseteq R\subseteq X'$, every arc of
$F=\partial^+_{\mathrm{real}}(R)$ therefore has both endpoints in
$X'\setminus X$. Moreover, $F$ is non-empty because it is an $S$-$T_1$ cut
and the standing assumptions guarantee an $S$-$T_1$ path.

Since $X$ and $X'$ are consecutive members of the minimum cut sequence, the
set $X'\setminus X$ is one of the layers in
\Cref{lem:layerArcRaisesCut}. Applying that lemma to the $S$-$T_1$ cut
$F$, we get
$|F|\geq \lambda_{\widehat D}(S,T_1)+1$.
Since $F$ is a minimum $S_X^\star$-$T_{X'}^\star$ cut, we have that
$\lambda_{\widehat D}(S_X^\star,T_{X'}^\star)
=
|F|
>
\lambda_{\widehat D}(S,T_1)$.
\end{proof}

\begin{proof}[Proof of \Cref{lem:nestedWithin}]
We first verify that $Z$ is a local chip. As in the proof of \Cref{obs:localCutIncrease},
\[
S_X^\star
=
X\cup \head(\partial^+_{\mathrm{real}}(X))
\subseteq Z\subseteq X'.
\]
Also $S\subseteq S_X^\star$, and $Z$ is reachable from $S$ inside
$\widehat D[Z]$. Thus $Z$ is reachable from $S_X^\star$ inside
$\widehat D[Z]$.

The set $Z$ is disjoint from $T_{X'}^\star$. Indeed, if
$t\in Z\cap T_{X'}^\star$, then some real arc from $t$ to
$V(\widehat D)\setminus X'$ belongs both to
$\partial^+_{\mathrm{real}}(Z)$ and to
$\partial^+_{\mathrm{real}}(X')$, contrary to the assumption. The cut
bound, the condition $\widehat D_{\mathrm{real}}[Z]\models\pi$, and the
condition $Z\cap T_2=\emptyset$ hold because $Z$ is an $\sttqpk$-chip.

We next check strictness and the request condition in the local instance.
We shall use the following {\bf prefix claim}. If $v\in X'\setminus Z$ and there
is a $v$-$T_1$ path in $\widehat D-Z$, then there is a
$v$-$T_{X'}^\star$ path in $\widehat D-Z$. Indeed, since
$X'\cap T_1=\emptyset$, the path must leave $X'$. Since $X'$ is
auxiliary-respecting, the first arc of the path leaving $X'$ is real. The
prefix up to the tail of this arc gives the desired path to
$T_{X'}^\star$.

Let $y\in \head(\partial^+_{\mathrm{real}}(Z))$. Since $Z\subset X'$ and
$\partial^+_{\mathrm{real}}(Z)$ is disjoint from
$\partial^+_{\mathrm{real}}(X')$, we have $y\in X'\setminus Z$. Since $Z$
is strict as an $S$-$T_1$ reachability set, there is a $y$-$T_1$ path in
$\widehat D-Z$. The prefix claim gives a $y$-$T_{X'}^\star$ path in
$\widehat D-Z$.

Now let $r\in Q^\star$. If $r\in Q\cap X'$, then the chosen path
$\avoidpath{Z}{X'}{T_1}{r}$ is an $r$-$T_1$ path in $\widehat D-Z$, and
the prefix claim gives an $r$-$T_{X'}^\star$ path in $\widehat D-Z$.
If $r\in \Delta_Z$, then $r$ is the first vertex of $X'$ on
$\avoidpath{Z}{X'}{T_1}{q}$ for some $q\in Q\setminus X'$. The suffix of
this path from $r$ to $T_1$ avoids $Z$, and the prefix claim again gives
an $r$-$T_{X'}^\star$ path in $\widehat D-Z$. Thus every vertex of
$Q^\star$ reaches $T_{X'}^\star$ in $\widehat D-Z$.

It remains to prove importance. Suppose for a contradiction that there is an
$(S_X^\star,\allowbreak T_{X'}^\star,\allowbreak T_2,\allowbreak Q^\star,\allowbreak \pi,\allowbreak k)$-chip $Z'$ such that
$Z\subset Z'$ and
$|\partial^+_{\mathrm{real}}(Z')|
\le
|\partial^+_{\mathrm{real}}(Z)|
$. 
We show that $Z'$ is an $\sttqpk$-chip in $\widehat D$.

First, $Z'$ is reachable from $S$ inside $\widehat D[Z']$, since
$S_X^\star$ is reachable from $S$ inside $\widehat D[Z]$, and $Z'$ is
reachable from $S_X^\star$ inside $\widehat D[Z']$.

We claim that $Z'\subseteq X'$. Otherwise, take a path in
$\widehat D[Z']$ from $S_X^\star$ to a vertex of $Z'\setminus X'$, and let
$(p,h)$ be the first arc on this path that leaves $X'$. This arc is real,
because $X'$ is auxiliary-respecting. Hence
$p\in T_{X'}^\star$, contradicting $Z'\cap T_{X'}^\star=\emptyset$.
Since $X'\cap T_1=\emptyset$, it follows that $Z'\cap T_1=\emptyset$.

We shall also use the following {\em  extension claim}. If $v$ has a
$v$-$T_{X'}^\star$ path in $\widehat D-Z'$, then $v$ has a $v$-$T_1$ path
in $\widehat D-Z'$. Indeed, let $t\in T_{X'}^\star$ be the endpoint of such
a path. By \Cref{obs:exitFromXprime}, there is a $t$-$T_1$ path whose first
arc is a real arc leaving $X'$ and whose remaining vertices lie outside
$X'$. Since $Z'\subseteq X'$ and $t\notin Z'$, this path avoids $Z'$.
Concatenating the two paths proves the claim.

We now show that $Z'$ is strict in the original instance.
Let $y\in \head(\partial^+_{\mathrm{real}}(Z'))$. Since $Z'\subseteq X'$
and $Z'\cap T_{X'}^\star=\emptyset$, we have $y\in X'$. The local
strictness of $Z'$ gives a $y$-$T_{X'}^\star$ path in $\widehat D-Z'$,
which extends to a $y$-$T_1$ path in $\widehat D-Z'$.

The cut bound, the condition $\widehat D_{\mathrm{real}}[Z']\models\pi$,
and the condition $Z'\cap T_2=\emptyset$ hold from the local chip
conditions.

It remains to check the requests in $Q$. Let $q\in Q$, and consider the
chosen path $\avoidpath{Z}{X'}{T_1}{q}$. If this path does not meet $X'$,
then it avoids $Z'$, since $Z'\subseteq X'$, and gives a $q$-$T_1$ path in
$\widehat D-Z'$.

If $q\in X'$, then $q\in Q^\star$. The local request condition gives a
$q$-$T_{X'}^\star$ path in $\widehat D-Z'$, and the extension claim gives a
$q$-$T_1$ path in $\widehat D-Z'$.

Finally, suppose $q\notin X'$ and $\avoidpath{Z}{X'}{T_1}{q}$ enters
$X'$. Let $r$ be its first vertex in $X'$. The prefix from $q$ to $r$
avoids $X'\setminus\{r\}$, and hence avoids $Z'$. By definition,
$r\in\Delta_Z\subseteq Q^\star$. The local request condition gives an
$r$-$T_{X'}^\star$ path in $\widehat D-Z'$, and the extension claim turns it
into an $r$-$T_1$ path in $\widehat D-Z'$. Concatenating with the prefix
from $q$ to $r$ gives the required path for $q$.

Thus $Z'$ is an $\sttqpk$-chip in $\widehat D$ that properly contains $Z$
and has at most as many real outgoing arcs as $Z$. This contradicts the importance of $Z$.

Finally,
$
\lambda_{\widehat D}(S_X^\star,T_{X'}^\star)
>
\lambda_{\widehat D}(S,T_1)
$ 
follows from \Cref{obs:localCutIncrease}.
\end{proof}

\begin{proof}[Proof of \Cref{lem:routedInitialNestedWithin}]
Boundary disjointness gives $Z\cap T_{X_1}^\star=\emptyset$ and places
every real boundary head of $Z$ in $X_1\setminus Z$. Apply the prefix and
extension arguments from \Cref{lem:nestedWithin} with outer region $X_1$
and unchanged source $S$. Truncating witnesses at their first exit from
$X_1$ gives local strictness and requests for
$Q^\star=(Q\cap X_1)\cup\Delta_Z$. Any local dominator stays inside
$X_1$, because it avoids the tails of all real exits and $X_1$ is
auxiliary-respecting. Extending its witnesses by \Cref{obs:exitFromXprime}
and prepending the chosen first-entry prefixes restores all original
requests. The other chip conditions are unchanged, so a local dominator
would be an original dominator.

For the cut increase, put $\ell=\lambda_{\widehat D}(S,T_1)$.
A minimum real $S$-$T_{X_1}^\star$ cut $F$ exists, since $Z$ supplies one.
It is also an $S$-$T_1$ cut, as every $S$-$T_1$ path reaches a real exit
tail of $X_1$. If $|F|\le\ell$, equality holds. Its source-reachable side
lies in $X_1$ and avoids $T_{X_1}^\star$, a non-empty subset of $X_1$
because $\ell\ge1$. This gives a compact source side properly contained
in $X_1$, contradicting the choice of $X_1$.
\end{proof}

\begin{proof}[Proof of \Cref{lem:routedFinalNestedWithin}]
Since $X_q$ is auxiliary-respecting, no auxiliary arc leaves $X_q$. Since
$X_q\subseteq Z$ and 
$\partial_{\mathrm{real}}^+(Z)\cap
\partial_{\mathrm{real}}^+(X_q)=\emptyset$, 
the head of every real arc leaving $X_q$ belongs to $Z$. Hence
$S_{X_q}^\star\subseteq Z$. Every vertex of $S_{X_q}^\star$ is reachable from $S$ inside
$\widehat D[Z]$, and $Z$ is reachable from $S_{X_q}^\star$ inside
$\widehat D[Z]$. Since the target remains $T_1$, strictness follows from
\Cref{obs:routedStrictnessCriterion}. The cut bound, the condition
$\widehat D_{\mathrm{real}}[Z]\models\pi$, the forbidden-set condition, and
the request condition are inherited from the original chip. Thus $Z$ is an
$(S_{X_q}^\star,T_1,T_2,Q,\pi,k)$-chip.

Suppose that there is an
$
(S_{X_q}^\star,T_1,T_2,Q,\pi,k)
$-chip $Z'$ such that $Z\subset Z'$ and
\[
|\partial_{\mathrm{real}}^+(Z')|
\le
|\partial_{\mathrm{real}}^+(Z)|.
\]
Since every vertex of $S_{X_q}^\star$ is reachable from $S$ inside
$\widehat D[Z]$, the set $Z'$ is reachable from $S$ inside
$\widehat D[Z']$. The remaining chip conditions with source $S$ are
inherited from the local chip, again using
\Cref{obs:routedStrictnessCriterion} for strictness. Hence $Z'$ is an
original $(S,T_1,T_2,Q,\pi,k)$-chip that dominates $Z$, contradicting the
importance of $Z$. Therefore $Z$ is important in the local instance.

It remains to prove the cut increase. Let
\[
\ell\coloneqq \lambda_{\widehat D}(S,T_1).
\]
Suppose, for a contradiction, that
\[
\lambda_{\widehat D}(S_{X_q}^\star,T_1)\le \ell .
\]
Since $S\subseteq S_{X_q}^\star$, every $S_{X_q}^\star$-$T_1$ cut is also
an $S$-$T_1$ cut. Hence equality holds. Let $F$ be an inclusion-wise
minimal $S_{X_q}^\star$-$T_1$ cut of size $\ell$, and let $Y$ be the set of
vertices reachable from $S_{X_q}^\star$ in $\widehat D-F$. Then
\[
F=\partial_{\mathrm{real}}^+(Y).
\]
Since $F$ is a minimum $S$-$T_1$ cut, the definition of the minimum cut
sequence gives
\[
F\subseteq \bigcup_{i=1}^q \partial_{\mathrm{real}}^+(X_i).
\]
Let $a=(u,v)\in F$. Then
$a\in\partial_{\mathrm{real}}^+(X_i)$ for some $i\in[q]$. Since
$X_i\subseteq X_q\subseteq Y$, we have $u\in X_q$. If $v\in X_q$, then
$v\in Y$, contradicting $a\in\partial_{\mathrm{real}}^+(Y)$. If
$v\notin X_q$, then
\[
v\in \head(\partial_{\mathrm{real}}^+(X_q))
\subseteq S_{X_q}^\star
\subseteq Y,
\]
again contradicting $a\in\partial_{\mathrm{real}}^+(Y)$. This contradiction
shows that
\[
\lambda_{\widehat D}(S_{X_q}^\star,T_1)>
\lambda_{\widehat D}(S,T_1).\qedhere
\]
\end{proof}

\subsection{Gluing and the attachment boundary}
\label{app:crossingReductionDetails}

This subsection contains the proofs of the elementary observations used in
the crossing reduction.

\begin{observation}\label{obs:splitAlongGlue}
Let $H$ be a digraph, let $(X,Y)$ be a separation of the undirected graph
underlying $H$, and let $B\coloneqq X\cap Y$. View $H[X]$ and $H[Y]$ as
boundaried digraphs with the same labeled boundary $B$. Then
$H=H[X]\oplus H[Y]$,
where the gluing takes place along $B$.
\end{observation}

\begin{proof}[Proof of \Cref{obs:splitAlongGlue}]
Let $H_X\coloneqq H[X]$ be the boundaried digraph with boundary $B$ and an
arbitrarily chosen injective labeling $\beta$ whose image is $[|B|]$.
Similarly, let $H_Y\coloneqq H[Y]$ be the boundaried digraph with boundary
$B$ and the same labeling $\beta$.

Since $(X,Y)$ is a separation of the underlying undirected graph of $H$,
there is no arc between $X\setminus Y$ and $Y\setminus X$ in either
direction. Hence every arc of $H$ belongs to $H[X]$ or to $H[Y]$.
Therefore $H_X\oplus H_Y$ is isomorphic to $H$.
\end{proof}

\begin{observation}[Attachment decomposition]\label{obs:attachmentDecomposition}
Let $\widehat D$ be a routed digraph, and let $X,Z\subseteq V(\widehat D)$.
Let $B\coloneqq B_X(Z)$ and let
\[
L\coloneqq L_X(Z)=(X\cap Z)\cup B .
\]
Then
\[
\widehat D_{\mathrm{real}}[Z]
=
\widehat D_{\mathrm{real}}[L]
\oplus
\widehat D_{\mathrm{real}}[Z\setminus X],
\]
where the gluing takes place along $B$.
\end{observation}

\begin{proof}[Proof of \Cref{obs:attachmentDecomposition}]
The vertex sets $L_X(Z)$ and $Z\setminus X$ cover $Z$, and their
intersection is exactly $B_X(Z)$. By the definition of $B_X(Z)$, there is
no real arc, in either direction, between
\[
L_X(Z)\setminus B_X(Z)=X\cap Z
\]
and
\[
(Z\setminus X)\setminus B_X(Z).
\]
Equivalently, $(L_X(Z),Z\setminus X)$ is a separation of the underlying
undirected graph of $\widehat D_{\mathrm{real}}[Z]$. The claim follows from
\Cref{obs:splitAlongGlue} applied to
$H=\widehat D_{\mathrm{real}}[Z]$.
\end{proof}

\begin{proof}[Proof of \Cref{obs:boundedAttachmentBoundary}]
Every vertex of $B_X(Z)$ lies in $Z\setminus X$ and has a real adjacency to
a vertex of $X\cap Z$. Hence
\[
B_X(Z)\subseteq
\head(\partial^+_{\mathrm{real}}(X))
\cup
\tail(\partial^-_{\mathrm{real}}(X)).
\]
Both sets on the right-hand side lie in $\outf_{\widehat D_{\mathrm{real}}}(X)$,
which proves the first claim. Therefore
\[
|B_X(Z)|
\le
|\partial^+_{\mathrm{real}}(X)|
+
|\partial^-_{\mathrm{real}}(X)|.
\]
Since the real part of $\widehat D$ is $\alpha$-balanced,
\[
|\partial^-_{\mathrm{real}}(X)|
\le
|\partial^+_{\mathrm{real}}(X)|+\alpha .
\]
Using $|\partial^+_{\mathrm{real}}(X)|\le k$, we obtain
\[
|B_X(Z)|\le 2k+\alpha .
\]
\end{proof}

\subsection{Properties of the reduced instance}
\label{app:attachment-proofs}

\begin{proof}[Proof of \Cref{obs:basicReducedData}]
The first item follows directly from $S\subseteq X\cap Z$ and
\[
S_X(Z)=S\cup V(R)\subseteq (X\cap Z)\cup V(R)=C_X(Z).
\]

We prove the second item. By \Cref{def:safeBlockedExits}, both
$\safeExit{Z}{X}$ and $\blockedExit{Z}{X}$ are subsets of $X\setminus Z$.
By \Cref{def:contactVertices}, also $\contact{Z}{X}\subseteq X\setminus Z$.
Hence these three sets are disjoint from
\[
C_X(Z)=(X\cap Z)\cup V(R),
\]
since $V(R)$ consists of the representative vertices, with boundary
$B\subseteq Z\setminus X$.

It remains only to consider $\tail(A_{\mathrm{aux}}')$ and $T_2\cap X$.
The set $T_2\cap X$ is disjoint from $X\cap Z$ because $Z$ is an
$\sttqpk$-chip, and it is disjoint from $V(R)$ because
$B\subseteq Z\setminus X$ and the remaining vertices of $R$ are new
representative vertices. Thus $(T_2\cap X)\cap C_X(Z)=\emptyset$.

Now let $(u,v)\in A_{\mathrm{aux}}'$. If $(u,v)$ is an old auxiliary arc
kept from $\widehat D$, then $u\in T_2\cap X$ by guardedness. If
$(u,v)\in\Gamma_X(Z)$, then $u\in X\setminus Z$ by definition. In both
cases $u\notin C_X(Z)$. Therefore
$\tail(A_{\mathrm{aux}}')\cap C_X(Z)=\emptyset$, and the second item
follows.

For the third item, since $Z$ is an $\sttqpk$-chip, no vertex of $Q$ lies
in $Z$. Hence $Q\cap X\subseteq X\setminus Z$. Every vertex in
\[
\entry{Z}{X}{T_1}{Q\setminus X}
\cup
\entry{Z}{X}{T_1}{\head(\partial_{\mathrm{real}}^+(Z))\setminus X}
\]
is, by definition, an entry vertex into $X$ on a path in $\widehat D-Z$.
Thus it lies in $X$ and does not lie in $Z$. Therefore
$Q_X(Z)\subseteq X\setminus Z$.

The fourth item follows from the fact proved for the second item that
no auxiliary arc has its tail in $C_X(Z)$.
\end{proof}

\begin{proof}[Proof of \Cref{obs:reducedChipRealSubgraph}]
Put $B\coloneqq B_X(Z)$. By definition, the real part of the reduced graph is
\[
D_{\mathrm{real}}'
=
\bigl(\widehat D_{\mathrm{real}}[X\cup B]-A_B\bigr)\oplus R.
\]
The reduced chip is
\[
C_X(Z)=(X\cap Z)\cup V(R).
\]
After gluing, the boundary vertices of $R$ are identified with the vertices
of $B$. Hence the original side of the induced real graph on $C_X(Z)$ is
precisely the subgraph induced on
\[
(X\cap Z)\cup B=L_X(Z),
\]
with the boundary-internal arcs $A_B$ removed. Therefore
\[
\widehat D_X(Z)_{\mathrm{real}}[C_X(Z)]
=
\bigl(\widehat D_{\mathrm{real}}[L_X(Z)]-A_B\bigr)\oplus R.
\]

It remains to check the CMSO condition. Let
\[
D_1\coloneqq \widehat D_{\mathrm{real}}[L_X(Z)]-A_B
\qquad\text{and}\qquad
D_2\coloneqq \widehat D_{\mathrm{real}}[Z\setminus X].
\]
The graph $D_1$ is boundary-clean, in the sense that it has no arcs with
both endpoints in the boundary $B$. By \Cref{obs:attachmentDecomposition}
and by the definition of the gluing operator,
\[
\widehat D_{\mathrm{real}}[Z]=D_1\oplus D_2.
\]
Indeed, the only arcs removed from
$\widehat D_{\mathrm{real}}[L_X(Z)]$ have both endpoints in $B$, and these
arcs are present in $D_2$.

Now assume $\widehat D_{\mathrm{real}}[Z]\models\pi$. Then
$D_1\oplus D_2\models\pi$. Since $R$ is the chosen
$\equiv_\pi$-representative of $D_2$ with the same labeled boundary $B$,
replacing $D_2$ by $R$ in the boundary-clean context $D_1$ preserves the
truth of $\pi$. Hence $D_1\oplus R\models\pi$ and by the first part of the proof, this is the same as $\widehat D_X(Z)_{\mathrm{real}}[C_X(Z)]\models\pi$.
\end{proof}

\begin{proof}[Proof of \Cref{obs:shortcutEndpointsRoutedInterface}]
A shortcut $(u,v)$ leaves $X$ through a real arc, since $X$ is
auxiliary-respecting, so $u\in\intf_{\widehat D_{\mathrm{real}}}(X)$.
Its return arc is either real, placing $v$ in the same interface, or
auxiliary, placing $v$ in $X\cap\auxv(\widehat D)$. Thus both endpoints
lie in $\rintf_{\widehat D}(X)$.

Exits and contacts are incident with real arcs crossing $X$, so they lie
in the real interface. The two sets of entry vertices come from paths
starting
outside $X$. \Cref{obs:routedPrefPath} puts their first entries
in the routed interface as well.
\end{proof}

\subsection{\texorpdfstring{\arcDoct{} NP-hardness details}{Arc DOCT NP-hardness details}}
\label{app:doct-hardness-proofs}

We obtain the NP-hardness in two steps. The first is a simple translation from multidigraphs to simple digraphs, which allows us to produce a multidigraph in our main reduction (see \Cref{thm:arcDoctNPComplete}). 

\begin{lemma}\label{lem:multiToSimpleDOCT}
There is a polynomial-time parameter-preserving reduction from
\textsc{Arc DOCT} on Eulerian multidigraphs to \textsc{Arc DOCT} on simple
Eulerian digraphs.
\end{lemma}

\begin{proof}[Proof of \Cref{lem:multiToSimpleDOCT}]
Let $(D,k)$ be an instance where $D$ is an Eulerian multidigraph. We build a
digraph $D'$ by replacing every arc $a=(u,v)$ of $D$ by a directed
path $u,x_a,y_a,v$ 
of length three, where $x_a$ and $y_a$ are new vertices used only for this
arc. The digraph $D'$ is simple by construction. Moreover, $D'$ is Eulerian since each old vertex
has the same indegree and outdegree as in $D$, and each new subdivision vertex has indegree and outdegree one.

Since every new vertex has indegree and outdegree one, the directed cycles of $D'$ are exactly the cycles obtained from directed cycles of $D$ by replacing each arc $a$ by its path. This replacement triples the length of a cycle and so preserves its parity. If $F$ is an arc directed odd cycle transversal of $D$, then $\{(u,x_a)\mid a=(u,v)\in F\}$ is an arc directed odd cycle transversal of $D'$ of the same size. Conversely, if $F'$ is an arc directed odd cycle transversal of $D'$, then the set of arcs $a$ of $D$ whose path contains an arc of $F'$ is an arc directed odd cycle transversal of $D$ of size at most $|F'|$. Hence $(D,k)$ is a yes-instance if and only if $(D',k)$ is a yes-instance. 
\end{proof}

\begin{theorem}\label{thm:arcDoctNPComplete}
{\arcDoct} is NP-complete on Eulerian digraphs.
\end{theorem}

\begin{proof}[Proof of \Cref{thm:arcDoctNPComplete}]
Membership in $\np$ follows because, after deleting a guessed arc set, one
can test in polynomial time whether any strongly connected component has
non-bipartite underlying undirected graph. This test is correct because a strongly connected digraph contains a directed odd cycle if and only if its underlying undirected graph is not bipartite.

We prove hardness by reduction from \textsc{Multiway Cut}. The input is an
undirected graph $G$, a terminal set $T$, and an integer $k'$. The question
is whether there is an edge set $S'\subseteq E(G)$ of size at most $k'$ such
that $G-S'$ contains no path between two distinct terminals. We use that
\textsc{Multiway Cut} is NP-hard already for $|T|=3$.

Let $T=\{t_1,\ldots,t_m\}$, where $m\ge 3$, and fix the cyclic order
  $t_1,t_2,\ldots,t_m,t_1$.
Indices are read modulo $m$, so $t_{m+1}=t_1$.

We construct an Eulerian multidigraph $D$ as the union of two multidigraphs
$D_1$ and $D_2$ that share exactly the terminal set $T$.

First construct $D_1$. Start with the vertex set $V(G)$. For every edge
$uv\in E(G)$, add a new vertex $x_{uv}$ and the four arcs
\[
  (u,x_{uv}),\quad
  (x_{uv},u),\quad
  (v,x_{uv}),\quad
  (x_{uv},v).
\]
Thus every edge of $G$ becomes a bidirected path of length two through its
private subdivision vertex.

Next construct $D_2$. Start with the terminal vertices $T$. For each
$i\in[m]$, add a new vertex $y_i$. Add $k'+1$ parallel copies of each of the
three arcs
\[
  (t_i,t_{i+1}),\qquad
  (t_i,y_i),\qquad
  (y_i,t_{i+1}).
\]
Set
\[
  k=3k'+2 .
\]

We first verify that $D$ is Eulerian. In $D_1$, every old vertex $v\in V(G)$
gets one incoming and one outgoing arc for every edge incident with $v$ in
$G$, and every subdivision vertex $x_{uv}$ has indegree two and outdegree
two. In $D_2$, every terminal $t_i$ has $2(k'+1)$ outgoing arcs, namely the
copies of $(t_i,t_{i+1})$ and $(t_i,y_i)$, and $2(k'+1)$ incoming arcs, namely the
copies of $(t_{i-1},t_i)$ and $(y_{i-1},t_i)$. Each vertex $y_i$ has indegree
$k'+1$ and outdegree $k'+1$. Since $D_1$ and $D_2$ share only terminals and
both are balanced at every terminal, their union $D$ is Eulerian.

For each $i\in[m]$, let
\[
  A_i=(t_i,t_{i+1}),\qquad
  B_i=(t_i,y_i),\qquad
  C_i=(y_i,t_{i+1})
\]
denote the three route types between $t_i$ and $t_{i+1}$, where each route
type consists of $k'+1$ parallel copies of the corresponding arc. Thus the
direct route from $t_i$ to $t_{i+1}$ uses $A_i$, while the subdivided route
uses $B_i$ followed by $C_i$.

 Let $F$ be an inclusion-wise minimal arc directed odd cycle transversal of $D$ of size at most $k$. 
 We first show that we can make two assumptions without loss of generality.
 
 For
every route type in $D_2$, the set $F$ either deletes all copies of that
route type or deletes none of them. Indeed, if at least one copy of a route
type remains, then restoring any other deleted copy of the same route type
cannot create a new directed odd cycle. Any directed odd cycle using the
restored copy could use one of the copies that was already present instead.
This contradicts inclusion-wise minimality.

Next,  suppose $F$ deletes the route type
$B_i=(t_i,y_i)$. Remove all $k'+1$ copies of $B_i$ from $F$, and add all
$k'+1$ copies of $C_i=(y_i,t_{i+1})$. This does not increase the size of
$F$. It also preserves the property of being an arc directed odd cycle
transversal. Indeed, in the whole constructed multidigraph, the only arcs
entering $y_i$ are the copies of $B_i$, and the only arcs leaving $y_i$ are
the copies of $C_i$. Hence every directed cycle using a copy of $B_i$ also
uses a copy of $C_i$, and every directed cycle using a copy of $C_i$ also
uses a copy of $B_i$. Therefore, after replacing the deletion of $B_i$ by
the deletion of $C_i$, every directed odd cycle is still hit. Repeating this
operation for all $i$, we may assume that no route type $B_i$ is deleted.
This operation preserves the property that each route type is either deleted
completely or not deleted at all.

\begin{claim}\label{lem:d2_works}
If $D$ has an arc directed odd cycle transversal of size at most $k$, then
it has one, say $F$, such that $F\cap A(D_2)$ consists exactly of the two
route types entering some terminal $t_i$. Equivalently, for some $i$, the
set $F\cap A(D_2)$ consists of all copies of
  $A_{i-1}=(t_{i-1},t_i)$
and 
  $C_{i-1}=(y_{i-1},t_i)$.
In particular, $|F\cap A(D_1)|\le k'$ and 
  $|F\cap A(D_2)|=2(k'+1)=2k'+2$.  
\end{claim}

\begin{claimproof}
Start with a solution of size at most $k$, make it inclusion-wise minimal,
and apply the two normalization steps above. Thus, in $D_2$, the solution
deletes only whole route types, and it deletes no route type of the form
$B_i=(t_i,y_i)$.

Since each deleted route type has size $k'+1$ and
  $k=3k'+2<3(k'+1)$, 
the solution deletes at most two route types in $D_2$.

We show that the deleted route types must be $A_i$ and $C_i$ for one common
index $i$. Suppose not. Since no $B_i$ is deleted, the direct route from
$t_i$ to $t_{i+1}$ remains available exactly when $A_i$ is not deleted, and
the subdivided route remains available exactly when $C_i$ is not deleted.
Hence both routes from $t_i$ to $t_{i+1}$ are destroyed only if both $A_i$
and $C_i$ are deleted. By assumption this does not happen for any $i$.
Therefore, for every $i$, at least one directed route from $t_i$ to
$t_{i+1}$ remains in $D_2-F$. Thus $D_2-F$ contains a directed closed walk
going once around the terminal cycle.

Moreover, since $m\ge 3$ and at most two route types are deleted, there is
some index $j$ for which neither $A_j$ nor $C_j$ is deleted. For this
segment, both the direct route
  $t_j\to t_{j+1}$ 
and the subdivided route
  $t_j\to y_j\to t_{j+1}$ 
remain available. Switching between these two routes changes the length of
the closed terminal walk by one, and hence changes its parity. Thus one of
the two choices gives a directed closed walk of odd length in $D_2-F$.

Every directed closed walk of odd length contains a directed odd cycle. This
contradicts the assumption that $F$ is an arc directed odd cycle
transversal. Therefore, for some index $i$, the solution deletes both $A_i$
and $C_i$. Since at most two route types are deleted in $D_2$, these are
exactly the deleted route types in $D_2$. They are precisely the two route
types entering $t_{i+1}$. Renaming $i+1$ as the terminal index gives the
claimed form.

These two route types cost $2(k'+1)=2k'+2$ arcs. Since the total budget is
$k=3k'+2$, at most $k'$ arcs remain for $D_1$.
\end{claimproof}

We now prove equivalence of the two instances.

\medskip
\noindent
$\Rightarrow$.
Assume first that $(G,T,k')$ is a yes-instance. Let
$S'\subseteq E(G)$ be a multiway cut of size at most $k'$. Consider the
connected components of $G-S'$. The component containing $t_i$ receives rank
$i$. Components containing no terminal receive an arbitrary rank in $[m]$.

We construct an arc set $F$. First delete from $D_2$ the two route types
entering $t_1$, namely all copies of $A_m=(t_m,t_1)$ and all copies of
$C_m=(y_m,t_1)$. This costs $2k'+2$ arcs. Next, for every edge $uv\in S'$,
let $C_u$ and $C_v$ be the components of $G-S'$ containing $u$ and $v$. If
$\operatorname{rank}(C_u)>\operatorname{rank}(C_v)$, delete the arc
$(u,x_{uv})$. If
$\operatorname{rank}(C_v)>\operatorname{rank}(C_u)$, delete the arc
$(v,x_{uv})$. If the two ranks are equal, delete either $(u,x_{uv})$ or
$(v,x_{uv})$. This adds at most $|S'|\le k'$ arcs, so $|F|\le 3k'+2=k$.

We claim that $D-F$ has no directed odd cycle. The graph $D_1-F$ is a
subgraph of the bipartite graph with bipartition given by the old vertices
and the subdivision vertices $x_{uv}$. Every directed cycle of $D_1-F$ is a closed walk in this bipartite graph and therefore has even length. On the other hand, the graph $D_2-F$ is acyclic. After deleting the route types
entering $t_1$, all remaining arcs of $D_2$ go forward along
  $t_1,t_2,\ldots,t_m$ 
possibly through the corresponding vertex $y_i$, and no arc returns to
$t_1$.

It remains to exclude directed odd cycles using arcs of both $D_1$ and
$D_2$. In $D_2-F$, every directed path between terminals goes from $t_i$ to
$t_j$ with $i<j$. In $D_1-F$, ranks never decrease along a directed path.
Indeed, inside a component of $G-S'$ all arcs of $D_1$ remain available in
both directions, while across an edge of $S'$ the remaining direction, if
any, goes from the endpoint in the lower-ranked component to the endpoint in
the higher-ranked component. Therefore $D_1-F$ contains no directed path
from $t_j$ to $t_i$ with $i<j$. A directed cycle that uses a non-empty path
inside $D_2-F$ would have to increase the terminal index and later return to
a smaller one through $D_1-F$, which is impossible. Thus $F$ is an arc
directed odd cycle transversal of $D$.

\medskip
\noindent
$\Leftarrow$.
Assume conversely that $D$ has an arc directed odd cycle transversal of size
at most $k$. By \Cref{lem:d2_works}, choose such a transversal $F$ so that
$F\cap A(D_2)$ consists exactly of the two route types entering one
terminal. Relabel the cyclic order so that this terminal is $t_1$. Thus
$F\cap A(D_2)$ deletes all copies of $A_m=(t_m,t_1)$ and all copies of
$C_m=(y_m,t_1)$, and
  $|F\cap A(D_1)|\le k'$.

We first show that, for all $1\le i<j\le m$, there is no directed path from
$t_j$ to $t_i$ in $D_1-F$. Suppose such a path $P$ exists. In $D_2-F$, all
route types on the segments from $t_i$ to $t_j$ remain available. Hence
there is a directed $t_i$-$t_j$ path in $D_2-F$, and its parity can be
chosen freely by switching one segment between the direct route of length
one and the subdivided route of length two. Choose its parity opposite to
the parity of $P$. Together with $P$, this gives a directed closed walk of
odd length in $D-F$, and hence a directed odd cycle in $D-F$. This
contradicts the choice of $F$.

Define
\[
  S'=
  \{uv\in E(G)\mid
  F\text{ contains at least one arc incident with }x_{uv}\}.
\]
Then $|S'|\le |F\cap A(D_1)|\le k'$. We claim that $S'$ is a multiway cut
in $G$. Suppose not. Then $G-S'$ contains a path between two distinct
terminals, say $t_i$ and $t_j$ with $i<j$. Since no edge of this path lies
in $S'$, all arcs through the corresponding subdivision vertices remain in
$D_1-F$. Thus this path lifts to a directed $t_j$-$t_i$ path in $D_1-F$.
This contradicts the previous paragraph. Therefore $S'$ is a multiway cut
of size at most $k'$.

We have proved NP-hardness for Eulerian multidigraphs. Applying
\Cref{lem:multiToSimpleDOCT} gives NP-hardness for simple Eulerian digraphs.
\end{proof}

\section*{Acknowledgements}
This work was supported by UKRI EPSRC grant (EP/V044621/1). We thank anonymous reviewers for helpful suggestions on improving the exposition of the paper. 

\section*{AI Methodology}
The technical contributions of this work were developed independently of generative AI.
Claude Opus 4.8 and GPT-5.6 were used for author-directed assistance with proofreading, exposition, LaTeX formatting, figures, and revisions of the full and proceedings versions. All such use was closely supervised.


\bibliographystyle{alpha}
\bibliography{main}

\end{document}